\documentclass[seceq]{ptptex}
\usepackage{subeqn}
\usepackage{graphicx}
\usepackage{psfrag}
\usepackage{amsthm}

\usepackage{syntonly}
\usepackage{amssymb}

\def\e{{\rm e}}

\theoremstyle{definition}

\newtheorem{theorem}{Theorem}[section]

\newtheorem{corollary}[theorem]{Corollary}

\newtheorem{proposition}[theorem]{Proposition}

\newtheorem{lemma}[theorem]{Lemma}

\def\eq{\hspace{0mm}&=&\hspace{0mm}}
\def\espace{\hspace{0mm}&&\hspace{0mm}}
\def\eequiv{\hspace{0mm}&\equiv&\hspace{0mm}}
\def\eesim{\hspace{0mm}&\simeq&\hspace{0mm}}

\newcommand{\vt}[1]{\mbox{\boldmath$#1$}}
\def\vecu{\mbox{\boldmath$u$}}
\def\vecq{\mbox{\boldmath$q$}}

\def\hf{\frac{1}{2}}

\def\bseq{\begin{subequations}} \def\eseq{\end{subequations}}
\def\bea{\begin{eqnarray}} \def\eea{\end{eqnarray}}
\def\bsea{\begin{subeqnarray}} \def\esea{\end{subeqnarray}}
\let\nn=\nonumber
\def\beann{\begin{eqnarray*}} \def\eeann{\end{eqnarray*}}

\let\a=\alpha   \let\de=\delta
 
\let\z=\zeta 
 
  \let\la=\lambda

\def\Thet{{\rm \Theta}}

 \let\Ds=\displaystyle

\newcommand{\dsum}[2]{\sum_{\stackrel{\scriptstyle #1}{#2}}}
\newcommand{\dprod}[2]{\prod_{\stackrel{\scriptstyle #1}{#2}}}

\def\0{\over } \def\1{\vec }     \def\2{{1\over2}} \def\4{{1\over4}}
\def\5{\bar }  \def\6{\partial } \def\7#1{{#1}\llap{/}}

\def\<{\langle } \def\>{\rangle }

\let\ti=\tilde

\def\i{{\rm i}}

\def\d{{\rm d}}
\def\e{{\rm e}}

 \def\sech{\mbox{\,sech}}

\pubinfo{2004}
\title{
\bf 
$N$-Soliton Collision in the Manakov Model}

\author{
Takayuki \textsc{Tsuchida}\footnote{E-mail: tsuchida@poisson.ms.u-tokyo.ac.jp}
}
\inst{
Graduate School of Mathematical Sciences, University of Tokyo, 
\\
Tokyo 153-8914, Japan
}

\recdate{March 14, 2003}
\abst{
We investigate soliton collisions in 
the Manakov model, 
which is a 
system of coupled nonlinear Schr\"{o}dinger equations 
that is 
integrable via the inverse scattering method. 
Computing the 
asymptotic forms of the general $N$-soliton solution 
in the limits $t \to \mp \infty$, 
we elucidate a mechanism 
that factorizes 
an $N$-soliton collision into 
a nonlinear superposition of 
$N \choose 2$ 
pair collisions 
with arbitrary order. 
This removes the misunderstanding 
that multi-particle 
effects exist in the Manakov model 
and provides 
a new 
``set-theoretical'' 
solution to the 
quantum Yang-Baxter equation. 
As a by-product, we also obtain a new nontrivial relation 
among determinants and extended determinants. 
}
\begin{document}

\maketitle

\section{Introduction}
\setcounter{equation}{0}
\label{}

In recent years, there has been 
%increasing 
a surge of interest in 
%various 
some systems of coupled nonlinear Schr\"{o}dinger (coupled NLS)
%-type systems. 
%recently. 
equations 
%owing to 
%for 
because of 
their relevance 
%wide applicability 
%Systems of coupled NLS equations 
%They are encountered in various branches of physics such as 
in nonlinear optics.\cite{Meny,Agra,Akhm} \ Among 
%deep water waves \cite{}, and ... 
%have attracted increasing interest in recent years. 
%In 
%those 
such systems, this paper 
%The main aim of this poster is to make clear 
focuses on 
%an integrable 
the following system of coupled NLS equations: 
%a system of coupled NLS equations, 
%namely, the following system (the Manakov model?): 
%which is known to be completely integrable: 
%investigate soliton collisions in the system of coupled NLS equations 
%(for short, Manakov model \cite{Manakov}): 
%, for short):
\begin{equation}
%\begin{array}{l}
%\displaystyle
 {\rm i}  \vt{q}_{t} + \vt{q}_{xx} 
  + 2 |\!|\vt{q}|\!|^2 \vt{q} = \vt{0},
%_{\vphantom \int},
%\\
%\displaystyle
\hspace{5mm} \vt{q} = (q_1, q_2, \cdots, q_m).
%\end{array}
\label{cNLS}
\end{equation}
Here $|\!|\vt{q}|\!|^2 
\equiv \vecq \cdot \vecq^\dagger = \sum_{j=1}^m |q_j|^2$, 
%Throughout this paper, we use 
%${}^\dagger$ 
%and 
where the superscript $\dagger$ denotes Hermitian conjugation. 
The subscripts $t$ and $x$ denote the partial 
differentiation with respect to these variables. 
%differentiations by $t$ and $x$, respectively. 
It is 
%already 
well known that (\ref{cNLS}) is a completely integrable 
system.\cite{Manakov,Makhankov} \ We call (\ref{cNLS}) the Manakov model, 
%Henceforce 
since the two-component ($m=2$) case of (\ref{cNLS}) was solved 
for the first time by Manakov\cite{Manakov} 
%via 
using the inverse scattering method (ISM). 
The extension of the ISM to the general $m$-component 
%more than two-component 
%a general $m$ 
case is 
%almost 
straightforward.\cite{Makhankov} \ Nevertheless, 
%at the level of the ISM 
%the Lax representation or 
%inverse scattering method. 
%However, 
%if 
% concretely, 
%and their asymptotic behavior, 
%concrete behaviors of solitons, 
%solutions, 
the value of $m$ is extremely important 
when we 
%investigate 
consider 
soliton solutions. 
%and 
The $m=2$ case is 
%straightforwardly 
%obviously 
%obtained simply as 
merely 
a special case of the general $m$ 
%-component 
case. 
%more than two 
%Indeed, 
%Moreover, from a mathematical point of view, 
The most 
interesting 
%the only essential
%ly 
%important 
%case is such 
is the case 
%where 
in which the total number of 
solitons, say $N$, is equal to the number of components, 
%of $\vt{q}$, i.e.\ 
$m$. 
Indeed, in the $N=m$ case, 
the coefficient vectors for 
%each 
the 
%sech-
hyperbolic-secant-type envelope of 
%each 
solitons [\hspace{1pt}see, 
e.g., 
%{\it e.g.\ }, 
$\vt{u}_1$ in (\ref{pol})], 
%for the 
%with 
%profile 
%of each soliton, 
%solitons 
which we refer to 
%call 
after normalization as 
%the 
{\it polarization vectors}, 
are just 
sufficient 
%enough 
to span the vector space 
${\mathbb C}^m$, in which $\vt{q}$ exists. 
%lives. 
%In considering multi-soliton solutions, 
%all 
Soliton solutions in the other cases ($N > m$ or $N < m$) 
are 
%contained 
obtained from those in this case through a special choice of 
soliton parameters 
%a reduction 
or the operation of a unitary transformation. 
%obtained through either a special case or of this 
%takes its value. 
%in which $\vt{q}$ takes its value. 
%seems to be important. 
%Therefore, 
In this paper, 
%For this reason, 
%we do not specify the value of $m$ and 
%discuss 
we consider 
%deal with 
the most general case, in which 
%where 
%the total number of solitons, 
$N$ and $m$ 
are arbitrary positive integers. 
% so that 
%to include this 
Then, the most interesting case, $N=m$, 
%case 
is automatically included. 
%as a simple reduction, 
%is an arbitrary positive integer and 
%where the total number of solitons, 
%, say 
%$N$, is also arbitrary. 
%The most interesting case is straightforwardly 
%in this paper. 
%and 
%assume that 
%Thus $m$ is 
%

Although the integrability of the Manakov model (\ref{cNLS})
has been 
established formally through application of 
%via 
%within the framework of 
the 
%inverse scattering method 
ISM, 
% or its variants, 
%extensions, 
%concrete behavior of the multi-soliton solutions 
multi-soliton dynamics in the model 
%is not 
%well 
%fully understood and 
%still remains 
remain to be clarified. 
%We infer that 
There are 
%We have (at least) 
two reasons for this. 
One reason 
%of the reasons 
%for this comes from 
is 
%that, owing to 
that 
the vector nature of (\ref{cNLS}), which supports 
%(admits?) 
%a rich 
the internal degrees of 
%internal 
freedom 
%(structure?) 
of solitons, leads to complicated behavior. 
%As a consequence, 
Indeed, even a 
%collision laws of 
%scattering rule of 
%asymptotic 
%properties 
%behaviour 
%the 
two-soliton collision 
is highly nontrivial,\cite{Rad,Shche,Yang,Elgin} as 
%in 
%%solutions of 
%the Manakov model (\ref{cNLS}) 
%(cf.\ Theorem~\ref{two-rule}). 
%Indeed, the rule is not 
%%longer 
%``linear'' in a certain sense 
%as was in the the single-component NLS equation \cite{}. 
%Indeed, 
%it not only displaces 
not only does a displacement of 
the soliton centers 
%in 
%%nontrivial 
%dependence on 
%depending nontrivially 
%is detemined 
depend on the initial polarization vectors 
but also 
%changes 
the polarization vectors themselves are changed. 
%change as a result of the two-soliton collision. 
%are 
%not invariant under 
%changed by 
%rotate 
%Those changes 
%final polarization vectors are 
%rotation 
%change 
%of the polarization vectors 
%
%in a nontrivial way. 
%determined 
%nontrivially 
%by their initial 
%dependent on 
%In the multi-soliton case, 
%while the polarization vectors play a role in the collision laws later. 
%which 
%this leads to a change of the results of later two-soliton collisions. 
%influences 
%changes the effect of later 
%in time. 
%In other words, 
%it is clear that 
Therefore, in this model, 
%Hence, 
%Thus, 
the effect of an $N$-soliton collision can never be 
%never be 
written 
%expressed 
as the {\it algebraic} \/sum of 
%those of 
the effects of 
%two-soliton 
pair collisions (at least, 
%seen in 
%with respect to 
%in terms of 
if we employ only 
%original 
%usual 
%the conserved physical quantities). 
the initial soliton parameters). 
%To 
%factorize 
%
%(remove the following sentence?) 
%Even if 
%%we can write it as a sum 
%something similar is possible, 
%%we can write the effect of an $N$-soliton collision as a 
%%``sum'' of two-soliton collisions in a certain order, 
%%even if possible, 
%we need to 
%%extend the notion of ``factorization'' to one 
%%including 
%consider 
%%the corresponding 
%nonlinear superposition 
%%(in other words, composition) 
%%a composite mapping) 
%of pair collisions in a certain order. 
%-wise collision mappings. 
%all the 
%Therefore, 
%According to 
%Because of 
This is quite different from the 
%single-component 
NLS case 
[\hspace{1pt}i.e.\ (\ref{cNLS}) with $m=1$], 
in which the effect of an $N$-soliton collision 
%is 
in fact can be 
written as the {\it algebraic} \/sum of 
%two-soliton 
%those 
the effects of pair collisions 
%(thus 
(in which 
the order of the pair collisions does not 
matter).\cite{ZS,Novikov,FT} \ The other reason is that 
%in arbitrary order (cf.\ the commutativity of algebraic sum). 
%which is of course regardless of order. 
%the the equation 
%\cite{}. 
%``linear'' in a certain sense 
%Owing to the nontrivial properties of the 
%collision laws of two solitons, 
%this is a challenging problem. 
%It is a challenging problem 
%which makes it 
%very tough 
%to clarify 
%investigate 
%the asymptotic behavior of 
%general multi-soliton solutions. 
%that Mankov wrote 
Manakov 
%wrote 
gave a rather misleading description 
of an $N$-soliton collision in 
%the Manakov's 
%his paper 
Ref.~\citen{Manakov}. 
%celebrated 
%Let us 
We quote the corresponding part, 
the first two sentences of the last paragraph of 
%section 2, 
\S 2 in 
%from 
%the Manakov's paper 
Ref.~\citen{Manakov}:
\vspace{1.5mm}
\begin{center}
\hspace{5mm}
\begin{minipage}{14cm}
%\begin{enumerate}
%``
Comparison of relations (17) and (18) indicates that an $N$-soliton 
collision does not, in general, reduce to a pair collision. 
This is clear, for example, from the fact that the expression 
for $\vt{S}_k^+$ contains $\vt{S}_j^+$ with $j>k$, which depend on 
the initial parameters of all the remaining solitons.
%'' 
%In the case when all the solitons have identical po
%\end{enumerate}
\end{minipage}
\hspace{5mm}
($\clubsuit$)
\end{center}
\vspace{1.5mm}
%
%Here 
The relations 
%equations 
(17) referred to here 
%in \cite{Manakov} 
are given by 
\bea
\vt{S}_N^+ 
\eq 
\Biggl\{
\prod_{n<N_{\vphantom \int}} \alpha_{11}^{\; -1} (\z_N,\z_n) \Biggr\} 
\,
%\hspace{1pt}
\hat{\alpha}^{T} (\z_N, \z_1, \vt{S}_1^{-}) 
\ldots \hat{\alpha}^{T} (\z_N, \z_{N-1}, \vt{S}_{N-1}^{-}) \vt{S}_N^{-}, 
\nn \\
\vt{S}_i^+ 
\eq 
\Biggl\{
\prod_{k>i} \alpha_{11} (\z_i,\z_k) 
\Biggr\}  \Biggl\{
\prod_{n<i} \alpha_{11}^{\; -1} (\z_i, \z_n) \Biggr\} 
\, 
%\hspace{1pt}
\hat{\alpha}^{\ast} (\z_i^\ast, \z_{i+1}, \vt{S}_{i+1}^{+}) 
\ldots \hat{\alpha}^{\ast} (\z_i^\ast, \z_{N}, \vt{S}_{N}^{+}) 
\nn \\
\espace \mbox{} \cdot
\hat{\alpha}^{T} (\z_i, \z_1, \vt{S}_1^{-}) 
\ldots \hat{\alpha}^{T} (\z_i, \z_{i-1}, \vt{S}_{i-1}^{-}) \vt{S}_i^{-},
\hspace{10mm} i=1, 2, \cdots, N-1,
\nn
\eea
%which are derived 
%%via an elegant method based on 
%by using deep knowledge on the inverse scattering method.~\fotenote{aa}
while 
%equations 
the relations (18) are given by 
%the $N=2$ case of (17), i.e.\
\bea
\vt{S}_{2_{\vphantom \int}}^+ 
\eq \alpha_{11}^{\; -1} 
	(\z_2,\z_1) \, \hat{\alpha}^{T} (\z_2, \z_1, \vt{S}_1^{-}) 
	\vt{S}_2^{-}, 
\nn \\
\vt{S}_1^+ 
\eq \alpha_{11} (\z_1,\z_2) \, \hat{\alpha}^{\ast} 
	(\z_1^\ast, \z_{2}, \vt{S}_{2}^{+})  \vt{S}_1^{-}.
\nn
\eea 
%
%We should note that 
%merely $N=2$ case of equations (17). 
%i.e.\

%Let us briefly 
%first 
%explain 
%summarize 
We now 
briefly explain the situation considered 
%as well as 
and the notation used in Ref.~\citen{Manakov}. 
%Equations 
The equations (17) of Ref.~\citen{Manakov} 
%give 
represent 
the solution 
%of 
% the case of 
for the collision 
%problem 
of $N$ solitons 
%(we name them 
(to which we refer as solitons-$1$, $2$, 
%$\cdots$, 
$\cdots$, $N$), 
while the equations (18) 
%give that of 
represent that for two solitons. 
% is considered 
We note that the 
equations (18) are obtained from (17) 
%simply 
by setting $N=2$. 
The 
equations (17) were 
%originally 
derived 
%obtained 
%via an elegant method based on 
through 
%using 
%deep thinking and 
%some 
%intuition based on the 
%careful 
analysis based on and intuition gained from the 
%insight
%knowledge 
%inverse scattering method.
ISM. 
%\footnote{
%%Although 
%%On the one hand, 
%The derivation of (17) 
%%is quite ingenious and seems to be correct. 
%was done in \cite{Manakov} 
%by an 
%%quite 
%ingenious method 
%%manner 
%%way
%%, which 
%and seems to be correct. 
%However, 
%%in the authors' opinion, 
%it 
%%does not 
%%seems to be 
%%very 
%is neither 
%very rigorous 
%%enough 
%nor understandable for 
%%one 
%the reader who is not familiar with 
%%does not have a deep knowledge in 
%%the inverse scattering method 
%the ISM. 
%%For this reason, in this paper, 
%This is a 
%%one 
%%main 
%reason why 
%in this paper we 
%%use 
%take 
%%more 
%a more straightforward way 
%%method 
%%approach 
%to get 
%%know 
%%obtain 
%the asymptotic behavior of solitons. 
%%$N$ 
%We will start from an exact formula for the 
%general $N$-soliton solution and 
%take the limit $t \to \mp \infty$. 
%%employ a more 
%} 
%There, it was 
It is 
assumed that a soliton 
%with 
designated by 
a larger number 
moves faster along the $x$-axis. 
%Thus, 
%Namely, 
That is, 
soliton-$i$ {\it overtakes\hspace{1pt}}\footnote{Throughout 
%\hspace{1pt}
this paper, we use the term ``overtake'' if only the relative velocity is 
positive. Thus it can 
%also 
be used 
for head-on collisions, etc.} 
solitons-$1$, $2$, $\cdots$, $i-1$ and is 
%{\it 
overtaken 
%} 
by solitons-$i+1$, $i+2$, $\cdots$, $N$ 
as time $t$ 
%goes by. 
passes from $- \infty$ to $+ \infty$. 
%Thus, as $t \to \infty$, the solitons are disc
The velocity of soliton-$i$ 
%as well as 
and its amplitude 
%is 
are determined by the complex parameter 
$\z_i$, which is time independent. 
%time-independent. 
%
%The subscripts 
%
The quantities 
$\vt{S}_i$ are column vectors with 
two complex components, corresponding to 
the choice of $m=2$ 
%case considered 
in Ref.~\citen{Manakov}. 
The vector norm $|\!|\vt{S}_i|\!| \,
\bigl[\hspace{1pt}= (\vt{S}_i^\dagger \cdot \vt{S}_i)^{\hf}\bigr]$ 
%(= (\vt{S}_i^\dagger \cdot \vt{S}_i)^{\hf})$ 
%which 
determines 
%both 
the 
%initial 
center position of soliton-$i$, while 
the normalized vector $\vt{S}_i/|\!|\vt{S}_i|\!|$ 
gives its polarization vector 
%up to 
after the operation of Hermitian 
conjugation. 
The superscripts $+$ and $-$ 
denote the final state ($t \to + \infty$) and the initial state 
($t \to - \infty$), respectively. 
$\alpha_{11}$
%(\z_i, \z_j)$ 
is a scalar function 
and $\hat{\alpha}$ 
%(\z_i, \z_j, \vt{S}$ 
is 
%a scalar function and 
a $2 \times 2$ 
%two-by-two 
matrix function. 
%dependent on 
%of 
%in their arguments. 
We do not give 
%omit 
their explicit forms, which 
%of those functions, 
%since they 
are not important in 
%dispensable to 
%not essential in 
%necessary 
the following discussion. 
%here and thus we omit them. 
%
The superscripts $T$ and $\ast$ denote 
the operations of transposition and 
complex conjugation, respectively. 

Although the assertion 
%meaning of 
($\clubsuit$) 
is 
%a little bit 
%rather 
somewhat ambiguous, 
% and some interpretations may be possible. Among others, 
the 
%a 
most natural and reasonable 
%naive 
interpretation 
seems to be 
%is 
the following:  
% one: 
%is that 
%
%The final state of each soliton depends on
%the final states of others. 
%\\
%
%Equations (17) and (18) can be interpreted as follows. 

\vspace{1.5mm}
\begin{center}
\begin{minipage}{14cm}
%\quad
Let us try to 
%factorize 
%consider a factorization of 
explain (17) 
%into pair 
%-wise 
%collisions of which 
%whose 
%laws are given 
%described 
by assuming that the $N$ solitons collide pairwise in 
accordance with (18). 
%Then, 
%In terms of 
%According to (18), 
Then, the first equation 
%for $\vt{S}_{N}^{+}$ 
in (17) can 
%read 
be understood 
as follows. 
%means that
Soliton-$N$ first overtakes soliton-$N-1$ 
%with 
in its initial state, 
i.e.\ soliton-$N-1$, which has not 
collided with other solitons. Next, soliton-$N$ overtakes 
soliton-$N-2$, which has not collided with 
other solitons.$\; \ldots \;$Finally, 
%... 
%soliton-$N$ 
it overtakes soliton-$1$, which has not collided with other solitons. 
%with its initial state.

%\indent
$\quad$
Similarly, the second equation 
%for $\vt{S}_{i}^{+}$ 
in (17) 
%is 
can be understood 
%read 
as follows. 
%means that 
Soliton-$i$ 
%first 
overtakes solitons-$i-1$, $i-2$, $\cdots$, $1$ in this order, 
%all 
none of which 
%have not 
has collided with other solitons. 
%soliton-$(i-2)$, $\cdots$, 
%then overtakes 
%and 
%soliton-$1$, and it 
%Then 
Next, soliton-$i$ 
%overtakes 
%solitons-$(i-2)$, $\cdots$, 
%%..., then overtakes soliton-
%$1$, and it 
is overtaken by solitons-$N$, $N-1$, $\cdots$, $i+1$ 
%with 
in their final states, i.e.\ those which will not collide 
with other solitons. 
%after that. 
%..., finally by soliton-

$\quad$
If we attempt 
%try 
to 
%describe 
%illustrate 
diagram these 
%the 
events, 
%this situation 
%by a diagram, 
%figure, 
we 
%straightforwardly 
%instantly 
immediately 
encounter 
%easily see 
a contradiction. 
%Meanwhile it seems that 
This {\it indicates} 
%``indicates'' 
\/that an $N$-soliton collision cannot be 
%explained by 
%understood as 
%constructed from 
described as 
a sequence 
%chain 
of pair collisions, 
%(17) cannot be factorized into pair collisions, 
since 
%because 
%Noting that 
%Meanwhile, 
the matrices $\hat{\alpha}$ for different sets of arguments 
do not commute in general. 
%the above statement 
%
\end{minipage}
\end{center}
\vspace{2.5mm}
%Of course, 
%Obviously, 
%this interpretation is not mathematically rigorous. 
%Although 
The logic 
%in 
of the above 
%This 
interpretation is not mathematically rigorous, but 
%seemingly 
it seems to be correct 
%owing to 
if the complex 
%nontrivial 
structure of (17) is taken into account. 
%enough to 
%cannot be 
%complete a mathematical proof. 
%prove 
It might also be 
possible to interpret the assertion 
($\clubsuit$) in a different way. 
In any 
%either 
case, 
it appears 
%one probably 
%%may 
%believes 
%establish 
that an $N$-soliton collision in the Manakov model (\ref{cNLS}) 
does not reduce to a pair collision, and thus 
%admits 
that 
some multi-particle effects exist 
in the Manakov model. 
%(see, e.g.\ \cite{}). 
%of the first sentence in ($\clubsuit$)
However, in fact this is not true. 

The main goal 
%purpose 
of this paper is 
%, however, 
to clear up 
%remove 
%resolve 
this misunderstanding. 
%and as well as to 
%by showing 
We 
%show 
explicitly 
demonstrate 
a mechanism that 
%which 
%that 
%by 
%explicitly 
%how 
factorizes an $N$-soliton collision in the Manakov model (\ref{cNLS}) 
%is factorized 
into a {\it nonlinear} \/superposition of 
%a sequence of maps 
pair collisions. 
%in arbitrary order. 
Here, we have used the term ``nonlinear'' 
%superposition'' 
%since 
to express the fact 
%indicate 
that 
the considered superposition is no longer additive. 
%Let us 
For 
the sake of 
definiteness, 
%clarity, 
we explain 
%here 
%should explain 
%how to consider the superposition of pair collisions 
in advance what we 
%are going to 
prove 
%by using 
in terms of the Manakov 
%above 
notation, which also 
%simultaneously 
gives the definition of {\it factorization} \/in this paper. 
%it 
%specifically 
%in the 
%%Manakov's 
%above context. 
%First, 
We first interpret 
%regard 
%rewrite 
the equations (18) 
%can be rewritten 
as forming a nonlinear mapping with two complex parameters, 
$f(\z_2, \z_1)$, which 
maps the initial state $\{\vt{S}_2^-, \vt{S}_1^- \}$ into 
the final state $\{\vt{S}_2^+, \vt{S}_1^+ \}$. 
Then, we can use the mapping $f(\z_j, \z_k)$ 
%can be used 
to evaluate in an $N$-soliton collision 
the 
%contribution 
effect of 
%is applied to 
%the 
the two-soliton collision 
%that 
whereby 
soliton-$j$ 
%with 
in a state $\vt{S}_j$ 
overtakes soliton-$k$ 
%with 
in a state $\vt{S}_k$. 
%It 
%should 
%is to be noted that 
%
%We note that 
%%$f(\z_1,\z_2)$ 
%the mapping is not symmetric with respect to interchange of 
%the subscripts $j$ and $k$. 
For a given 
%Suppose that 
%the initial state of $N$ solitons 
%$\{ \vt{S}_N^-, \vt{S}_{N-1}^-, \cdots, \vt{S}_1^- \}$ and 
%some 
order of $N \choose 2$ pair collisions, 
%is given. 
%we are 
%Then, 
%regardless of 
%whatever order was given, 
%In correspondence to 
%%the 
%order of pair collisions, 
we consider the corresponding composition 
of 
the 
$N \choose 2$ mappings: 
%of the form 
$f(\z_j, \z_k), \; 
N \ge j > k \ge 1$. 
%1 \le k < j \le N$. 
Then, regardless of the order of 
the pair 
collisions,\footnote{Here, we are referring to 
%We 
%assume that 
%mean 
the order 
%of collisions 
%in which 
%every pair collision takes place 
%between 
in which 
%two 
{\it neighboring} \/solitons collide pairwise. 
%only pairs of neighboring solitons collide. 
%consider pair collisions 
%between 
%of neighboring solitons only. 
%sions of neighboring solitons take place. 
%can collide. 
That is, unlike the $m=1$ case (scalar NLS), 
%case, 
%In other words, 
we do not 
consider 
%the case where 
%any 
virtual collisions 
%of 
%two 
between 
non-neighboring solitons.} 
%can collide 
%beyond 
%over others.} 
%soliton(s).
%which is possible for the scalar NLS. 
%Therefore, 
%In this respect, the factorization considered in this paper 
%is much restricted than that for the scalar NLS.}, 
%\hspace{1pt}\footnote{Throughout }, 
%whatever order was given, 
the composed mapping 
%it 
maps the initial state 
$\{ \vt{S}_N^-, \vt{S}_{N-1}^-, \cdots, \vt{S}_1^- \}$ 
%correctly 
exactly 
%into 
to the 
%correct 
%actual 
final state 
%of $N$ solitons, 
%i.e.\ 
$\{ \vt{S}_N^+, \vt{S}_{N-1}^+, \cdots, \vt{S}_1^+ \}$ 
given 
%determied 
by the 
equations (17). 
%the given order of collisions is. 
%each of which 
%describles a pair collision correctly. 
%of the form $f(\z_j, \z_k)$. 
%In this extended sense, 
%%Within this extended definition, 
%%we can factorize 
%an $N$-soliton collision 
%in the Manakov model is factorized into pair collisions. 
%
%, which has been 
%believed to be an characteristic nature of integrable systems. 
%
%prove finally that superposition of $N \choose 2$ mappings of the form 
%$f(\z_j, \z_k)$ 
%
%If we consider factorization of an $N$-soliton 
%collision into pair collisions and assume that 
%more than two solitons do not collide at one 
%%the same 
%point, 
%
%What we are going to prove is that 
%
%taking into account the nontrivial structure of (17), 
%one can believe that the conclusion given in ($\clubsuit$) 
%
%is seemingly correct it is, in combination with the nontrivial structure 
%of (17), seemingly correct 

To prove this factorization, 
%tion theorem 
%property 
%of an $N$-soliton collision 
%explicitly, 
we do not employ 
%part with 
the Manakov results, 
equations (17) and (18), 
for the following two reasons. 
One reason is that, 
although 
%the derivation of (17) 
it is 
%seems to be 
ingenious and seems to be 
correct, 
%is quite ingenious and 
the derivation of (17) in Ref.~\citen{Manakov} 
%was done 
%by an 
%quite 
%ingenious 
%method 
%manner 
%way
%, which 
%and seems to be correct. 
%However, 
%in the authors' opinion, 
%it 
%does not 
%seems to be 
%very 
is neither 
very 
rigorous 
%enough 
nor understandable 
%for 
%one 
to the reader 
%who is 
not familiar with 
%does not have a deep knowledge in 
the ISM. 
%inverse scattering method. 
%For this reason, in this paper, 
%This is a 
%one 
%main 
%reason why 
The other reason is that the equations (17) are 
not tractable for our purpose. 
%, at least, for the authors. 
%Therefore, 
In this paper, we 
%use 
%take 
employ 
%more 
a more straightforward 
%way 
%method 
approach 
to 
%get 
%know 
obtain 
another formula for the asymptotic behavior of 
$N$ solitons. 
%employ a more 
%
We start from an explicit formula for 
the $N$-soliton solution of the matrix NLS 
%nonlinear Schr\"{o}dinger 
equation 
%which was 
derived 
%via 
using the ISM 
%inverse scattering method 
%reported 
in Ref.~\citen{Tsuchida1}. 
%through the ISM. 
Through 
%considering 
a simple reduction, we 
%derive 
obtain an explicit formula 
for the general $N$-soliton solution of the Manakov model 
(\ref{cNLS}).\cite{PhD} \ To make the paper self-contained, we 
first 
%discuss the 
set $N=2$ 
%case 
and 
%present 
%discuss 
%the 
%investigate 
compute the asymptotic forms of the two-soliton solution 
%as 
in the $t \to \mp \infty$ limits in 
%using 
our notation. 
%The asymptotic forms 
%They 
These solutions 
define the collision laws of 
%a pair 
two solitons 
%(the mapping $f(\z_j, \z_k)$ in the above context) 
%collision 
in the Manakov model, 
%(\ref{cNLS}), 
which are essentially the same as those 
given by the 
equations (18). 
%Essentially, 
%In fact, 
%they coincide with equations (18). 
%They define the collision laws 
%Then, 
Next, we consider the general $N$ case and 
%take the limit $t \to \mp \infty$ and 
%investigate 
%discuss 
compute the asymptotic forms of the $N$-soliton solution 
%as 
in the $t \to \mp \infty$ limits. 
To express the polarization vectors 
%which 
appearing 
in the asymptotic forms concisely, 
we extend the definition of a determinant 
in such a way that 
%so that 
%employ an extension of the 
%use an extended version of 
%propose an extention 
%use an extended definition 
%determinants 
%generalize the definition of a determinant 
%to such that 
%whose 
%in which 
the last 
%each entry of the last 
%one 
column of an extended determinant 
%is composed 
consists of 
%not scalars but 
vectors. 
%It 
This determinant represents a vector defined 
%by 
%as 
in terms of 
the Laplace 
expansion with respect to the last column. 
%a vector. 
%composed of vectors. 
We 
%prove 
%present 
find a beautiful relation which casts the Hermitian product 
between 
%of 
%between such 
such extended 
%the extended 
determinants into 
the form of 
a product of conventional determinants. 
%Finally, 
%with the help of 
Using 
%the presented 
this relation and the Jacobi formula for determinants, 
we prove that an $N$-soliton collision in the Manakov 
model (\ref{cNLS}) 
%is 
can be factorized into 
%the composition 
a nonlinear superposition 
of $N \choose 2$ pair collisions in arbitrary order. 
%
%We investigate soliton collisions in the 
%%system of coupled NLS equations (for short, 
%Manakov model \cite{Manakov}.
%\\
%
%We naturally extend the notion of ``factorization'' to one 
%including nonlinear superposition (in other words, composite 
%mapping) of pair-wise collision maps. 
%\\
%Obviously, 
%Needless to say, 
%This is 
%%almost 
%%practically 
%equivalent
This 
%means
%demonstrates
%indicates
reveals\footnote{To 
%shows
%more 
%be precise, 
deduce 
property (b) 
from the factorization rigorously 
in the present 
%our 
setting, 
%we need to make 
%some discussions 
%a 
some discussion is needed. 
%for proving 
%demonstrate 
%the 
%there is one subtle point in our approach. 
%it is not evident 
%we 
%need to 
%should 
%also prove that an $N$-soliton collision gives a bijection 
%described by 
%is expressible as 
%in terms of 
%from an arbitrary 
%ily chosen 
%initial state into the 
%corresponding final state for any $N$, 
%the space of all possible initial states into 
%those 
%the space of 
%all possible 
%that of final states. 
%in the space of initial states into the space of final states. 
%Within our framework, 
%In the present work, 
%this 
%which 
%show that is not evident in our approach. 
%within our framework. 
%since we do not express the final state in terms of the initial state. 
%the initial 
%at first, 
%but it can be proved easily after the factorization is demonstrated. 
%a bijective mapping. 
%and 
%We 
%which will be given in 
%discuss this point 
%after the factorization is proved.
%We will do it 
This is given 
%will be done 
%this 
%explain this point 
%leave it to 
in 
%section 
\S 5.} 
%demonstrating 
%to showing 
%that both of 
the following two 
%distinct 
properties of 
%for 
the Manakov model: 
%hold: 
%(a), (b): 
%for soliton collisions in the Manakov model: 
\vspace{1.5mm}
\begin{enumerate}
\item[(a)]
An $N$-soliton collision 
%in the Manakov model (\ref{cNLS}) 
is 
%factorized into 
%equivalent to 
%recovered 
%reproduced by 
composed of a 
nonlinear superposition 
of $N \choose 2$ pair collisions of 
%a 
certain 
%arbitrary 
order. 
\item[(b)] 
\vspace{1.5mm}
%The result of 
The composition 
%composite 
%nonlinear superposition 
of $N \choose 2$ 
%pair-collision 
mappings 
%which 
corresponding to 
pair collisions in 
an $N$-soliton collision 
%for explaining an $N$-soliton collision by pair collisions 
%expresses by 
%ing 
%corresponding to 
%representing 
%
%in 
%an $N$-soliton collision. 
%each 
%for a two-soliton collision 
%pair collisions 
%gives 
yields 
%becomes 
the same mapping 
%regardless of 
%result 
%independent of 
for every 
possible 
%admissible 
%the 
order of composition. 
%of pair collisions 
%corresponding to an $N$-soliton collision. 
%for all possible orders. 
%of pair collisions. 
%in a certain 
\end{enumerate}
\vspace{1.5mm}
%
%We should note that 
We remark 
%should 
%note 
%recall 
that 
%, since 
solitons are not mass points, and do not have 
%, not on 
compact support. 
Rather, they are structures 
%but objects 
with 
%having 
%have 
infinitely long 
%spread 
%ly decreasing 
tails. 
%of infinite length. 
%each pair 
%As a result, 
Even in 
a 
%any 
two-soliton collision, it 
%needs 
%it 
takes 
%place over 
an infinite time 
%infinitely long time, 
%so that 
for the solitons 
%can 
to completely recover their 
%original 
own 
shapes. 
%completely. 
%so that 
%Thus 
%In this respect, 
Taking this into consideration, it is more 
%natural 
meaningful 
to understand this 
%the 
factorization 
%should be understood 
conceptually 
%rather 
than phenomenologically. 
%In this respect, 
%sense
%We note that 
It is unlikely that 
%The 
properties (a) and (b) are 
%essentially 
%apparently distinct and 
%unlikely to be 
%somewhat 
related 
%with each other 
directly. 
%On the other hand, 
%As a matter of fact, 
%In fact, 
%demonstrating 
We note that 
proving 
%the property 
(a) is equally difficult for every order of pair collisions 
and 
%there is 
%no essential difference 
%between 
%is engendered 
%caused 
%by different 
%by the order. 
%its 
%in the proof. 
%of pair collisions. 
%Thus, 
%As a consequence of it, 
%For this reason, 
%with a simple combinatorial discussion, 
%construct a logic for 
%will 
%can 
%we demonstrate 
%ing 
%both 
%(a) and (b) simultaneously in this paper. 
%together. 
%take the way of 
%prove 
%simultaneously. 
%demonstrating 
%that (a) holds true for every possible order of pair collisions. 
%the factorization for 
%However, this does not mean that 
%It is 
%the 
%a 
%well known 
%fact 
that proving (b) for 
%general 
arbitrary $N$ 
%is reduced 
reduces to 
%that 
proving it 
%only 
for $N=3$, that is, the following: 
%namely, 
%: 
%the $N=3$ case: 
\vspace{1.5mm}
\begin{enumerate}
\item[(b')]
The 
%composite 
composition 
%nonlinear superposition 
of three 
%pair-collision 
mappings 
%which corresponds to 
corresponding to pair collisions in a three-soliton collision 
does not depend on 
%is independent of 
the (conceptional) 
%(conceptual) 
order of pair collisions. 
%is the same for two possible (conceptual) orders of pair collisions. 
%the order of composition. 
\end{enumerate}
\vspace{1.5mm}
The property (b') is 
%, in general, 
%usually 
called the Yang-Baxter property. 
%Specifically, 
%More specifically, 
%in our case, 
%through proving the factorization, 
%this 
The validity of this 
%the Yang-Baxter 
property 
means that 
the collision laws of two solitons 
in the Manakov model (\ref{cNLS}) 
%for the Manakov model, 
%we obtain 
give 
%define 
a ``set-theoretical'' 
%{\it set-theoretical} \/
solution 
%of 
to the quantum Yang-Baxter equation.\cite{Dri} \ To the best 
of the author's knowledge, 
%(see Drinfeld \cite{Dri}). 
%in the sense of Drinfeld 
%terminology 
%which is 
%, to the author's knowledge, 
%seems to be 
this solution 
%seems to be 
is new. 
%result 
%(see \cite{}) defined by 
%Here 
We 
%should 
mention that 
%Veselov and collaborators \cite{} have recently extracted 
%obtained 
another ``set-theoretical'' solution 
to the quantum Yang-Baxter equation 
%(``Yang-Baxter map'' in their terminology) 
%was 
%obtained 
%discussed 
is studied 
%recently 
by Veselov\cite{Ves} 
%and collaborators 
%have recently extracted 
%from the collision laws of two solitons in 
through investigation of 
%studying 
the matrix 
KdV equation. 
%\cite{}. 

%To date, 
%Up to the present, some 
%There exist 
A few 
interesting 
ideas\cite{Zak,Kul,Gon} have 
%already 
been 
%known 
proposed 
to explain 
%generically 
in a general manner 
%demonstrate 
%explain 
the pairwise nature of soliton collisions in 
%a generic 
integrable systems. 
%\cite{}. 
However, 
those 
%these 
ideas 
%approaches 
seem to be 
%an 
too intuitive in 
%the 
their present forms 
%(at least in present form) 
%a discussion 
%rather than 
to 
complete 
%provide 
a mathematically 
rigorous proof. 
%of it. 
%factorization. 
For instance, it is not 
%easy 
%so trivial to prove 
obvious 
%evident 
%demonstrate 
that a pair collision is 
%phenominologically 
%not 
unaffected by 
the other solitons 
%any other soliton 
%when 
%if 
given only that they are 
%which 
%it is 
%physically 
far away, 
%apart, 
%being far apart from the pair. 
%and/
or that 
%one can change 
the final result 
%unchanged. 
is invariant
when the order of 
%taking 
%different 
multiple limits 
is changed. 
%without destroying 
%changing 
%with 
%keeping the final result 
%%unchanged. 
%invariant. 
%unchanged. 
In addition, 
%Moreover, 
%Furthermore, 
unlike 
%our 
this 
%the present 
work, 
%unfortunately, 
those works 
%approaches 
%they 
%these 
%the ideas 
%%proposed in 
%from \cite{Zak,Kul,Gon} 
%those 
%ideas 
do not elucidate 
%explain 
%any explicit 
%the 
%any 
a nice 
mechanism of 
%for 
factorization. 
%nicely. 
%well. 
%explicitly. 
%, which is elucidated 
%factorizing an $N$-soliton collision into pair collisions, 
%as 
%we 
%will be 
%is elucidated 
%we will see 
%later 
%in this paper. 
% explained in this paper. 

Finally, we would like to 
%make some comments 
%remarks 
%bibliographic 
%about 
comment on the 
%relevant 
literature 
%about 
%on 
concerning 
the Manakov model (\ref{cNLS}). 
%relevant to the present work. 
%We mention that 
Multi-soliton solutions of the Manakov model 
%(\ref{cNLS}) 
have already been obtained 
%were obtained in the literature 
%by using 
%through 
using the Hirota method\cite{Ohta,Chow,Trubatch1} 
%\cite{Ohta,Rad,Chow,Trubatch1} 
(see also 
%an alternative 
%some 
Refs.~\citen{Nogami} and~\citen{Akhmed} 
for results obtained with 
%by 
another method). 
%,Park}. 
%Thus 
%In this sense, 
%We should stress that 
In this sense, 
%respect, 
although 
it 
is very useful, 
%seems to be 
%a 
%a simplest one, 
%most simplified one, 
%one, 
%of a simplest form, 
%the formula in this paper 
%it is not 
%should 
%we do not claim that 
%exact 
the 
%obtained 
explicit formula for the 
%general 
$N$-soliton solution 
obtained 
in this paper 
%in this paper, 
%(\ref{N-soliton}) with (\ref{U-def}) and (\ref{lambda}) 
may not be essentially new.\footnote{In any case, 
%However, 
%Nevertheless, 
our formula 
has the advantage of 
%being compact.
compactness in its own right.} \ 
The main contribution of 
%can not 
%regard it as 
%consider it to be 
%Thus 
%achievement
%this paper lies in 
%the present 
this work 
%lies in 
is 
%not the exact formula of the $N$-soliton solution 
%obtained above is not the main subject 
%but 
%an 
the 
%explicit 
elucidation of the 
%mechanism of 
pairwise nature of soliton collisions in the Manakov model. 
%in the Manakov model. 
%This 
%%finally 
%%can 
%%now 
%removes the longtime 
%%longstanding 
%misunderstanding that 
%%an $N$-soliton collision does not reduce to a pair collision \cite{}. 
%%suspicion that 
%%or 
%%This establishes that 
%multi-particle effects 
%%may 
%exist in the Manakov model. 
%\cite{}. 
%This paper is based on 
%The authors obtained 
The results of this paper 
were 
%presented in this paper were 
obtained by the author 
in the summer of 
%year 
2000 and presented at the autumn meeting of the Physical Society 
of Japan in 
%the 
that year. 
%After that, 
Very recently, he 
%they 
%the authors wish to 
%declare 
%mention that 
%they were not motivated to solve this problem 
%tackle this famous problem 
%write this paper 
%by recent 
encountered some papers\cite{Park,Ghosh,Kanna}\footnote{Actually, 
%\footnotetext{Actually, 
%We mention that 
the work of Park and Shin\cite{Park} is very similar to 
that 
%the 
%early work 
of Steudel.\cite{Steudel}}
%the authors of 
which 
%the authors 
%were interested in 
%devoted to 
%posed 
pose the same problem (but do not solve it completely). 
%relevant studies \cite{}. 
%(see ), which 
%
%Unfortunately, 
%them 
%All of such studies seem to be 
%those studies seem to be 
%%relevant but 
%not as complete as 
%%our 
%%the present 
%%study. 
%this work. 
%%and thus 
%%of other authors 
%%of determinants and 
%\end{itemize}
%\end{enumerate}
%Moreover, after the first submission of this paper, 
In particular, 
in 
%the paper 
Ref.~\citen{Kanna}, 
which 
%was 
actually 
%published 
appeared after the first submission of this paper, 
Kanna and Lakshmanan 
%published the paper \cite{}, in which they 
gave 
%presents 
%they gave 
a ``proof'' of the pairwise 
%``rigorous proof''
collision nature 
%of solitons 
%nature of soliton collisions 
%for 
%in 
of a three-soliton collision. 
%the three-soliton case. 
%in the Manakov model. 
Unfortunately, 
%{\bf 
the ``proof'' of Kanna-Lakshmanan\cite{Kanna} is 
{\it absolutely 
%totally 
%false
incorrect}. 
Indeed, 
%This is clear from, 
%besides 
in addition to a few 
%big 
%other 
fatal mistakes, 
%the fact that 
%{\bf 
their ``proof'' 
%itself 
as a whole is 
%based on 
a typical example of 
%a 
{\it 
circular 
%argument.
reasoning}. 

%The rest of this 
The remainder of this 
%This 
paper is organized as follows. 
In 
%section 
\S 2, we 
%start from the $N$-soliton solution 
%of the matrix nonlinear Schr\"{o}dinger equation 
%%obtained 
%reported in \cite{Tsuchida1}. 
%Through considering a simple reduction, 
derive an explicit formula 
for the 
%general 
$N$-soliton solution of the Manakov model.\cite{PhD} \ In 
%section 
\S 3, we 
%reported 
%from the results in \cite{Tsuchida1} 
%convenient
%We 
%%also 
%rewrite it as a form convenient to investigate 
%%for investigating 
%the asymptotic behavior as $t \to \mp \infty$. 
%forms. 
obtain the collision laws of two solitons. 
%consider the $N=2$ case and 
%investigate 
%in detail 
%the two-soliton collision. 
%derive the two-soliton collision laws. 
%of two solitons. 
In 
%section 
\S 4, we 
%consider the general $N$ case and 
%present 
compute the asymptotic forms of the 
%general 
$N$-soliton solution 
%the asymptotic behavior 
%as 
in the limits 
$t \to \mp \infty$. 
In 
%section 
\S 5, we elucidate a mechanism 
%which 
%show 
that factorizes an $N$-soliton collision 
%in the Manakov model 
%can be 
%is equivalent to 
%expressed as 
into a nonlinear 
%nonlinear 
superposition of 
%$N \choose 2$ 
pair 
%pair 
collisions 
%in arbitrary order. 
%, regardless of 
%%the 
%collision order. 
%We also demonstrate 
and 
%also 
%demonstrate 
discuss 
the Yang-Baxter property. 
%(b'). 
Section 6 is devoted to concluding remarks. 
%, including some comments on the literature. 
%the conclusion. 
%of the paper. 
%The conclusion is presented in section 6. 
%
%%This paper is 
%%a report 
%%based on 
%The results presented in this paper were 
%obtained by the authors 
%in the summer of 
%%year 
%2000. The authors wish to 
%%declare 
%mention that 
%they were not motivated to solve this problem 
%%tackle this famous problem 
%%write this paper 
%by recent relevant studies \cite{}. 
%%(see ), which 
%%Unfortunately, 
%All of 
%%them 
%such studies seem to be 
%%relevant but 
%not as complete as 
%%our 
%this 
%%study. 
%work. 
%%of other authors 

\section{Explicit formula for the general $N$-soliton solution}
\setcounter{equation}{0}
\label{Exact}
%\noindent
%
In this section, 
%through 
considering a reduction 
of a formula 
%for results 
given in 
%ref.\ 
Ref.~\citen{Tsuchida1}, 
we 
%consider a reduction for the 
derive an explicit formula for the 
general $N$-soliton solution 
of the Manakov model (\ref{cNLS}). 

In 
%ref.\ 
Ref.~\citen{Tsuchida1}, under vanishing boundary conditions, 
we applied the ISM to 
%considered 
nonlinear evolution equations associated with 
%the inverse scattering method to 
%and its hierarchy associated with 
%a generalized version 
the generalized 
%following generalization of the 
Zakharov-Shabat eigenvalue problem:
\begin{equation}
%\beq
\left[
\begin{array}{c}
\Psi_1  \\
\Psi_2  \\
\end{array}
\right]_x = 
\left[
\begin{array}{cc}
-\i \z I & Q \\
 -Q^\dagger &  \i \z I \\
\end{array}
\right] 
\left[
\begin{array}{c}
\Psi_1  \\
\Psi_2  \\
\end{array}
\right].
\label{eigen}
\end{equation}
%\eeq
%
Here, 
%$\vt{\psi}$ is a $2m$-component column vector and 
$\z$ is the spectral parameter, $I$ is the $m \times m$ unit matrix, 
and 
%The potential 
%variable 
$Q$ is an 
$m \times m$ matrix-valued 
potential function. 
%which takes its value in $m \times m$ matrices. 
The first two 
%nontrivial 
of the 
nonlinear evolution equations associated with 
%the eigenvalue problem 
(\ref{eigen}) are the matrix NLS 
%hierarchy associated with 
equation,
%nonlinear Schr\"{o}dinger 
\begin{equation}
{\rm i} Q_t + Q_{xx} + 2Q Q^\dagger Q = O, 
\label{mNLS}
\end{equation}
and the matrix complex mKdV equation,
%while the next equation is 
%
\begin{equation}
%\beq
Q_t + Q_{xxx} +3 (Q_x Q^{\dagger} Q + Q Q^{\dagger} Q_x) = O.
\label{mcmKdV}
%\eeq
\end{equation}
%We established 
We mention that integrable 
space-discretizations of 
%semi-
%equations 
(\ref{mNLS}) and (\ref{mcmKdV}) were 
found 
%proposed 
recently\cite{Tsuchida2002} 
(see also the relevant work in Refs.~\citen{Trubatch1} and~\citen{GI2,
Tsuchida1999,Vakh}). 
%early attempts 
%We 
%applied performed 
%Through 
%the ISM 
%inverse scattering method (ISM) 
%for the 
%eigenvalue problem (\ref{eigen}) 
%we 
%and 
%obtained in \cite{Tsuchida1} 
The 
%bright 
%most 
general 
%bright 
$N$-soliton solution of 
%for the associated 
%equations 
(\ref{mNLS}) or (\ref{mcmKdV}) 
%through the ISM. 
%They are expressed as \cite{Tsuchida1}
%The formula 
%for the 
%general 
%$N$-soliton solution 
is 
%given by 
expressed as\cite{Tsuchida1}
%
%We can construct soliton solutions of the matrix AKNS hierarchy. 
%the $N$-soliton solution of the matrix AKNS hierarchy is expressed as
 \bea
  Q(x,t) \eq 
  -2 \i 
\hspace{1pt}
%\, 
  (\hspace{1pt} \underbrace{ \hspace{1pt}
%\, 
	I \; I \; \cdots \; I \hspace{1pt}
%\,
	}_{N} \hspace{1pt} )
  \; S^{-1} \;
  \left(
  \begin{array}{c}
   C_1 (t)^{\dagger} \hspace{1pt} \e^{-2\i \z_1^\ast x}   \\
   C_2 (t)^{\dagger} \hspace{1pt} \e^{-2\i \z_2^\ast x}   \\
     \vdots  \\
   C_N (t)^{\dagger} \hspace{1pt} \e^{-2\i \z_N^\ast x}   \\
  \end{array}
  \right),
 % \label{N-soliton}
 %\nn
  \label{nso}
  \eea
where the $mN \times mN$ matrix $S$ is given by
% \beq
\begin{equation}
  S_{jk} = \de_{jk}I -\sum_{l=1}^N 
  \frac{\e^{2\i (\z_l -\z_j^{\ast})x}}
  {(\z_l -\z_k^{\ast})(\z_l-\z_j^{\ast})} C_j(t)^\dagger \hspace{1pt}
	C_l(t),
  \hspace{5mm} 1 \le j, k \le N.
  \label{S-def}
%  \eeq
\end{equation}
%The eigenvalues 
Here, 
$\z_j$ are discrete 
%distinct 
eigenvalues 
%which lie 
in the upper-half plane of $\z$ (${\rm Im}\, \z_j > 0$), 
each of which determines a bound state 
%by 
in 
the potential $Q$. 
% \; (j=1, 2, \cdots, N)$ 
%correspond to simple poles 
%of $\z$ 
%of a reflection coefficient 
%
The quantities 
$C_j(t)$ are $m \times m$ nonzero matrices 
whose 
%of which 
time 
dependences are given by 
%\beq
\begin{equation}
C_j (t) = C_j (0) \e^{4 \i \z_j^2 t},
\hspace{5mm} j =1, 2, \cdots, N,
\label{C-time}
%\eeq
\end{equation}
for the matrix NLS equation (\ref{mNLS}), and
\[
C_j (t) = C_j (0) \e^{8 \i \z_j^3 t},
\hspace{5mm} j =1, 2, \cdots, N,
\]
for the matrix complex mKdV equation (\ref{mcmKdV}), respectively. 

Let us consider a reduction of the $N$-soliton solution 
of the matrix NLS equation 
%(\ref{mNLS}) 
to 
%those for 
that of the Manakov model. 
%(\ref{cNLS}). 
%It is obvious that the choice,
We restrict the 
%$m \times m$ 
matrix $Q$ 
%takes 
to the 
%special 
%following form: 
form 
%(\ref{Q_sym}), 
 %
% \beq
\begin{equation}
 Q = 
 \left[
 \begin{array}{cccc}
  q_1 & q_2 &\cdots & q_m \\
  0 & 0 & \cdots & 0 \\
 \vdots & \vdots  &\ddots & \vdots\\
  0 & 0 & \cdots & 0 \\
 \end{array}
 \right] \equiv 
\left[
\begin{array}{c}
 \vt{q}  \\
 O  \\
\end{array}
\right],
 \label{Q_sym}
% \eeq
\end{equation}
%
%The $2m \times 2m$ Lax matrices for the reduced hierarchy can be 
%compressed into $(m+1) \times (m+1)$ matrices:
%\[
% U = \left[
%\begin{array}{cccc}
% -\i \z & q_1 &\cdots & q_m \\
% -q_1^\ast & \i \z & &  \\
% \vdots & & \ddots &  \\
% -q_m^\ast & & & \i \z \\
%\end{array}
%\right].
%\]
%Since we have applied the ISM to the eigenvalue problem (\ref{eigen}) 
%with a square matrix $Q$, 
%necessary to take into account the 
%
%It is easy to 
%By reflecting 
%in the scattering data, 
so that the matrix NLS equation (\ref{mNLS}) is reduced to the Manakov model 
(\ref{cNLS}). 
%when 
In this case, the matrices $C_j (t)^\dagger$
%$\{ C_1, \cdots, C_N \}$ 
must 
%take 
have the 
%following 
same form as $Q$, from their definition,\cite{Tsuchida1,PhD} 
%\beq
\begin{equation}
C_j (t)^\dagger =
\left[
\begin{array}{cccc}
 c_j^{(1)} & c_j^{(2)} & \cdots & c_j^{(m)} \\
 0 & 0 &  \cdots & 0 \\
 \vdots & \vdots &\ddots & \vdots \\
 0 & 0 &\cdots & 0 \\
\end{array}
\right] 
\equiv \i \left[
\begin{array}{c}
\vt{c}_j (t) \\
 O \\
\end{array}
\right]
, \hspace{5mm} j=1, 2, \cdots, N.
%\label{}
\label{C-form}
%\eeq
\end{equation}
%where $\{ \a_j^{(i)} \}$ are complex functions of time $t$. 
%It is 
%practically sufficient to note that
%easy to check 
Conversely, 
%that, 
if $C_j (t)^\dagger$ 
%are of 
take the form (\ref{C-form}), 
$Q(x,t)$ given by the formula (\ref{nso}) with (\ref{S-def}) 
%obeys 
fits the form (\ref{Q_sym}). 
%The result is summarized as follows:
%In order to make the ISM applicable to the reduced systems, 
%we have to reflect internal symmetry of $Q$ and $R$
%in the scattering data. 
Then, 
%The nontrivial part of 
the formula 
%for $N$-soliton solutions 
%(\ref{nso}) with (\ref{S-def}) 
%is 
can be compressed into 
%written in 
a compact form,\cite{PhD}
%we obtain the $N$-soliton solution of the cNLS equations (\ref{ccNLS}):
%
%\bseq
\[
%(q_1, q_2, \cdots, q_m) 
\vecq (x,t) = 2 \sum_{j=1}^N \sum_{k=1}^N 
(T^{-1})_{jk} \e^{-2\i \z_k^\ast x} \vt{c}_k (t),
\]
where the $N \times N$ matrix $T$ is given by 
\[
T_{jk} = \de_{jk} - \sum_{l=1}^N 
 \frac{\e^{2 \i (\z_l - \z_j^\ast) x}}{(\z_l-\z_k^\ast)(\z_l-\z_j^\ast)}
\vt{c}_j (t) \cdot \vt{c}_l (t)^\dagger, 
\hspace{5mm} 1 \le j, k \le N.
\]
Thanks to (\ref{C-time}) and (\ref{C-form}), 
the time dependence of $\vt{c}_j (t)$ is given by 
\[
\vt{c}_j (t) = \e^{-4 \i \z_j^{* 2} t} \vt{c}_j (0), 
\hspace{5mm} j =1, 2, \cdots, N. 
%\equiv \vt{d}_j \e^{-4 \i \z_j^{* 2} t}.
\]
%
%\label{n_formula}
%\eseq
%
%Let us 
The above set of 
equations gives 
%above three 
%set of equations gives 
%is 
%given 
a formula for the general 
%We rewrite the 
$N$-soliton solution 
%(\ref{n_formula}) 
of the Manakov model (\ref{cNLS}) under 
%the 
vanishing boundary conditions. 
%to 
%as a form 
%We then 
Let us rewrite 
%it 
%so that it 
%will be 
%becomes 
this 
%to 
into a form 
%which is more 
convenient to investigate 
% suitable for 
%investigating 
%studying 
the asymptotic behavior. 
We first rewrite it as 
%the $N$-soliton solution as 
%(\ref{n_formula}) as 
\bea
\vecq (x,t) = 2 \sum_{j=1}^N \sum_{k=1}^N 
(W^{-1})_{jk} \e^{-\i [ (\z_k + \z_k^\ast) x 
	+ 2(\z_k^{2} + \z_k^{\ast 2}) t]} \vt{c}_k (0),
\nn 
\eea
where the $N \times N$ matrix $W$ is given by 
\bea
W_{jk} \eq \de_{jk} 
\e^{-\i [(\z_j -\z_j^\ast) x + 2(\z_j^{2} - \z_j^{\ast 2})t]} 
 - \sum_{l=1}^N 
 \frac{\vt{c}_j (0) \cdot \vt{c}_l (0)^\dagger}
{(\z_l-\z_k^\ast)(\z_l-\z_j^\ast)} 
\e^{\i [ (\z_l - \z_l^\ast) x + 2(\z_l^2 -\z_l^{\ast 2}) t] }
\nn \\
\espace 
\mbox{} \times
\e^{\i [ (\z_l + \z_l^\ast) x + 2(\z_l^2 +\z_l^{\ast 2}) t] }
\e^{- \i [ (\z_j + \z_j^\ast) x + 2(\z_j^2 +\z_j^{\ast 2}) t] },
\hspace{5mm} 1 \le j, k \le N.
\nn
\eea
%
%Moreover, to simplify this formula, 
Next, we introduce the 
%following 
parametrization
\bea
&& \z_j = \xi_j + \i \eta_j \; \,  (\xi_j \in {\mathbb R}, \; \eta_j > 0),
\nn \\
&&  \vt{c}_j (0) = 2 \eta_j \e^{-\a_j} \vt{u}_j 
	\; \,  (\a_j \in {\mathbb R}, \; |\! | \vt{u}_j |\! | =1),
\nn 
\eea
and employ the following abbreviations:
\bea
&& 
\tau_j \equiv -\i [ (\z_j - \z_j^\ast) x + 2 (\z_j^{2} -\z_j^{\ast 2}) t]
= 2 \eta_j (x + 4 \xi_j t), 
\nn \\
&&
\Thet_j \equiv (\z_j + \z_j^\ast) x + 2 (\z_j^{2} +\z_j^{\ast 2}) t
= 2 \xi_j x + 4 (\xi_j^2 - \eta_j^2) t.
\nn 
\eea
%Finally, 
%Then, 
In this way, 
we obtain 
%get 
%a 
the 
simplest formula for 
the general $N$-soliton solution of the Manakov model (\ref{cNLS}),
%is 
%finally 
%written as
%\beq
\begin{equation}
%(q_1, q_2, \cdots, q_m) 
\vecq (x,t) = 2 \sum_{j=1}^N \sum_{k=1}^N 
( U^{-1} )_{jk} \e^{-\i \Thet_k} \vecu_k,
\label{N-soliton}
%\eeq
\end{equation}
where the $N \times N$ matrix $U$ is given by 
\bea
&& U_{jk} = \frac{\e^{\tau_j + \a_j}}{2\eta_j} 
\de_{jk} + \sum_{l=1}^N 
% \frac{2 \eta_j (\vecu_l \cdot \vecu_j^\dagger)}
% {(\z_j-\z_k^\ast)(\z_j-\z_l^\ast)}
\la_{jkl} \e^{ -(\tau_l + \a_l) + \i (\Thet_l - \Thet_j)},
%\nn \\
%&&
\hspace{5mm}
1 \le j, k \le N, 
\label{U-def}
\eea
%
%$\a_j$ are real constants and $\vecu_j$ are constant unit vectors:
%
%\bea
%&&
%\vecu_j = (u_j^{(1)}, u_j^{(2)}, \cdots, 
%u_j^{(m)}),
%\hspace{10mm}
%|\!| \vecu_j |\!| =1.
%\nn
%\eea
%In the following, we employ the abbreviation:
%and
with
%\beq
\begin{equation}
\la_{jkl}
= - \frac{2 \eta_l (\vecu_j \cdot \vecu_l^\dagger)}
{(\z_l-\z_k^\ast)(\z_l-\z_j^\ast)}.
\label{lambda}
%\eeq
\end{equation}
%
%Setting 
If we set $N=1$ in the above formula, 
we obtain the one-soliton solution, 
%of the following form: 
%of the Manakov model (\ref{cNLS}):
%
\bea
 \vecq (x,t) = 2 \eta_1 
\sech (\tau_1 + \a_1) \e^{-\i \Thet_1} \vecu_{1}.
\label{pol}
\eea
%From this expression, 
%Thus we conjecture 
%This indicates that 
Therefore, 
%Thus, 
we understand 
%can determine 
the significance of 
%interpret 
each parameter/coordinate 
%in the above formula 
as follows: 
%for the $N$-soliton solution has the following significance: 
%meaning:
%
\bea
\hspace{10mm}
&&
\mbox{$2 \eta_j$ :
amplitude of soliton-$j$},
\nn \\
&&
\mbox{$-4\xi_j$ : 
velocity of soliton-$j$'s envelope},
\nn \\
&&
\mbox{$\tau_j$ :
coordinate for observing soliton-$j$'s envelope},
\nn \\
&&
\mbox{$\Thet_j$ : 
coordinate for observing soliton-$j$'s carrier waves},
\nn \\
&&
\mbox{$\vecu_j$ : 
%(unit) 
polarization vector of soliton-$j$ \ ($ |\! | \vt{u}_j |\! | =1$)}.
%\footnotemark[1]
\nn  
\eea
To be 
%more 
precise, 
in the case of two or more solitons, 
the {\it real} \/polarization vectors 
%
%each polarization vector in 
%the presence of other solitons 
%multi-soliton solutions 
are not invariant and 
%changed by a 
change under soliton collision. 
%in time. 
%if we consider multi-soliton 
%Indeed, the polarization vector of each soliton 
%rotates, with its unit length unchanged, 
%without changing its length 
%when the 
%correspoinding s
%soliton collides with another. 
%with its unit length unchanged. 
%while keeping its unit length. 
%In fact, 
The vector $\vecu_j$ defines 
%denotes 
%represents 
the {\it bare} \/polarization 
of soliton-$j$, 
%the 
which is realized 
%achieved 
%accomplished 
%the one 
%polarization 
when 
%the the soliton-$j$ 
it 
%is 
becomes the rightmost soliton.
%, which 
This point is demonstrated below. 
%will be seen later. 
%(cf.\ section~\ref{}). 
%in the ``bare'' state where the 
%In what follows, 
%Throughout this paper, 
%To 
In the following, 
%Troughout
%what follows, 
%we do not consider any 
%collision-free situation. 
we assume 
%We shall exclude any 
%in what follows we assume 
%by assuming in what follows 
that all the soliton velocities are distinct, 
so that every soliton collides with all others. 
%any paired solitons can collide. 

\section{Two-soliton collision}
\setcounter{equation}{0}
\label{Two-soliton}
%\noindent

In this section, we 
%investigate 
compute the asymptotic forms of 
the two-soliton solution 
%as 
in the limits 
$t \to \mp \infty$, 
%This defines 
%obtain 
%by 
which 
%we can 
define the 
%pairwise 
collision laws of two solitons 
%for paired solitons 
in the Manakov model (\ref{cNLS}). 

%collision 
We first 
%calculate the exact form of 
write out the 
%exact 
two-soliton solution 
given by (\ref{N-soliton}) with $N=2$. 
%explicitly. 
%When $N=2$, 
According to (\ref{U-def}), 
the matrix $U$ in this case 
%is given by 
takes the form
%given by 
%is written explicitly as
%written as 
%the following two by two matrix:
\bea
U = 
\left[
\begin{array}{cc}
\Ds \frac{\e^{\tau_1 + \a_1} }{2 \eta_1} + \sum_{l=1}^2 \la_{11l} 
	\e^{-(\tau_l+ \a_l) + \i (\Thet_l -\Thet_1)} 
& \Ds  \sum_{l=1}^2 \la_{12l} \e^{-(\tau_l + \a_l) + \i (\Thet_l - \Thet_1)}
\\
\Ds   \sum_{l=1}^2 \la_{21l} \e^{-(\tau_l + \a_l) + \i (\Thet_l - \Thet_2)}
 & \Ds \frac{\e^{\tau_2 + \a_2} }{2 \eta_2} + \sum_{l=1}^2 \la_{22l} 
	\e^{-(\tau_l+ \a_l) + \i (\Thet_l -\Thet_2)} \\
\end{array}
\right].
\nonumber
\eea
Then, the two-soliton solution 
%, (\ref{N-soliton}) with $N=2$, 
is given by 
%written as 
%obtained: 
%calculated as 
%\bseq
\bea
\hspace{-3mm}
\vecq (x,t) \eq 
\frac{2}{\det U} \left\{
\left[ 
\frac{\e^{\tau_2 + \a_2}}{2 \eta_2}
	+ \sum_{l=1}^2 \la_{22l} 
	\e^{- (\tau_l + \a_l)+ \i (\Thet_l - \Thet_2)} 
	- \sum_{l=1}^2 \la_{21l} 
	\e^{- (\tau_l + \a_l)+ \i (\Thet_l - \Thet_2)} \right]
	\e^{-\i \Thet_1} \vecu_1 \right.
\nn \\
\espace \left.  \mbox{}+ 
\left[ 
\frac{\e^{\tau_1 + \a_1}}{2 \eta_1}
	+ \sum_{l=1}^2 \la_{11l} 
	\e^{- (\tau_l + \a_l)+ \i (\Thet_l - \Thet_1)} 
	- \sum_{l=1}^2 \la_{12l} 
	\e^{- (\tau_l + \a_l)+ \i (\Thet_l - \Thet_1)} \right]
	\e^{-\i \Thet_2} \vecu_2 \right\}, 
%\hspace{5mm}
\label{two-soliton}
\eea
with
%where 
%we can calculate 
%the determinant of $U$ 
%$\det U$ is given by 
\bea
\det U \eq \frac{\e^{\tau_1 + \a_1}}{2 \eta_1}
	\frac{\e^{\tau_2 + \a_2}}{2 \eta_2} 
+ \frac{\e^{\tau_1 + \a_1}}{2\eta_1} \sum_{l=1}^2 \la_{22l} 
	\e^{- (\tau_l + \a_l)+ \i (\Thet_l - \Thet_2)}
+ \frac{\e^{\tau_2 + \a_2}}{2\eta_2} \sum_{l=1}^2 \la_{11l} 
	\e^{- (\tau_l + \a_l)+ \i (\Thet_l - \Thet_1)} 
\nn \\
\espace \mbox{} 
+ \e^{-(\tau_1 + \a_1)}\e^{-(\tau_2 + \a_2)}
\sum_{\{l_1 ,l_2 \}= \{ 1,2 \}}
\left|
\begin{array}{cc}
\la_{11 l_1} & \la_{12 l_1} \\
\la_{21 l_2} & \la_{22 l_2} \\
\end{array}
\right|.
\label{detU-2}
\eea
%\label{two-soliton}
%\eseq
%
Here, we have simplified the expression of 
%the determinant of $U$ 
$\det U$ 
%with the help of 
%by 
using the relations
%\beq
\begin{equation}
\left|
\begin{array}{cc}
\la_{11l} & \la_{12l} \\
\la_{21l} & \la_{22l} \\
\end{array}
\right| = 0, \hspace{5mm} l=1,2,
\label{la22}
%\eeq
\end{equation}
which can be proved 
straightforwardly. 
%by direct computations. 

%Let us 
%solutions at $t \to \pm \infty$. 
%In the case $N=2$, 
Next, we assume that 
\[ 
\xi_1 (= \mbox{Re}\, \z_1)> \xi_2 (= \mbox{Re}\, \z_2)
\]
and 
%discuss 
investigate the asymptotic behavior of 
$\vt{q} (x,t)$ 
%in the the limit 
as $t \to \mp \infty$. 
This is accomplished by 
%picking 
identifying 
the dominant terms 
%in these regions 
%from 
in the numerator of (\ref{two-soliton}) and 
its 
%the 
denominator (\ref{detU-2}). 
%We should 
We here 
%remark 
note the relation 
$\tau_1/\eta_1 = \tau_2/\eta_2 + 8(\xi_1- \xi_2)t$. 

\begin{figure}[t]
 \begin{center}
    \includegraphics[width=17cm]{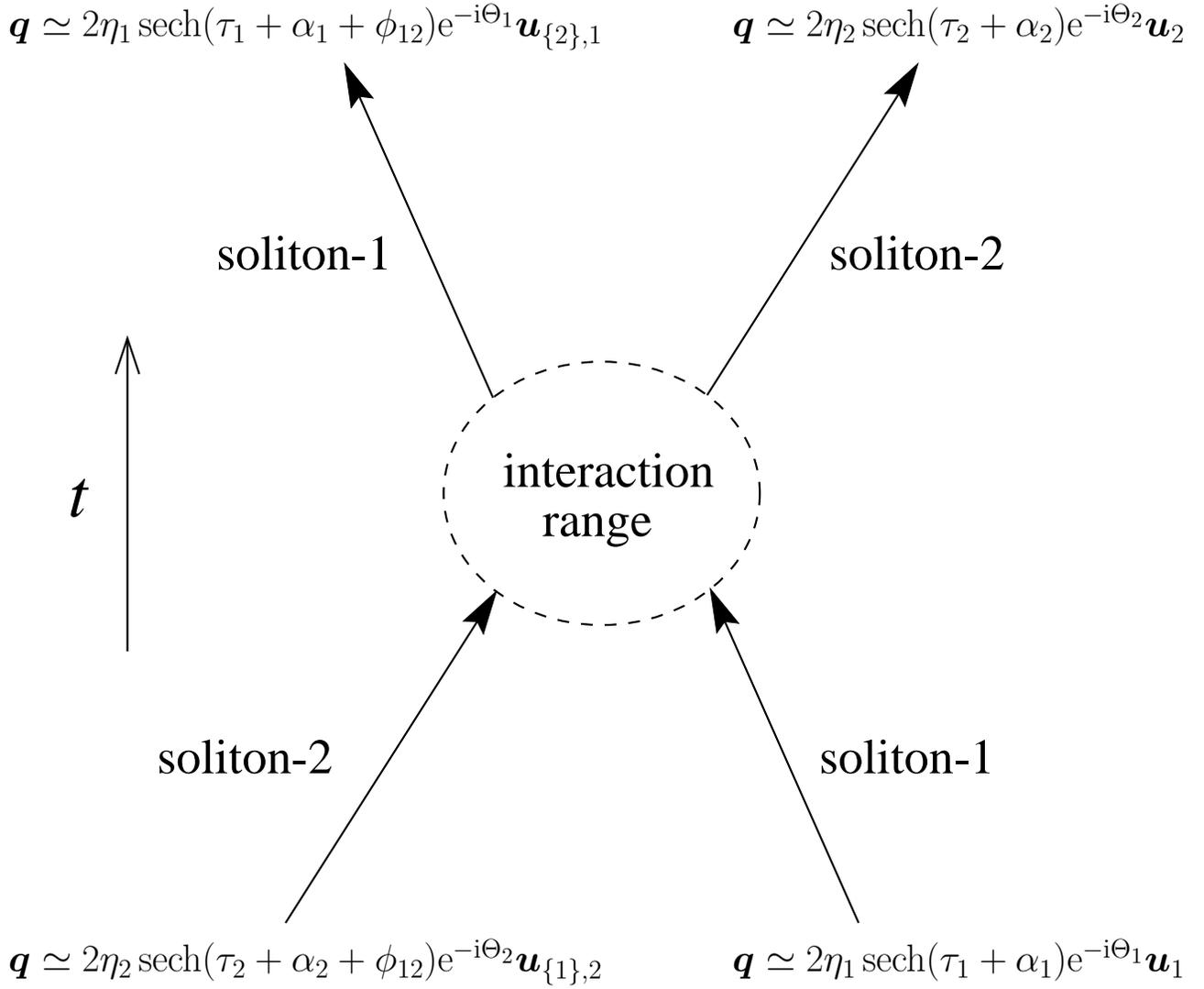}
%\psfrag{t}[]{xx}
\\  \vspace*{10mm}
\caption{Two-soliton collision.} 
    \label{two_sol}
%  \nonumber
%\noindent
 \end{center}
\end{figure}

In the limit $t \to - \infty $, we have 
\[
\frac{\tau_1}{\eta_1} \ll \frac{\tau_2}{\eta_2}.
\]
In this case, we 
%need to 
%discuss asymptotic forms of $\vt{q}$ in 
have to consider separately the 
%following 
two regions ($1^-$) and ($2^-$) defined below. 
%separately. 
%where $\vecq$ takes a nonzero 
It is easily seen that 
$\vecq \simeq {\mathbf 0}$ in 
%the 
all other regions. 
\begin{enumerate}
\item[($1^-$)] ${\rm finite}~\tau_1, \;\;\;
\tau_2 \to +\infty$
%$\tau_1 (\sim {\rm finite}) \ll \tau_2 (\to +\infty)$
%

%The terms including the factor $\e^{\tau_2}$ are dominant. 
In this case, 
the dominant terms are those which contain the factor $\e^{\tau_2}$. 
Then, 
using the relation $\la_{111} = 1/(2 \eta_1)$, we obtain
\bea
\vecq \eesim \frac{ 2
 \frac{\e^{\tau_2 + \a_2}}{2 \eta_2} \e^{-\i \Thet_1} \vecu_1}
{\frac{\e^{\tau_1 + \a_1}}{2 \eta_1} \frac{\e^{\tau_2 + \a_2}}{2 \eta_2} 
	+ \frac{\e^{\tau_2 + \a_2}}{2 \eta_2} 
\la_{111} \e^{-(\tau_1 + \a_1)} }
\nn \\ \eq 
2 \eta_1 \sech (\tau_1 + \a_1) \e^{-\i \Thet_1} \vecu_1.
\label{2sol-1-}
\eea
\item[($2^-$)] $\tau_1 \to - \infty, \;\;\; {\rm finite}~\tau_2$ 
%$\tau_1 (\to - \infty) \ll \tau_2 (\sim {\rm finite})$ 

Here, the dominant terms are those which contain 
%including 
the factor $\e^{-\tau_1}$. 
%are dominant. 
Then, 
%Thus 
we obtain
\bea
\vecq \eesim \frac{ 2 \left[
(\la_{221} - \la_{211}) \e^{-(\tau_1 +\a_1)} \e^{-\i \Thet_2} \vecu_1
+ (\la_{111} - \la_{121}) \e^{-(\tau_1 +\a_1)} \e^{-\i \Thet_2} \vecu_2
\right]}
{ \frac{\e^{\tau_2 + \a_2}}{2 \eta_2} \la_{111} \e^{-(\tau_1 + \a_1)} 
+ \e^{-(\tau_1 + \a_1)} \e^{-(\tau_2 + \a_2)} 
\dsum{ \{ l_1 ,l_2 \}}{= \{1,2\}} 
\left|
\begin{array}{cc}
\la_{11 l_1} & \la_{12 l_1} \\
\la_{21 l_2} & \la_{22 l_2} \\
\end{array}
\right| }.
\label{tochu}
\eea
%
%We define 
In terms of $\phi_{12}$ defined by 
\bea
\e^{-2 \phi_{12}} \equiv
\frac{1}{\la_{111} \la_{222}}
\sum_{ \{ l_1 ,l_2 \}= \{1,2\}} 
\left|
\begin{array}{cc}
\la_{11 l_1} & \la_{12 l_1} \\
\la_{21 l_2} & \la_{22 l_2} \\
\end{array}
\right|,
\label{phi12-def}
\eea
%and 
we can rewrite the asymptotic form (\ref{tochu}) as 
%in terms of $\phi_{12}$ as 
%
\bea
\vecq \eesim 2 \eta_2 \sech (\tau_2 + \a_2 + \phi_{12}) \e^{-\i \Thet_2} 
\times
\e^{\phi_{12}} 
\left[ \left( \frac{\la_{221} - \la_{211}}{\la_{111}} \right)\vecu_1 
+ \left( 1- \frac{\la_{121}}{\la_{111}} \right)\vecu_2 \right]. 
%\nn \\
%\espace
\label{2sol-2-}
\eea
%
%Here 
%we notice that 
%We note that 
Here, $\phi_{12}$ is always taken as 
%well-defined set to be 
real, since (\ref{phi12-def}) 
%is 
can be rewritten as 
[\hspace{0pt}cf.\ (\ref{la22}) and (\ref{lambda})]
\bea
\e^{-2 \phi_{12}} 
\eq
\frac{1}{\la_{111} \la_{222}}
\left|
\begin{array}{cc}
\la_{111} + \la_{112} & \la_{121} + \la_{122} \\
\la_{211} + \la_{212} & \la_{221} + \la_{222} \\
\end{array}
\right|
\nn \\
\eq
(2 \eta_1) (2\eta_2) 
\left|
\begin{array}{cc}
 \i \frac{ \vecu_1 \cdot \vecu_1^\dagger }{\z_1 - \z_1^\ast} 
	&  \i \frac{\vecu_1 \cdot \vecu_2^\dagger}{\z_2 - \z_1^\ast} \\
\i \frac{\vecu_2 \cdot \vecu_1^\dagger}{\z_1 - \z_2^\ast} 
	& \i \frac{\vecu_2 \cdot \vecu_2^\dagger}{\z_2 - \z_2^\ast} \\
\end{array}
\right|
\left|
\begin{array}{cc}
 \i \frac{2 \eta_1}{\z_1 - \z_1^\ast} & \i \frac{2 \eta_1}{\z_1 - \z_2^\ast} \\
 \i \frac{2 \eta_2}{\z_2 - \z_1^\ast} & \i \frac{2 \eta_2}{\z_2 - \z_2^\ast} \\
\end{array}
\right|
\nn \\
\eq
 \left| \frac{\z_1 - \z_2}{\z_1 - \z_2^\ast} \right|^2
\left\{
1 + \frac{(\z_1 -\z_1^\ast)(\z_2 -\z_2^\ast)}
{|\z_1 - \z_2^\ast|^2} \left| \vecu_1 \cdot \vecu_2^\dagger \right|^2
\right\}
%\nn \\ &>& 
(> 0).
\nn
\eea
\end{enumerate}
In the limit $t \to + \infty$, we have 
\[
\frac{\tau_1}{\eta_1} \gg \frac{\tau_2}{\eta_2}.
\]
In this case, we have to
%discuss asymptotic forms of $\vt{q}$ in 
consider separately 
the 
%following 
two regions ($2^+$) and ($1^+$) defined below. 
%separately. 
%where $\vecq$ takes nonzero asymptotics. 
It is easily seen that $\vecq \simeq {\mathbf 0}$ in 
%the 
all other regions. 
\begin{enumerate}
\item[($2^+$)] $\tau_1 \to + \infty,\;\;\; {\rm finite}~\tau_2$
%$\tau_1 (\to + \infty) \gg \tau_2 (\sim {\rm finite})$
%

In this case, the dominant terms are those which contain 
%including 
the factor $\e^{\tau_1}$. 
%are dominant. 
Then, using the relation $\la_{222} = 1/(2 \eta_2)$, we obtain
\bea
\vecq \eesim \frac{ 2
 \frac{\e^{\tau_1 + \a_1}}{2 \eta_1} \e^{-\i \Thet_2} \vecu_2}
{\frac{\e^{\tau_1 + \a_1}}{2 \eta_1} \frac{\e^{\tau_2 + \a_2}}{2 \eta_2} 
	+ \frac{\e^{\tau_1 + \a_1}}{2 \eta_1} 
\la_{222} \e^{-(\tau_2 + \a_2)} }
\nn \\ \eq 
2 \eta_2 \sech (\tau_2 + \a_2) \e^{-\i \Thet_2} \vecu_2.
\label{2sol-2+}
\eea
\item[($1^+$)] ${\rm finite}~\tau_1,\;\;\; \tau_2 \to - \infty$
%$\tau_1 (\sim {\rm finite}) \gg \tau_2 (\to - \infty)$

%Here, 
In this case, 
the dominant terms are those which contain 
%including 
the factor $\e^{-\tau_2}$. 
%are dominant. 
Then, 
with the help of (\ref{phi12-def}), we obtain
\bea
\vecq \eesim \frac{ 2 \left[
(\la_{222} - \la_{212}) \e^{-(\tau_2 +\a_2)} \e^{-\i \Thet_1} \vecu_1
+ (\la_{112} - \la_{122}) \e^{-(\tau_2 +\a_2)} \e^{-\i \Thet_1} \vecu_2 
\right]}
{ \frac{\e^{\tau_1 + \a_1}}{2 \eta_1} \la_{222} \e^{-(\tau_2 + \a_2)} 
+ \e^{-(\tau_1 + \a_1)} \e^{-(\tau_2 + \a_2)} 
\dsum{ \{ l_1,l_2 \}}{= \{1,2\}} 
\left|
\begin{array}{cc}
\la_{11 l_1} & \la_{12 l_1} \\
\la_{21 l_2} & \la_{22 l_2} \\
\end{array}
\right| }
\nn \\
\eq 2 \eta_1 \sech (\tau_1 + \a_1 + \phi_{12}) \e^{-\i \Thet_1} 
\times
\e^{\phi_{12}} 
\left[ \left( 1- \frac{\la_{212}}{\la_{222}} \right)\vecu_1 
+ \left(\frac{\la_{112} - \la_{122}}{\la_{222}} \right)\vecu_2 \right]. 
%\nn \\
%\espace
\label{2sol-1+}
\eea
\end{enumerate}
%
%Summing up 
Taking the sum of 
%eqs.\ 
(\ref{2sol-1-}) and (\ref{2sol-2-}), or 
%eqs.\ 
(\ref{2sol-2+}) and (\ref{2sol-1+}), with a slight simplification, 
we arrive at 
%obtain 
the following theorem. 
%In the limit 
%
\vspace{5mm}
\begin{theorem}
\label{two-rule}
{\it
The asymptotic forms of the two-soliton solution of the 
Manakov model $(\ref{cNLS})$ are 
%given 
as follows $($see also Fig.~$\ref{two_sol})$:
\\
as $t \to - \infty$, 
%the asymptotic form is given by 
}
\bea
\vecq \eesim
2 \eta_1 \sech (\tau_1 + \a_1) \e^{-\i \Thet_1} \vecu_1 
+ 2 \eta_2 \sech (\tau_2 + \a_2 + \phi_{12}) \e^{-\i \Thet_2}
\vecu_{\{1\}, 2};
%\label{}
\nonumber
\eea
{\it 
%while 
%in the limit 
as $t \to + \infty$, 
%the asymptotic form is given by 
}
\bea
 \vecq \eesim
2 \eta_1 \sech (\tau_1 + \a_1 + \phi_{12}) \e^{-\i \Thet_1} 
 \vecu_{\{2\},1}
+ 2 \eta_2 \sech (\tau_2 + \a_2) \e^{-\i \Thet_2}
\vecu_{2}.
\nonumber
%\label{}
\eea
{\it
Here 
$\phi_{12}$ and $\vecu_{\{ 1 \}, 2}, \, \vecu_{\{ 2 \}, 1}$ are given by 
\bea
%\e^{-2 \phi_{\{1\}, 2}} = \e^{-2 \phi_{\{2\}, 1}} 
%&&
%\e^{-2 \phi_{\{1\},2}} = \e^{-2 \phi_{\{2\},1}} 
%\nn \\
%\equiv
\e^{-2 \phi_{12}} 
\eq \left|\frac{\z_1 - \z_2}{\z_1 - \z_2^\ast} \right|^2
\left\{ 
1 + \frac{(\z_1 -\z_1^\ast)(\z_2 -\z_2^\ast)}
{|\z_1 - \z_2^\ast|^2} \left| \vecu_1 \cdot \vecu_2^\dagger \right|^2
\right\},
%\nn \\ &>& 
%> 0,
%\label{}
\nonumber
\eea
and 
\bea
\vecu_{\{1\},2} 
%\vecu_{1 2} 
\eq \e^{\phi_{12}}
\frac{\z_1^\ast -\z_2^\ast}{\z_1 -\z_2^\ast}
\left\{
\vecu_2  - \frac{\z_1 -\z_1^\ast}{\z_1 -\z_2^\ast}
\left( \vecu_2 \cdot \vecu_1^\dagger \right) \vecu_1
\right\}_{\vphantom \int},
%\label{}
\nonumber
\\
\vecu_{\{2 \}, 1} 
%\vecu_{2 1} 
\eq \e^{\phi_{12}}
\frac{\z_2^\ast -\z_1^\ast}{\z_2 -\z_1^\ast}
\left\{
\vecu_1  - \frac{\z_2 -\z_2^\ast}{\z_2 -\z_1^\ast}
\left( \vecu_1 \cdot \vecu_2^\dagger \right) \vecu_2
\right\}.
%\label{}
\nonumber
\eea
%respectively.
}
\end{theorem}
\vspace{5mm}
%\noindent

Theorem~\ref{two-rule} defines the 
%two-soliton 
%pairwise 
collision 
laws of two solitons in the Manakov model, 
which we 
%will 
use in 
%section~\ref{Facto} to factorize an $N$-soliton 
\S\ref{Facto} to factorize an $N$-soliton 
collision into pair collisions. 
Here, we mention 
%list 
%note 
some important properties of the collision laws:
%two-soliton collisions 
%%From Theorem~\ref{two-rule}, 
%%in the Manakov model 
%%whose 
%of which laws are 
%presented in 
%%expressed by 
%Theorem~\ref{two-rule}: 
%described 
%given 
%are as follows:
%\begin{enumerate}
\begin{itemize}
%
%\item[(i)]
\item
%Any 
%The 
A two-soliton 
collision 
%does not 
changes 
neither 
the amplitudes of 
the 
solitons 
%length of polarization vectors 
%as well as 
nor 
%and 
the modulus of the Hermitian 
%inner 
product 
%between 
of the 
%some 
polarization vectors. 
%, i.e. 
%\vspace{2mm}
%\\
%\hspace*{4mm}
%Recalling that 
In fact, 
%using the relation 
recalling that 
$ |\!| \vecu_{1} |\!| = |\!| \vecu_{2} |\!| = 1 $, 
we can prove by direct computations that 
%\vspace{2mm}
%\\ 
%\hspace*{50mm}
\[
%\Rightarrow
\left| 
%\hspace{-2pt}
\! 
\left| \vecu_{\{1\},2} \right|\! \right| =
\left| \! \left| \vecu_{\{2\},1} \right|\! \right| = 1, 
%\\ \hspace*{50mm}
\hspace{5mm}
\left| \vecu_1 \cdot \vecu_{\{1\}, 2}^\dagger \right| = 
\left| \vecu_{\{2\},1} \cdot \vecu_{2}^\dagger \right|,
%\\ 
\hspace{5mm}
\left| \vecu_{1} \cdot \vecu_{2}^\dagger \right|
= \left| \vecu_{\{2\}, 1}\cdot \vecu_{\{1\}, 2}^\dagger \right|. 
\]
%\vspace{2mm}
%
%\item[(b)]
%\\
%All of 
%These relations can be proved by a straightforward calculation. 
%Thus 
Although we omit here 
%here the 
%their 
the tiresome proof, 
%here, 
%the proof. 
%However, 
the most important relation 
$\left| \! \left| \vecu_{\{1\},2} \right|\! \right| =
\left| \! \left| \vecu_{\{2\},1} \right|\! \right| = 1$ 
%will be 
is shown 
%later 
in \S\ref{Facto} 
%section~\ref{Facto} 
%in the next section 
in a more general context. 
%where we of deal with for the $N$-soliton solutions. 
%It is clear from 
This relation shows that the 
%two-soliton 
collision is elastic 
if we observe it 
%%terms of 
%in 
with 
%the 
conserved density, $|\!|\vecq|\!|^2 = \sum_{j=1}^m |q_j|^2 $. 
%or the space ${\mathbb C}^m$. 
%
%\item[(ii)]
\item
% $\,$
As a result of the 
%two-soliton 
collision, 
the polarization vectors rotate 
nontrivially 
%($\vecu_{1} \mapsto \vecu_{\{2\},1}$ and $\vecu_{\{1\},2} \mapsto \vecu_{2}$) 
on the unit sphere in 
%the space 
${\mathbb C}^m$. 
Thus, if we 
%keep our eyes on 
observe the collision 
%in 
with respect to 
each component $q_k$, 
%the collision looks 
it appears as if it were inelastic. 
%This makes the asymptotic behavior 
%of the $N$-soliton solution as $t \to \mp \infty$ very complicated. 
%It is clear from this relation that 
%On the other hand, the collision is elastic 
%if we observe it 
%%is observed 
%in 
%%terms of 
%the conserved density $|\!|\vecq|\!|^2 = \sum_{k=1}^m |q_k|^2 $ 
%or regard it merely as the rotation in 
%%the space 
%${\mathbb C}^m$. 
%
%{\it Remark:}
%\item[(iii)]
\item
We have expressed $\phi_{12}$, 
$\vecu_{\{1\},2}$ and $\vecu_{\{2\},1}$ in terms of 
$\vecu_1$ and $\vecu_2$. 
Then, 
%and obtained 
the collision laws 
are 
%become 
%scattering 
%which are 
%invariant under the interchange $ 1 \leftrightarrow 2$. 
symmetric with respect to interchange of 
the subscripts $1$ and $2$. 
This 
%version 
form of the collision laws is 
%crucial 
very useful in 
%considering 
studying 
%investigating 
the factorization of an $N$-soliton collision into 
pair collisions. 
For any fixed 
unit 
vector $\vecu_1$, we can invert the mapping 
$\vecu_{2} \mapsto \vecu_{\{1\},2}$ 
%is invertible because 
using the following relation for the projection 
operator $\vt{u}_1^\dagger \vt{u}_1$ 
(cf.\ $(\vt{u}_1^\dagger \vt{u}_1)^2 = \vt{u}_1^\dagger \vt{u}_1$):
%$I- (\z_1 -\z_1^\ast)/(\z_1 - \z_2^\ast) 
\[
\left(
I  - \frac{\z_1 -\z_1^\ast}{\z_1 -\z_2^\ast} \vecu_1^\dagger \vecu_1
\right)
\left(
I  + \frac{\z_1 -\z_1^\ast}{\z_1^\ast -\z_2^\ast} \vecu_1^\dagger \vecu_1
\right) =I.
\]
%\newpage
%
Then, we can express $\phi_{12}$, $\vecu_{2}$ and $\vecu_{\{2\},1}$ 
in terms of $\vecu_1$ and $\vecu_{\{1\},2}$ 
%which is a formula obtained by 
as Manakov did in 
%original 
%his paper 
Ref.~\citen{Manakov}: 
%, namely, 
%
\bseq
\bea
%\e^{-2 \phi_{\{1\}, 2}} = \e^{-2 \phi_{\{2\}, 1}} 
%&&
%\e^{-2 \phi_{\{1\},2}} = \e^{-2 \phi_{\{2\},1}} 
%\nn \\
%\equiv
%\vspace*{15mm} 
\e^{-2 \phi_{12}} 
\eq \left|\frac{\z_1 - \z_2}{\z_1 - \z_2^\ast} \right|^2
\left\{ 
1 - \frac{(\z_1 -\z_1^\ast)(\z_2 -\z_2^\ast)}
{|\z_1 - \z_2|^2} 
\left| \vecu_1 \cdot \vecu_{\{1\}, 2}^\dagger \right|^2
\right\}^{-1}_{\vphantom \int_{\vphantom \int}},
%\nn \\ &>& 
%> 0,
%\label{}
%\nonumber 
\\
%\eea
%and 
%\bea
\vecu_{2} 
%\vecu_{1 2} 
\eq \e^{- \phi_{12}}
\frac{\z_1 -\z_2^\ast}{\z_1^\ast -\z_2^\ast}
\left\{
\vecu_{\{1\},2} 
+ \frac{\z_1 -\z_1^\ast}{\z_1^\ast -\z_2^\ast}
\left( \vecu_{\{1\},2} \cdot \vecu_1^\dagger \right) \vecu_1
\right\}_{\vphantom \int},
\label{}
%\nonumber
\\
\vecu_{\{2 \}, 1} 
%\vecu_{2 1} 
\eq \e^{-\phi_{12}}
\frac{\z_2^\ast -\z_1}{\z_2 -\z_1}
\left\{
\vecu_1  - \frac{\z_2 -\z_2^\ast}{\z_2 -\z_1}
\left( \vecu_1 \cdot \vecu_{\{1\},2}^\dagger \right) \vecu_{\{1\},2} 
\right\}.
\label{}
%\nonumber
\eea
\label{YBsol}
\eseq
%
%However, 
\noindent
\hspace{-9pt}
%For 
Owing to the 
lack of 
%the 
symmetry 
under the exchange of $1$ and $2$ 
%$1 \leftrightarrow 2$ 
(even after 
%changing the notation of vectors), 
a change of the vector notation), 
%redifinition of the vectors), 
this 
%form 
new 
%version 
form of the collision laws 
%of the above form are 
is not useful in studying the factorization problem. 
%for lack of the symmetry 
%of 
%
%will then be 
%is lost to 
%%that 
%this rewritten version of 
%%the 
%collision laws 
%%scattering rule 
%%formula 
%and it is not useful in studying the factorization problem. 
%%for our purpose. 
%%investigating an $N$-soliton collision. 
%%\end{enumerate}
%%
%Nevertheless, 
%However, 
On the other hand, 
%this version is 
it is 
%also 
%important 
of prime importance 
%cricial 
in 
explicitly 
showing 
%seeing 
%that because it clearly 
%shows 
%presents a two-soliton collision in the 
%form of a mapping from an initial state into the final state. 
%presents the pair-collision mapping 
%from the initial state to the final state in the Manakov 
%mode. 
that a 
%the 
two-soliton collision 
%%an $N$-soliton collision is 
%described by 
%is expressible 
can be expressed as 
a mapping from 
%an initial state to the final state 
the initial state 
%into 
to the final state. 
%one 
%initial to final states 
Moreover, 
%from 
using (\ref{YBsol}), 
we can 
%prove 
easily 
check that 
%by direct computations that 
if $ |\!| \vecu_{1} |\!| = \left| \! \left| \vecu_{\{1\},2} \right|\! \right| 
 = 1 $, 
then 
$|\!| \vecu_{2} |\!| = 1$. 
%it also shows 
%we can straightforwardly show by using (\ref{YBsol}) 
This verifies that 
%and that 
for any 
%fixed 
unit vector $\vecu_{1}$, 
the mapping $\vecu_{2} \mapsto \vecu_{\{1\},2}$ 
in Theorem~\ref{two-rule} 
is a bijection on the unit sphere. 
%in ${\mathbb C}^m$. 
%between unit vectors. 
%We will use these 
%facts 
%in discussing 
%discuss 
%
We 
%will 
use this fact 
in \S 5 
%section 5 
%to prove 
%in 
%for proving 
to prove the Yang-Baxter property of the mapping (\ref{YBsol}). 
%in section 5. 
%two-soliton collision 
%of the mapping. 
\end{itemize}
\newpage
\noindent

\clearpage

\section{Asymptotic behavior of the $N$-soliton solution}
\setcounter{equation}{0}
\label{Asymp-N}

In this section, we compute the asymptotic forms of 
the $N$-soliton solution 
%as 
in the limits 
$t \to \mp \infty$ 
and simplify them as much 
%long 
as possible. 
%We use 
%used in this section are 
%The 
Extensions of some techniques 
%are used, which are extensions of those 
%used 
%for 
applied to the KdV equation are used.\cite{Kay,WT,GGKM2} 

We first rewrite the $N$-soliton solution (\ref{N-soliton}) 
%further 
%to a form suitable for 
%explicitly 
%further 
before considering 
%taking 
the limits $t \to \mp \infty$. 
We use the tilde 
%$\ti{}$ 
to denote 
cofactors. 
%For instance, 
%We recall that 
For instance, the cofactor $\ti{U}_{kj}$ is 
%a determinant 
obtained by deleting 
%removing 
the $k$-th row and the $j$-th column 
from the determinant of $U$ 
and multiplying it by $(-1)^{k+j}$. 
%matrix. 
%calculating the asymptotic behavior. 
Using the definition of $U$ 
given in 
(\ref{U-def}) and 
the 
multilinearity of 
%a 
determinants, we can 
%calculate 
rewrite 
%the numerator of 
(\ref{N-soliton}) as 
\bea
\espace \vecq (x,t) 
\nn \\
\eq \frac{2}{\det U} \sum_{j=1}^N \sum_{k=1}^N 
\ti{U}_{kj} \e^{-\i \Thet_k} \vecu_k
\nn \\
\eq \frac{2}{\det U} \Bigl[
	(\ti{U}_{11} +\cdots +\ti{U}_{1N}) \e^{-\i \Thet_1} \vecu_1 + \cdots
	+ (\ti{U}_{N1} +\cdots + \ti{U}_{NN}) \e^{-\i \Thet_N} \vecu_N
	\Bigr] 
\nn \\
\eq \frac{2}{\det U} \left[ \hspace{1pt}
%\,
	\left| \begin{array}{cccc}
            1 & 1 & \cdots & 1 \\
            U_{21} & U_{22} & \cdots & U_{2N} \\
            \vdots & \vdots & \ddots & \vdots \\
            U_{N1} & U_{N2} & \cdots & U_{NN} \\
	       \end{array} \right| \e^{-\i \Thet_1} \vecu_1 
                 + \cdots +
	\left| \begin{array}{cccc}
            U_{11} & U_{12} & \cdots & U_{1N} \\
            \vdots & \vdots &        & \vdots \\
            U_{N-1 \hspace{1pt} 1} & U_{N-1 \hspace{1pt} 2} 
	& \cdots & U_{N-1 \hspace{1pt} N} \\
            1 & 1 & \cdots & 1 \\
       \end{array} \right| \e^{-\i \Thet_N} \vecu_N 
	\right]
\nn \\ 
\eq \frac{2}{\det U} \left[ 
\sum_{n=0}^{N-1} \hspace{1pt}
%(-1)^n 
	\sum_{2 \le j_1 < \cdots < j_n \le N} 
	\left( \dprod{k=2}{k \neq j_1,\cdots, j_n}^N 
	\frac{\e^{\tau_k + \a_k}}{2 \eta_k} \right)
\sum_{l_1,\cdots, l_n = 1}^N 
	\e^{- (\tau_{l_1} + \a_{l_1}) + \i (\Thet_{l_1} -\Thet_{j_1})}
\right.
\nn \\
\espace 
\hspace{15mm} 
\mbox{}	\times
	\cdots \times
	\e^{- (\tau_{l_n} + \a_{l_n}) + \i (\Thet_{l_n} -\Thet_{j_n})}
	\left| \begin{array}{cccc}          
	   1 & 1& \cdots & 1 \\
           \la_{j_1 1 l_1} & \la_{j_1 j_1 l_1} & \cdots & \la_{j_1 j_n l_1} \\
	   \vdots & \vdots & \ddots & \vdots  \\
           \la_{j_n 1 l_n} & \la_{j_n j_1 l_n} & \cdots & \la_{j_n j_n l_n} 
        \end{array} \right| \e^{-\i \Thet_1} \vecu_1 
\nn \\
\espace
\hspace{15mm} 
\mbox{} + \cdots + 
\sum_{n=0}^{N-1} \hspace{1pt}
%(-1)^n 
\sum_{1 \le j_1 < \cdots < j_n \le N-1} 
	\left( \dprod{k=1}{k \neq j_1, \cdots, j_n}^{N-1} 
	\frac{\e^{\tau_k + \a_k}}{2 \eta_k} \right)
\sum_{l_1, \cdots, l_n = 1}^N 
	\e^{- (\tau_{l_1} + \a_{l_1}) + \i (\Thet_{l_1} -\Thet_{j_1})}
\nn \\
\espace
\hspace{15mm}
\left. 
\mbox{} \times 
\cdots \times \e^{- (\tau_{l_n} + \a_{l_n}) + \i (\Thet_{l_n} -\Thet_{j_n})}
	\left| \begin{array}{cccc}          
      \la_{j_1 j_1 l_1} & \cdots & \la_{j_1 j_{n} l_1} & \la_{j_1 N l_1} \\
      \vdots & \ddots & \vdots & \vdots  \\
      \la_{j_n j_1 l_n} & \cdots & \la_{j_n j_{n} l_n}  & \la_{j_n N l_n} \\
       1 &  \cdots & 1 & 1 
        \end{array} \right| \e^{-\i \Thet_N} \vecu_N
\right].
\label{num-den}
\eea
%
%Let us compute the numerator and the denominator of eq.\ (\ref{num-den}), 
%separately. 
%respectively. 
%
%Taking into account 
%
%Using the multilinearity of a determinant again, 
Similarly, we can rewrite the determinant of $U$ 
%$\det U$ 
%compute the denominator 
%of eq.\ (\ref{num-den}) 
%, $\det U$, 
as
\bea
\espace 
\det U 
%%\nn \\
%%\eq
%
%= \det \left[ \frac{\e^{\tau_l + \a_l}}{2\eta_l} 
%\de_{lk} - \sum_{j=1}^N 
%\e^{ -(\tau_j + \a_j) + \i (\Thet_j - \Thet_l)} \la_{lkj} \right]_{l, k=1}^N
\nn \\
\eq \prod_{k=1}^N \frac{\e^{\tau_k + \a_k}}{2 \eta_k } 
+ \sum_{j_1=1}^N \left(
\dprod{k=1}{k \neq j_1}^N \frac{\e^{\tau_k + \a_k}}{2 \eta_k} \right)
	\sum_{l_1=1}^N 	\e^{- (\tau_{l_1} + \a_{l_1})+ 
	\i (\Thet_{l_1} -\Thet_{j_1})} \la_{j_1 j_1 l_1} 
\nn \\
\espace 
\mbox{} + 
	\sum_{1 \le j_1 < j_2 \le N} \left(
	\dprod{k=1}{k \neq j_1, j_2}^N 
	\frac{\e^{\tau_k + \a_k}}{2 \eta_k} \right)
	\sum_{l_1,l_2=1}^N 
	\e^{- (\tau_{l_1} + \a_{l_1}) + \i (\Thet_{l_1} -\Thet_{j_1})}
	\e^{- (\tau_{l_2} + \a_{l_2}) + \i (\Thet_{l_2} -\Thet_{j_2})}
	\left| \begin{array}{cc}          
           \la_{j_1 j_1 l_1} & \la_{j_1 j_2 l_1} \\
           \la_{j_2 j_1 l_2} & \la_{j_2 j_2 l_2} 
        \end{array} \right|
\nn \\ 
\espace   \mbox{} + \cdots + 
%(-1)^N 
	\sum_{l_1, \cdots, l_N = 1}^N 
	\e^{- (\tau_{l_1} + \a_{l_1}) + \i (\Thet_{l_1} -\Thet_{1})}\cdots 
	\e^{- (\tau_{l_N} + \a_{l_N}) + \i (\Thet_{l_N} -\Thet_{N})}
	\left| \begin{array}{ccc}          
           \la_{1 1 l_1} & \cdots & \la_{1 N l_1} \\
	   \vdots & \ddots & \vdots  \\
           \la_{N 1 l_N} & \cdots & \la_{N N l_N} 
        \end{array} \right|
\nn \\
\eq 
\sum_{n=0}^N \hspace{1pt}
%(-1)^n 
	\sum_{1 \le j_1 < \cdots < j_n \le N} 
	\left(
	\dprod{k=1}{k \neq j_1, \cdots, j_n}^N 
	\frac{\e^{\tau_k + \a_k}}{2 \eta_k} \right)
\nn \\
\espace \hspace{0mm}
	\mbox{} \times
\sum_{l_1, \cdots, l_n = 1}^N 
	\e^{- (\tau_{l_1} + \a_{l_1}) + \i (\Thet_{l_1} -\Thet_{j_1})}\cdots 
	\e^{- (\tau_{l_n} + \a_{l_n}) + \i (\Thet_{l_n} -\Thet_{j_n})}
	\left| \begin{array}{ccc}          
           \la_{j_1 j_1 l_1} & \cdots & \la_{j_1 j_n l_1} \\
	   \vdots & \ddots & \vdots  \\
           \la_{j_n j_1 l_n} & \cdots & \la_{j_n j_n l_n} 
        \end{array} \right|.
\label{detU}
\eea
%
%We should 
Here, we note [\hspace{0pt}cf.\ the definition of 
$\la_{jkl}$ (\ref{lambda})] 
that the quantity
\[
\frac{\bigl( \vecu_{j'} \cdot \vecu_l^\dagger \bigr) }{\z_l - \z_{j'}^\ast}
\la_{jkl} = 
- \frac{\bigl( \vecu_{j'} \cdot \vecu_l^\dagger \bigr)
	2 \eta_l \bigl( \vecu_{j} \cdot \vecu_l^\dagger \bigr)}
{(\z_l - \z_{j'}^\ast)(\z_l - \z_k^\ast)(\z_l - \z_j^\ast)}
\]
is invariant under 
%symmetric with respect to 
interchange of the subscripts 
$j$ and $j'$. Thus, we have 
\[
\frac{\bigl( \vecu_{j'} \cdot \vecu_l^\dagger \bigr) }{\z_l - \z_{j'}^\ast}
\la_{jkl} 
-\frac{\bigl( \vecu_{j} \cdot \vecu_l^\dagger \bigr) }{\z_l - \z_{j}^\ast}
\la_{j' kl} =0.
\]
This 
%relation 
shows that 
%, 
if $\vecu_{j'} \cdot \vecu_l^\dagger \neq 0$ or 
$\vecu_{j} \cdot \vecu_l^\dagger \neq 0$, 
the two vectors 
$(\la_{j 1 l}, \la_{j 2 l}, \cdots, \la_{j N l})$ and 
$(\la_{j' 1 l}, \la_{j' 2 l}, \cdots, \la_{j' N l})$ are 
linearly dependent. 
In the case that 
%where 
$\vecu_{j'} \cdot \vecu_l^\dagger = \vecu_{j} \cdot \vecu_l^\dagger = 0$, 
according to 
%the definition of $\la_{jkl}$ 
(\ref{lambda}), 
both vectors become zero. 
Therefore, the determinants 
in 
%eqs.\ 
(\ref{num-den}) 
%and 
or (\ref{detU}) contribute 
only if $l_1, \cdots, l_n$ are all distinct. 
%identical. 
This fact is a generalization of the relations (\ref{la22}). 
%in section~\ref{Two-soliton}. 

%Let us investigate an asymptotic behavior of 
%the $N$-soliton solutions at $t \to \pm \infty$.
Next, we assume that 
\[ 
\xi_1 (= \mbox{Re}\, \z_1)> \xi_2 (= \mbox{Re}\, \z_2) > \cdots
> \xi_N (= \mbox{Re}\, \z_N),
\]
and investigate 
%discuss 
the asymptotic behavior of $\vt{q} (x,t)$ 
%as 
in the limits 
$t \to \mp \infty$. 
This is accomplished by 
%picking 
identifying 
the dominant terms 
%from 
%which terms in the numerator/denominator of (\ref{num-den}) 
in 
the numerator of (\ref{num-den}) and its denominator (\ref{detU}). 
We 
%should 
here 
%remark 
note the relations 
$\tau_j/\eta_j = \tau_k/\eta_k + 8(\xi_j- \xi_k)t$. 

\vspace{2.5mm}
%\begin{enumerate}
%\item[($-$)] 
In the limit $t \to - \infty$, we have 
\[
\frac{\tau_1}{\eta_1} \ll \frac{\tau_2}{\eta_2} \ll \cdots \ll
\frac{\tau_N}{\eta_N}.
\]
In this case, we have to consider 
%in this limit 
the following $N$ regions ($1^-$)--($N^-$) separately. 
It is easily seen that 
$\vt{q} \simeq \vt{0}$ in 
%the 
all other regions. 
%Let us calculate the asymptotic form of $\vecq$ in each region 
%from (\ref{num-den}). 
\begin{enumerate}
\item[($1^-$)] 
${\rm finite}~\tau_1,\;\;\; \tau_2, \cdots, \tau_N \to + \infty$
%$\tau_1 (\sim {\rm finite}) \ll \tau_2, \cdots, \tau_N (\to + \infty)$

In this case, 
the dominant terms are those which 
%including 
contain the factor $\e^{\tau_2 + \cdots + \tau_N}$. 
%are dominant. 
Then, 
using the relation $\la_{111} = 1/(2 \eta_1)$, we obtain
\bea
\vecq \eesim \frac{ 2
\left( \prod_{k=2}^N \frac{\e^{\tau_k + \a_k}}{2 \eta_k} \right)
	\e^{-\i \Thet_1} \vecu_1}
{\prod_{k=1}^N \frac{\e^{\tau_k + \a_k}}{2 \eta_k} 
 + \left( \prod_{k=2}^N \frac{\e^{\tau_k + \a_k}}{2 \eta_k} \right)
\e^{-(\tau_1 + \a_1)} \la_{111}}
\nn \\ \eq 
2 \eta_1 \sech (\tau_1 + \a_1) \e^{-\i \Thet_1} \vecu_1.
\label{1-}
\eea
%
%\item[$\vdots$]
\item[($n^-$)] 
$\tau_1, \cdots, \tau_{n-1} \to - \infty,\;\;\;
        {\rm finite}~\tau_n, \;\;\; 
	\tau_{n+1}, \cdots,\tau_N \to + \infty, 
\hspace{8mm} n=2, \cdots, N-1$
%\item[($n^-$)] $\tau_1, \cdots, \tau_{n-1} (\to - \infty) \ll 
%	\tau_n (\sim {\rm finite}) \ll \tau_{n+1}, \cdots,
%	\tau_N (\to + \infty), \hspace{8mm} n=2, \cdots, N-1$

Here, 
the dominant terms are those which contain 
%including 
the factor $\e^{- \tau_1- \cdots - \tau_{n-1}
	+ \tau_{n+1} + \cdots + \tau_N}$. 
%are dominant. 
Then, 
%the dominant terms 
those in the numerator of (\ref{num-den}) are
\bea
\espace 
2 
%(-1)^{n-1} 
\sum_{j=1}^n \left( \prod_{k=n+1}^N 
\frac{\e^{\tau_k + \a_k} }{2 \eta_k} \right)
\dsum{\{ l_1, \cdots, l_{n-1} \} }
	{= \{ 1, \cdots, n-1 \} } 
	\e^{-(\tau_1 + \a_1)} \cdots \e^{-(\tau_{n-1} + \a_{n-1})}
	\e^{\i (\Thet_j -\Thet_n)} 
\nn \\
\espace 
\mbox{} \times	\left| \begin{array}{ccc}          
           \la_{1 1 l_1} & \cdots & \la_{1 n l_1} \\
	   \vdots &   & \vdots  \\
           \la_{j-1 \hspace{1pt} 1 \hspace{1pt} l_{j-1}} & \cdots 
	& \la_{j-1 \hspace{1pt} n \hspace{1pt} l_{j-1}} \\
	            1 & \cdots & 1 \\
           \la_{j+1 \hspace{1pt} 1 \hspace{1pt} l_{j}} & \cdots 
	& \la_{j+1 \hspace{1pt} n \hspace{1pt} l_{j}} \\
	   \vdots &   & \vdots  \\
           \la_{n 1 l_{n-1}} & \cdots & \la_{n n l_{n-1}} 
        \end{array} \right| 
	\e^{-\i \Thet_j} \vecu_j,
\nn
\eea
while 
%the dominant terms 
those in the denominator 
%its denominator 
(\ref{detU}) are 
%including the factor $\e^{- \tau_1- \cdots - \tau_{n-1}	
%+ \tau_{n+1} + \cdots + \tau_N}$ 
\bea
\espace \hspace{0mm}
%(-1)^{n-1} 
\left( \prod_{k=n}^N 
\frac{\e^{\tau_k + \a_k} }{2 \eta_k} \right)
\dsum{\{ l_1, \cdots, l_{n-1} \} }
	{= \{ 1, \cdots, n-1 \} } 
	\e^{-(\tau_1 + \a_1)} \cdots \e^{-(\tau_{n-1} + \a_{n-1})}
\left| \begin{array}{ccc}          
           \la_{1 1 l_1} & \cdots 
	& \la_{1 \hspace{1pt} n-1 \hspace{1pt} l_1} \\
	    \vdots & \ddots  & \vdots  \\
	    \la_{n-1 \hspace{1pt} 1 \hspace{1pt} l_{n-1}} & \cdots 
	& \la_{n-1 \hspace{1pt} n-1 \hspace{1pt} l_{n-1}} 
        \end{array} \right| 
\nn \\
\espace \hspace{0mm}
\mbox{}+ 
%(-1)^{n} 
\left( \prod_{k=n+1}^N 
\frac{\e^{\tau_k + \a_k} }{2 \eta_k} \right)
\dsum{\{ l_1, \cdots, l_{n} \} }
	{= \{ 1, \cdots, n \} } 
	\e^{-(\tau_1 + \a_1)} \cdots 
%\e^{-(\tau_{n-1} + \a_{n-1})}
	\e^{-(\tau_{n} + \a_{n})}
\left| \begin{array}{ccc}          
           \la_{1 1 l_1} & \cdots & \la_{1 n l_1} \\
	   \vdots & \ddots  & \vdots  \\
           \la_{n 1 l_{n}} & \cdots & \la_{n n l_{n}} 
        \end{array} \right|.
\nn
\eea
As a natural extension 
%generalization 
of (\ref{phi12-def}), 
we define 
%a quantity 
$\phi_{i_1 i_2 \cdots i_p}$ 
for distinct positive integers $
%\{ 
i_1, i_2, \cdots, i_p 
%\}
$ by 
\bea
\e^{-2 \phi_{ i_1 i_2 \cdots i_p}}
\equiv 
\frac{1}{\la_{i_1 i_1 i_1} \cdots \la_{i_p i_p i_p}}
\dsum{\{ l_1, \cdots, l_{p} \} }
	{= \{ i_1, \cdots, i_p \} } 
\left| \begin{array}{ccc}          
           \la_{i_1 i_1 l_1} & \cdots & \la_{i_1 i_p l_1} \\
	   \vdots & \ddots  & \vdots  \\
           \la_{i_p i_1 l_{p}} & \cdots & \la_{i_p i_p l_{p}} 
        \end{array} \right|.
\label{phi-def}
\eea
%which will be proved to be 
%
We should note that $\phi_{ i_1 i_2 \cdots i_p}$ is symmetric 
with respect to permutations of the subscripts $i_1, i_2, \cdots, i_p$. 
We 
%will 
prove 
%later 
below that 
$\phi_{ i_1 i_2 \cdots i_p}$ 
%is 
can always be taken as real. 
%since 
% to be 
%later that 
%the right-hand side of (\ref{phi-def}) is always positive. 
%later. 
%
In terms of $\phi_{i_1 i_2 \cdots i_p}$, 
we can express 
the asymptotic form of $\vecq$ 
in this region 
%is expressed 
as
\bea
\vecq \eesim 2 \eta_n \sech (\tau_n + \a_n 
	+ \phi_{12 \cdots n} - \phi_{1 2 \cdots \hspace{1pt} n-1} ) 
	\e^{-\i \Thet_n} 
	\e^{\phi_{12 \cdots n} + \phi_{1 2 \cdots\hspace{1pt} n-1} }
\nn \\
\espace 
\mbox{} \times	
\frac{1}{\la_{111} \cdots \la_{n-1 \hspace{1pt} n-1 \hspace{1pt} n-1}}
\sum_{j=1}^n
\dsum{\{ l_1, \cdots, l_{n-1} \} }
	{= \{ 1, \cdots, n-1 \} } 
\left| \begin{array}{ccc}          
           \la_{1 1 l_1} & \cdots & \la_{1 n l_1} \\
	   \vdots &   & \vdots  \\
           \la_{j-1 \hspace{1pt} 1 \hspace{1pt} l_{j-1}} & \cdots 
                    & \la_{j-1 \hspace{1pt} n \hspace{1pt} l_{j-1}} \\
	            1 & \cdots & 1 \\
           \la_{j+1 \hspace{1pt} 1 \hspace{1pt} l_{j}} & \cdots 
		    & \la_{j+1 \hspace{1pt} n \hspace{1pt} l_{j}} \\
	   \vdots &   & \vdots  \\
           \la_{n 1 l_{n-1}} & \cdots & \la_{n n l_{n-1}} 
        \end{array} \right|  \vecu_j.
%\nn \\ \espace
\label{n-}
\eea
%
%\item[$\vdots$]
\item[($N^-$)] 
$\tau_1, \cdots, \tau_{N-1} \to - \infty,\;\;\; {\rm finite}~\tau_N$
%$\tau_1, \cdots, \tau_{N-1} (\to - \infty) \ll \tau_N (\sim {\rm finite})$

Here, 
the dominant terms are those which contain 
%including 
the factor $\e^{- \tau_1- \cdots - \tau_{N-1}}$. 
%are dominant. 
%Repeating the same 
With calculations similar to 
those in 
the case ($n^-$), 
%case, 
we obtain the asymptotic form of $\vt{q}$ 
%in this region 
%It 
%, which is 
%simply 
given by 
(\ref{n-}) with 
%setting 
$n=N$. 
%as expected. 
%\\
\end{enumerate}
%
%\item[($+$)] 
\vspace{2.5mm}
In the limit $t \to + \infty$, we have 
\[
\frac{\tau_1}{\eta_1} \gg \frac{\tau_2}{\eta_2} \gg \cdots \gg
\frac{\tau_N}{\eta_N}.
\]
In this case, we have to consider the following $N$ regions ($N^+$)--($1^+$) 
separately. 
It is easily seen that $\vt{q} \simeq \vt{0}$ 
in all 
%the 
other regions. 
\begin{enumerate}
\item[($N^+$)] 
$\tau_1, \cdots, \tau_{N-1} \to + \infty,\;\;\; 
{\rm finite}~\tau_N$ 
%$\tau_1, \cdots, \tau_{N-1} (\to + \infty) \gg \tau_N (\sim {\rm finite})$

In this case, 
the dominant terms are those which contain 
%including 
the factor $\e^{\tau_1 + \cdots + \tau_{N-1}}$. 
%are dominant. 
Then, 
using the relation $\la_{NNN} = 1/(2 \eta_N)$, we obtain
\bea
\vecq \eesim \frac{ 2
\left( \prod_{k=1}^{N-1} 
	\frac{\e^{\tau_k + \a_k}}{2 \eta_k} \right) \e^{-\i \Thet_N} \vecu_N}
{\prod_{k=1}^N \frac{\e^{\tau_k + \a_k}}{2 \eta_k} + 
 \left( \prod_{k=1}^{N-1} \frac{\e^{\tau_k + \a_k}}{2 \eta_k} \right)
	\e^{-(\tau_N + \a_N)} \la_{NNN}}
\nn \\ \eq 
2 \eta_N \sech (\tau_N + \a_N) \e^{-\i \Thet_N} \vecu_N.
\label{N+}
\eea
\item[($n^+$)] 
	$\tau_1, \cdots, \tau_{n-1} \to + \infty,\;\;\;
	{\rm finite}~\tau_n, \;\;\; \tau_{n+1}, \cdots,
	\tau_N \to - \infty, \hspace{8mm} n=2, \cdots, N-1$
%	$\tau_1, \cdots, \tau_{n-1} (\to + \infty) \gg
%	\tau_n (\sim {\rm finite}) \gg \tau_{n+1}, \cdots,
%	\tau_N (\to - \infty), \hspace{8mm} n=2, \cdots, N-1$

In this case, 
the dominant terms are those which contain 
%including 
the factor $\e^{ \tau_1+ \cdots + \tau_{n-1}
	- \tau_{n+1} - \cdots - \tau_N}$. 
%are dominant.
Then, those in the numerator of (\ref{num-den}) are 
%the dominant terms in the numerator are
\bea
\espace 
2 
%(-1)^{N-n} 
\sum_{j=n}^N \left( \prod_{k=1}^{n-1}  
\frac{\e^{\tau_k + \a_k} }{2 \eta_k} \right)
\dsum{\{ l_1, \cdots, l_{N-n} \} }
	{= \{ n+1, \cdots, N \} } 
	\e^{-(\tau_{n+1} + \a_{n+1})} \cdots \e^{-(\tau_{N} + \a_{N})}
	\e^{\i (\Thet_j -\Thet_n)} 
\nn \\
\espace 
\mbox{} \times	\left| \begin{array}{ccc}          
           \la_{n n l_1} & \cdots & \la_{n N l_1} \\
	   \vdots &   & \vdots  \\
           \la_{j-1 \hspace{1pt} n \hspace{1pt} l_{j-n}} & \cdots 
                    & \la_{j-1 \hspace{1pt} N \hspace{1pt} l_{j-n}} \\
	            1 & \cdots & 1 \\
         \la_{j+1 \hspace{1pt} n \hspace{1pt} l_{j-n+1}} 
               & \cdots & \la_{j+1 \hspace{1pt} N \hspace{1pt} l_{j-n+1}} \\
	   \vdots &   & \vdots  \\
           \la_{N  n   l_{N-n}} & \cdots & \la_{N N  l_{N-n}} 
        \end{array} \right| 
	\e^{-\i \Thet_j} \vecu_j,
\nn
\eea
while 
%the dominant terms 
those in the denominator (\ref{detU}) are 
%including the factor $\e^{- \tau_1- \cdots - \tau_{n-1}	
%+ \tau_{n+1} + \cdots + \tau_N}$ 
\bea
\espace \hspace{0mm}
%(-1)^{N-n} 
\left( \prod_{k=1}^n 
\frac{\e^{\tau_k + \a_k} }{2 \eta_k} \right)
\dsum{\{ l_1, \cdots, l_{N-n} \} }
	{= \{ n+1, \cdots, N \} } 
	\e^{-(\tau_{n+1} + \a_{n+1})} \cdots \e^{-(\tau_{N} + \a_{N})}
\left| \begin{array}{ccc}          
           \la_{n+1 \hspace{1pt} n+1 \hspace{1pt} l_1} & \cdots 
		& \la_{n+1 \hspace{1pt} N \hspace{1pt} l_1} \\
	   \vdots & \ddots  & \vdots  \\
           \la_{N \hspace{1pt} n+1 \hspace{1pt} l_{N-n}} & \cdots 
		& \la_{N N  l_{N-n}} 
        \end{array} \right| 
\nn \\
\espace \hspace{0mm}
\mbox{}+ 
%(-1)^{N-n+1} 
\left( \prod_{k=1}^{n-1} 
\frac{\e^{\tau_k + \a_k} }{2 \eta_k} \right)
\dsum{\{ l_1, \cdots, l_{N-n+1} \} }
	{= \{ n, \cdots, N \} } 
	\e^{-(\tau_n + \a_n)} \cdots 
%\e^{-(\tau_{n} + \a_{n})}
	\e^{-(\tau_{N} + \a_{N})}
\left| \begin{array}{ccc}          
           \la_{n n l_1} & \cdots & \la_{n N l_1} \\
	   \vdots & \ddots  & \vdots  \\
           \la_{N n  l_{N-n+1}} & \cdots & \la_{N N  l_{N-n+1}} 
        \end{array} \right|.
\nn
\eea
In terms of $\phi_{i_1 i_2 \cdots i_p}$ defined by (\ref{phi-def}), 
we can express the asymptotic form of $\vecq$ in this region 
%is written 
as
\bea
\vecq \eesim 2 \eta_n \sech (\tau_n + \a_n 
	+ \phi_{n \hspace{1pt} n+1 \hspace{1pt}\cdots N} 
        - \phi_{n+1 \hspace{1pt} n+2 \hspace{1pt}\cdots N} ) 
	\e^{-\i \Thet_n} 
	\e^{\phi_{n \hspace{1pt} n+1 \hspace{1pt}\cdots N} 
		+ \phi_{n+1 \hspace{1pt} n+2 \hspace{1pt}\cdots N} }
\nn \\
\espace 
\mbox{} \times	
\frac{1}{\la_{n+1 \hspace{1pt} n+1 \hspace{1pt} n+1} \cdots \la_{NNN}}
\sum_{j=n}^N
\dsum{\{ l_1, \cdots, l_{N-n} \} }
	{= \{ n+1, \cdots, N \} } 
\left| \begin{array}{ccc}          
           \la_{n n l_1} & \cdots & \la_{n N  l_1} \\
	   \vdots &   & \vdots  \\
           \la_{j-1 \hspace{1pt} n \hspace{1pt} l_{j-n}} & \cdots 
		& \la_{j-1 \hspace{1pt} N \hspace{1pt} l_{j-n}} \\
	            1 & \cdots & 1 \\
         \la_{j+1 \hspace{1pt} n \hspace{1pt} l_{j-n+1}} & \cdots 
		& \la_{j+1 \hspace{1pt} N \hspace{1pt} l_{j-n+1}} \\
	   \vdots &   & \vdots  \\
           \la_{N n l_{N-n}} & \cdots & \la_{N N  l_{N-n}}
        \end{array} \right|  \vecu_j.
\hspace{7mm}
%\nn \\ \espace
\label{n+}
\eea
\item[($1^+$)] 
${\rm finite}~\tau_1,\;\;\; \tau_2, \cdots, \tau_N \to - \infty$
%$\tau_1 (\sim {\rm finite}) \gg \tau_2, \cdots, \tau_N (\to - \infty)$

In this case, 
the dominant terms are those which contain 
%including 
the factor $\e^{-\tau_2  - \cdots - \tau_N}$. 
%are dominant. 
%Repeating the same 
With calculations 
%as in 
similar to 
those in 
the case ($n^+$), 
we obtain the asymptotic form of $\vt{q}$ 
%in this region. It is simply 
given by (\ref{n+}) with 
%setting 
$n=1$. 
\\
\end{enumerate}
%\vspace{5mm}

We now 
%shall 
%Now, 
%can 
simplify the 
above 
%obtained 
asymptotic forms 
%of the $N$-soliton solution 
%in the above 
%by using 
%with 
%in terms of 
by 
using the following 
%abbreviations
definitions: 
%\beq
\begin{equation}
c_{jk} \equiv \frac{\i}{\z_k-\z_j^\ast}, \hspace{5mm} 
d_{jk} \equiv \frac{\i }{\z_k - \z_j^\ast}
\left( \vecu_j \cdot
    \vecu_k^\dagger \right).
\label{cd-def}
%\eeq
\end{equation}
%We should note the relation 
%$\la_{lkj} = - 2 \eta_j c_{kj} d_{lj}$ (see (\ref{lambda})). 
According to 
%From 
the definition of $\la_{jkl}$ 
(\ref{lambda}), we have 
$\la_{jkl} =  2 \eta_l c_{kl} d_{jl}$. 
Then, we can rewrite the definition of $\phi_{i_1 i_2 \cdots i_n}$ 
[(\ref{phi-def}) with $p \to n$] 
in 
%as 
a factorized form 
%see also 
[\hspace{0pt}cf.\ the paragraph below (\ref{detU})]: 
%(see 
\bea
\hspace{-10mm}
\e^{-2 \phi_{ i_1 i_2 \cdots i_n}}
\eq 
\prod_{l=1}^{n} (2 \eta_{i_l}) \times
\sum_{ l_1= i_1, \cdots, i_n} \cdots \sum_{ l_n= i_1, \cdots, i_n} 
\left| \begin{array}{ccc}          
   2\eta_{l_1} c_{i_1 l_1} d_{i_1 l_1} & \cdots & 2\eta_{l_1} c_{i_n l_1} d_{i_1 l_1} \\
    \vdots & \ddots  & \vdots  \\
  2\eta_{l_n} c_{i_1 l_{n}} d_{i_n l_n} & \cdots & 2\eta_{l_n} c_{i_n l_{n}} d_{i_n l_n}
        \end{array} \right| 
\nn \\
\eq \prod_{l=1}^{n} (2 \eta_{i_l})^2 \times
	\left| \begin{array}{ccc}
	  d_{i_1 i_1} & \cdots & d_{i_1 i_n} \\
	   \vdots & \ddots  & \vdots  \\
	  d_{i_n i_1} & \cdots & d_{i_n i_n} \\	  
        \end{array} \right| 
\times 
	\left| \begin{array}{ccc}
	  c_{i_1 i_1} & \cdots & c_{i_n i_1} \\
	   \vdots & \ddots  & \vdots  \\
	  c_{i_1 i_n} & \cdots & c_{i_n i_n} \\	  
        \end{array} \right|.
\label{phi-def2}
\eea
Here $ i_1, i_2, \cdots, i_n $ are distinct positive integers. 
For any nonzero vector $(y_{i_1}, \cdots, y_{i_n})$, we have 
%that 
\bea
%\espace
\left(
%\begin{array}{c}
 y_{i_1}, \cdots,  y_{i_n}
%\end{array}
\right)
\underline{
	\left( \begin{array}{ccc}
	  d_{i_1 i_1} & \cdots & d_{i_1 i_n} \\
	   \vdots & \ddots  & \vdots  \\
	  d_{i_n i_1} & \cdots & d_{i_n i_n} \\	  
        \end{array} \right) 
}
\left(
\begin{array}{c}
 y_{i_1}^\ast \\
 \vdots \\
 y_{i_n}^\ast
\end{array}
\right) 
%\nn \\
\eq \sum_{j, k = i_1, \cdots, i_n} 
\frac{\i}{\z_{k} -\z_{j}^\ast} \left( \vt{u}_{j} 
	\cdot \vt{u}_{k}^\dagger \right) y_{j} y_{k}^\ast 
\nn \\
\eq \sum_{j, k = i_1, \cdots, i_n} \int_0^\infty 
\e^{\i (\z_k -\z_j^\ast) z} \left( \vt{u}_{j} 
	\cdot \vt{u}_{k}^\dagger \right) y_{j} y_{k}^\ast \hspace{1pt} \d z 
\nn \\
\eq 
%\mbox{\huge $\int$} 
 \int_0^\infty
%\left| 
\Biggl|
\! 
\Biggl|
%\left| 
\sum_{j = i_1, \cdots, i_n} \e^{-\i \z_j^\ast z} y_j \vt{u}_j 
%\right| \! \right|
\Biggl| \! \Biggl|^2 \d z 
\nn \\
&>& 
%\hspace{-2mm} 
0.
\nn
\eea
%This proves that 
Thus, the eigenvalues of the underlined 
Hermitian matrix 
%
%\[
%	\left( \begin{array}{ccc}
%	  d_{i_1 i_1} & \cdots & d_{i_1 i_n} \\
%	   \vdots & \ddots  & \vdots  \\
%	  d_{i_n i_1} & \cdots & d_{i_n i_n} \\	  
%        \end{array} \right) 
%\]
%for distinct integers $ i_1, \cdots, i_n $ 
are all positive. 
%Thus its determinant must be 
This proves that the second term on the right-hand side 
of (\ref{phi-def2}) is 
%always 
positive. 
Considering the special case 
%where 
in which 
all the vectors 
$\vt{u}_{i_1}, \cdots, \vt{u}_{i_n}$ are identical, 
we can prove the same 
%fact 
for the third term 
%of 
in (\ref{phi-def2}). 
%parallel, 
%matrix $(c_{i_l i_k})$. 
%This shows that 
Therefore, the right-hand side of (\ref{phi-def2}) is 
positive, and 
%that 
$\phi_{i_1 i_2 \cdots i_n}$ 
%is 
can always be taken as real. 
%
%Through a technique of the factorization in  (\ref{phi-def2}), 
%As is similar to the factorization in 
In 
the same way as for 
%that in 
%a similar way 
%similar to as 
(\ref{phi-def2}), 
we can rewrite the second line 
%appearing 
%in 
of (\ref{n-}) or (\ref{n+}) 
%as a 
in the following 
factorized form: 
%in a factorized form as 
\bea
\espace
\frac{1}{\la_{i_1 i_1 i_1} \cdots \la_{i_{n-1} i_{n-1} i_{n-1}} }
\sum_{j=1}^n
\dsum{\{ l_1, \cdots, l_{n-1} \} }
	{= \{ i_1, \cdots, i_{n-1} \} } 
\left| \begin{array}{ccc}          
           \la_{i_1 i_1 l_1} & \cdots & \la_{i_1 i_n l_1} \\
	   \vdots &   & \vdots  \\
           \la_{i_{j-1} i_1 l_{j-1}} & \cdots & \la_{i_{j-1} i_n l_{j-1}} \\
	            1 & \cdots & 1 \\
           \la_{i_{j+1} i_1 l_{j}} & \cdots & \la_{i_{j+1} i_n l_{j}} \\
	   \vdots &   & \vdots  \\
           \la_{i_n i_1 l_{n-1}} & \cdots & \la_{i_n i_n l_{n-1}} 
        \end{array} \right|  \vecu_{i_j} 
\nn \\
\eq 
\prod_{l=1}^{n-1} (2 \eta_{i_l}) \times 
\sum_{j=1}^n 
\sum_{ l_1= i_1, \cdots, i_{n-1}} \cdots \sum_{ l_{n-1}= i_1, \cdots, i_{n-1}} 
\left| \begin{array}{ccc}          
  2\eta_{l_1} c_{i_1 l_1} d_{i_1 l_1} & \hspace{-1.5mm} \cdots \hspace{-1.5mm}
	&   
	2\eta_{l_1} c_{i_n l_1} d_{i_1 l_1} \\
	   \vdots &  & \vdots  \\
  2\eta_{l_{j-1}} c_{i_1 l_{j-1}} d_{i_{j-1} l_{j-1}} 
          & 
	\hspace{-1.5mm}\cdots \hspace{-1.5mm}
	&  2\eta_{l_{j-1}} c_{i_n l_{j-1}} d_{i_{j-1} l_{j-1}} \\
            1 & 
	\hspace{-1.5mm}\cdots \hspace{-1.5mm}
	& 1 \\
  2\eta_{l_j}   c_{i_1 l_{j}} d_{i_{j+1} l_{j}} 
          & 
	\hspace{-1.5mm}\cdots \hspace{-1.5mm}
	&   2\eta_{l_j} c_{i_n l_{j}} d_{i_{j+1} l_{j}} \\
	   \vdots &  & \vdots  \\
  2\eta_{l_{n-1}}   c_{i_1 l_{n-1}} d_{i_n l_{n-1}} 
      & 
	\hspace{-1.5mm}\cdots \hspace{-1.5mm}
	&   2\eta_{l_{n-1}} c_{i_n l_{n-1}} d_{i_n l_{n-1}}
        \end{array} \right| \vt{u}_{i_j}
\nn \\
\eq
\prod_{l=1}^{n-1} (2 \eta_{i_l})^2 \times \sum_{j=1}^n
	\left| \begin{array}{cccc}
	  d_{i_1 i_1} & \cdots & d_{i_1 i_{n-1}} & 0 \\
	   \vdots & & \vdots & \vdots \\
	  d_{i_{j-1} i_1} & \cdots & d_{i_{j-1} i_{n-1}} & 0 \\
          0 & \cdots & 0 & 1 \\
	  d_{i_{j+1} i_1} & \cdots & d_{i_{j+1} i_{n-1}} & 0 \\	  
	   \vdots & & \vdots & \vdots \\
	  d_{i_n i_1} & \cdots & d_{i_n i_{n-1}} & 0 \\	  
        \end{array} \right| 
\times 
	\left| \begin{array}{ccc}
	  c_{i_1 i_1} & \cdots & c_{i_n i_1} \\
	   \vdots &  & \vdots  \\
	  c_{i_1 i_{n-1}} & \cdots & c_{i_n i_{n-1}} \\	  
          1 & \cdots & 1 \\
        \end{array} \right| \vt{u}_{i_j}
\nn \\
\eq
\prod_{l=1}^{n-1} (2 \eta_{i_l})^2 \times 
	\left| \begin{array}{ccc}
	  c_{i_1 i_1} & \cdots & c_{i_n i_1} \\
	   \vdots &  & \vdots  \\
	  c_{i_1 i_{n-1}} & \cdots & c_{i_n i_{n-1}} \\	  
          1 & \cdots & 1 \\
        \end{array} \right| 
\times
	\left| \begin{array}{cccc}
	  d_{i_1 i_1} & \cdots & d_{i_1 i_{n-1}} & \vt{u}_{i_1} \\
	   \vdots & & \vdots & \vdots \\
	  d_{i_n i_1} & \cdots & d_{i_n i_{n-1}} & \vt{u}_{i_n} \\	  
        \end{array} \right|.
\label{u1n}
\eea
%{\it Remark.} 
The last determinant, 
%whose 
%of which 
%the 
%last column 
which contains vectors in 
%the 
its last column, 
%denotes the vector obtained 
represents a vector 
%is 
defined 
in terms of 
%by 
the 
%a formal 
Laplace expansion with 
respect to the last column. 

%For further simplification of 
We can 
%To 
simplify the asymptotic forms 
%more 
further 
by noting 
some 
%the following 
relations between 
%for 
%expressing some ratios of 
%the 
the 
%usual 
conventional determinants 
%which appear 
in (\ref{phi-def2}) 
%or
and (\ref{u1n}). 
%which consist of 
%%contain 
%$c_{lk}$ or unity. 
%of the $N$-soliton solutions, 
%For this purpose, 
%
We have 
%prove 
%%a lemma. 
%%need 
%%prove 
the following 
%%prepare a 
lemma: 
\vspace{5mm}
\begin{lemma}
\label{lem-det}
{\it
The following 
%three 
equalities 
%relations 
%for 
involving 
determinants 
%relations of 
hold:
%\beq
\begin{equation}
	\left| \begin{array}{ccc}
	  c_{i_1 i_1} & \cdots & c_{i_n i_1} \\
	   \vdots &   & \vdots  \\
	  c_{i_1 i_{n-1}} & \cdots & c_{i_n i_{n-1} } \\ 
	  1 & \cdots & 1 \\	  
        \end{array} \right| 
= \prod_{l=1}^{n-1} \frac{\z_{i_l}^\ast -\z_{i_n}^\ast}
	{\z_{i_l} -\z_{i_n}^\ast} 
\times 
	\left| \begin{array}{ccc}
	  c_{i_1 i_1} & \cdots & c_{i_{n-1} i_1} \\
	   \vdots & \ddots  & \vdots  \\
	  c_{i_1 i_{n-1}} & \cdots & c_{i_{n-1} i_{n-1} } \\ 
        \end{array} \right|, 
\label{cc1}
%\eeq
\end{equation}
%
%\beq
\begin{equation}
	\left| \begin{array}{ccc}
	  c_{i_1 i_1} & \cdots & c_{i_{n} i_1} \\
	   \vdots & \ddots  & \vdots  \\
	  c_{i_1 i_{n}} & \cdots & c_{i_{n} i_{n} } \\ 
        \end{array} \right|
= \frac{\i \prod_{l=1}^{n-1} (\z_{i_n} -\z_{i_l})}
       {\prod_{l=1}^{n} (\z_{i_n} -\z_{i_l}^\ast)}
\times 
	\left| \begin{array}{ccc}
	  c_{i_1 i_1} & \cdots & c_{i_n i_1} \\
	   \vdots &   & \vdots  \\
	  c_{i_1 i_{n-1}} & \cdots & c_{i_n i_{n-1} } \\ 
	  1 & \cdots & 1 \\	  
        \end{array} \right|,
\label{cc2}
%\eeq
\end{equation}
%
%\beq
\begin{equation}
	\left| \begin{array}{ccc}
	  c_{i_1 i_1} & \cdots & c_{i_{n} i_1} \\
	   \vdots & \ddots  & \vdots  \\
	  c_{i_1 i_{n}} & \cdots & c_{i_{n} i_{n} } \\ 
        \end{array} \right| 
= \frac{\i}{\z_{i_n} -\z_{i_n}^\ast}
\prod_{l=1}^{n-1} \left| 
	\frac{\z_{i_l}-\z_{i_n}}{\z_{i_l} -\z_{i_n}^\ast}
	\right|^2 \times
	\left| \begin{array}{ccc}
	  c_{i_1 i_1} & \cdots & c_{i_{n-1} i_1} \\
	   \vdots & \ddots  & \vdots  \\
	  c_{i_1 i_{n-1}} & \cdots & c_{i_{n-1} i_{n-1} } \\ 
        \end{array} \right|.
\label{cc3}
%\eeq
\end{equation}
}
\end{lemma}
\vspace{5mm}
\noindent
{\bf Proof.\ } 
We can prove (\ref{cc1}) by subtracting on the left-hand side 
the last column from 
%the first through $(n-1)$-th columns. 
each of the other columns and using the relation
%,
\[
c_{jk} - c_{nk} = \frac{\z_j^\ast -\z_n^\ast}{\z_k -\z_n^\ast} c_{jk}. 
\]
Similarly, (\ref{cc2}) is proved by subtracting on the left-hand side 
the last row from 
%the first through $(n-1)$-th columns. 
each of the other rows and using the relation
%, 
\[
c_{jk} - c_{jn} = \frac{\z_n -\z_k}{\z_n -\z_j^\ast} c_{jk}. 
\]
(\ref{cc3}) is a direct consequence of (\ref{cc2}) and (\ref{cc1}). 
\hfill $\Box$
\vspace{5mm}

\begin{figure}[t]
  \begin{center}
    \includegraphics[width=16cm]{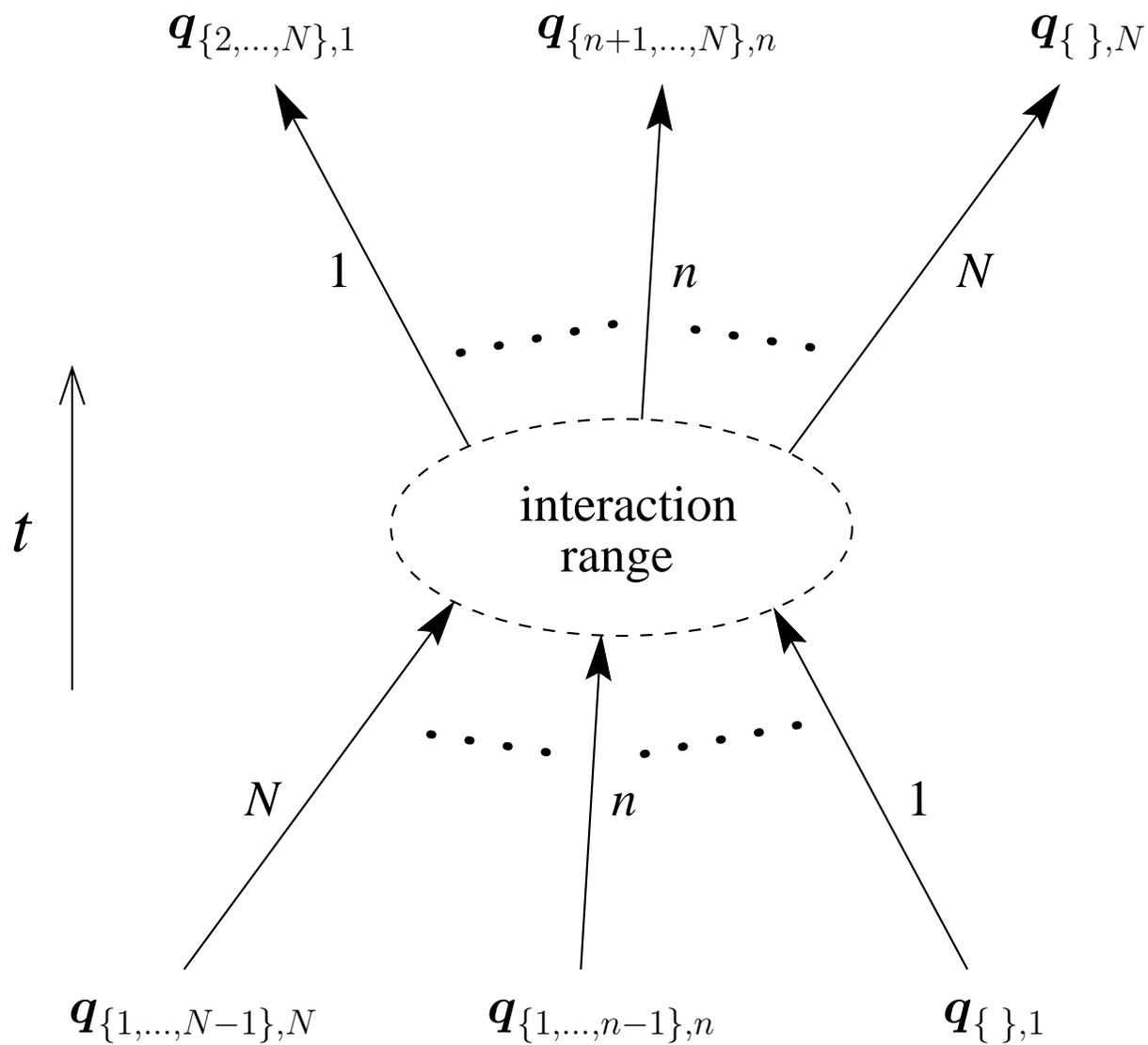}
  \\
    \vspace*{10mm}
    \caption{Asymptotic behavior of the $N$-soliton solution.}
    \label{Asym-N}
%  \nonumber
%\noindent
  \end{center}
\end{figure}

%Summing up 
Taking the sum of 
%eqs.\ 
(\ref{1-}) and (\ref{n-}) ($n=2, \cdots, N$), or 
%eqs.\ 
(\ref{N+}) and (\ref{n+}) ($n=1, \cdots, N-1$), with 
the help of (\ref{phi-def2}), (\ref{u1n}) and Lemma~\ref{lem-det}, 
we finally arrive at the following proposition. 
\vspace{5mm}
\begin{proposition}
\label{asym_N_sol}
{\it 
The asymptotic forms of the $N$-soliton solution of 
the Manakov model $(\ref{cNLS})$ are 
%given 
as follows:
\\
as $t \to - \infty$,
}
\bea
\vecq \eesim
\sum_{n=1}^N  
% \vecq_{\{1, \cdots, n-1\}, n},
2 \eta_n \sech \bigl(\tau_n + \a_n + \phi_{\{1, \cdots, n-1 \}, n} \bigr) 
\e^{-\i \Thet_n} \vecu_{\{1, \cdots, n-1\}, n};
\nn
\eea
{\it
%
%\noindent \point $\,$ 
as $t \to + \infty$,
}
\bea
\vecq \eesim
\sum_{n=1}^N 
%\vecq_{\{n+1, \cdots, N\}, n}.
2 \eta_n \sech \bigl( \tau_n + \a_n + \phi_{\{n+1, \cdots, N \}, n} \bigr) 
\e^{-\i \Thet_n} \vecu_{\{n+1, \cdots, N\}, n}.
\nn
\eea
{\it
Here $\phi_{ \{i_1, \cdots, i_{n-1}\},i_n}$ and 
$\vecu_{\{ i_1, \cdots, i_{n-1} \}, i_n}$ are defined for 
distinct positive integers $i_1, \cdots, i_{n-1}, i_n$ by 
%we have used the following abbreviations:
\bea
\e^{-2 \phi_{ \{i_1, \cdots, i_{n-1}\},i_n}} 
\eequiv 
\e^{-2 (\phi_{ i_1 \cdots i_{n-1} i_n} 
	- \phi_{ i_1 \cdots i_{n-1}})} 
\nn \\
\eq \prod_{l=1}^{n-1} 
\left| \frac{\z_{i_l} - \z_{i_n}}{\z_{i_l} - \z_{i_n}^\ast} \right|^2
\times
\frac{ \left|
\begin{array}{ccc}
d_{i_1 i_1} & \cdots & d_{i_1 i_n} \\
\vdots & \ddots & \vdots \\
d_{i_n i_1} & \cdots & d_{i_n i_n} \\
\end{array}
\right|}
{ d_{i_n i_n} \left|
\begin{array}{ccc}
d_{i_1 i_1} & \cdots & d_{i_1 i_{n-1}} \\
\vdots & \ddots & \vdots \\
d_{i_{n-1} i_1} & \cdots & d_{i_{n-1} i_{n-1}} \\
\end{array}
\right|}
\, 
%\hspace{1pt}
(> 0), 
\label{phi-def3}
\eea
and
\bea
\vecu_{\{ i_1, \cdots, i_{n-1} \}, i_n}
\eequiv 
\frac{\e^{\phi_{i_1 \cdots i_{n-1} i_n} 
	+ \phi_{i_1 \cdots i_{n-1}} }}
	{\la_{i_1 i_1 i_1} \cdots \la_{i_{n-1} i_{n-1} i_{n-1}}}
\sum_{j=1}^{n}
\dsum{\{ l_1, \cdots, l_{n-1} \} }
	{= \{ i_1, \cdots, i_{n-1} \} } 
\left| \begin{array}{ccc}          
           \la_{i_1 i_1 l_1} & \cdots & \la_{i_1 i_n l_1} \\
	   \vdots &   & \vdots  \\
       \la_{i_{j-1}  i_1  l_{j-1}} & \cdots & \la_{i_{j-1}  i_n l_{j-1}} \\
	            1 & \cdots & 1 \\
           \la_{i_{j+1}  i_1  l_{j}} & \cdots & \la_{i_{j+1}  i_n l_{j}} \\
	   \vdots &   & \vdots  \\
           \la_{i_n i_1 l_{n-1}} & \cdots & \la_{i_n i_n l_{n-1}} 
        \end{array} \right|  \vecu_{i_j}
\nn \\
\eq \e^{ \phi_{ \{i_1, \cdots, i_{n-1}\},i_n}} 
%\times
\prod_{l=1}^{n-1} 
\frac{\z_{i_l}^\ast - \z_{i_n}^\ast}{\z_{i_l} - \z_{i_n}^\ast}
\times 
\frac{\left|
\begin{array}{cccc}
d_{i_1 i_1} & \cdots & d_{i_1 i_{n-1}} & \vecu_{i_1} \\
\vdots & & \vdots & \vdots \\
d_{i_n i_1} & \cdots & d_{i_n i_{n-1}} & \vecu_{i_n} \\
\end{array}
\right|}
{ \left|
\begin{array}{ccc}
d_{i_1 i_1} & \cdots & d_{i_1 i_{n-1}} \\
\vdots & \ddots & \vdots \\
d_{i_{n-1} i_1} & \cdots & d_{i_{n-1} i_{n-1}} \\
\end{array}
\right|}.
\label{u-def3}
\eea
%
%respectively.
%as expected. 
%We can regard 
%In the case where 
When 
the set $\{ i_1, \cdots, i_{n-1} \}$ is empty, 
%compatible with 
%we read 
the definitions $(\ref{phi-def3})$ and 
$(\ref{u-def3})$ 
%can be 
%are read as 
should 
%be 
read 
%as 
%are considered as 
%being 
%valid even 
%(i.e.\ $n=1$): 
%\[
$\e^{-2 \phi_{\{ \; \}, i }} \equiv 1$ and 
%\hspace{5mm}
$\vt{u}_{\{ \; \}, i } \equiv \vt{u}_{i}$. 
%, respectively. 
%\]
}
\end{proposition}
\vspace{5mm}
%\noindent
%
%where
%\[
%%\hspace{15mm} 
%d_{jk} \equiv \frac{\i }{\z_k - \z_j^\ast}
%\bigl( \vecu_j \cdot
%    \vecu_k^\dagger \bigr).
%\]
%
%For brevity, 
%We employ the following abbreviation: 
We 
%will 
%show 
prove in the next section 
%later 
that 
%in the next section that 
$\vecu_{\{ i_1, \cdots, i_{n-1} \}, i_n}$ is always a unit 
%polarization 
vector, i.e.\ 
$\left| \! \left| 
	\vecu_{\{ i_1, \cdots, i_{n-1} \}, i_n} \right| \! \right| =1$. 
%Thus any 
This ensures 
%means 
%explicitly 
that 
%Thus, 
an $N$-soliton collision 
%as a whole 
does not change the amplitudes of solitons. 
%and 
%that 
The vector 
$\vecu_{\{1, \cdots, n-1 \}, n}$ 
%of the form $\vecu_{\{ i_1, \cdots, i_{n-1} \}, i_n}$ 
%in the above 
%asymptotic forms 
%is interpreted as 
%define 
gives 
%is 
the polarization vector of soliton-$n$ 
before an $N$-soliton collision, 
while $\vecu_{\{n+1, \cdots, N\}, n}$ 
gives that after the 
%$N$-soliton 
collision. 
Using the definition
%abbreviation, 
%$\vecq_{\{i_1, \cdots, i_{n-1} \}, i_n}$ defined by 
%
\bea
\vecq_{\{i_1, \cdots, i_{n-1} \}, i_n}
\eequiv
2 \eta_{i_n} \sech \bigl(\tau_{i_n} + \a_{i_n} + 
 \phi_{\{i_1, \cdots, i_{n-1} \}, i_n} \bigr) 
\e^{-\i \Thet_{i_n}} \vecu_{\{i_1, \cdots, i_{n-1} \}, i_n},
\label{q-def}
\eea
%By definition, $\vecq_{\{i_1, \cdots, i_{n-1} \}, i_n}$
%$\vecq_{\{i_1, \cdots, i_{n-1} \}, i_n}$ 
we can 
%compactly 
%illustrate 
diagram the asymptotic behavior 
%forms 
of the $N$-soliton solution in 
%a 
the simplest 
way (see Fig.~\ref{Asym-N}). 
%It is easily seen 
We should note that $\vecq_{\{i_1, \cdots, i_{n-1} \}, i_n}$ 
is 
%invariant under 
symmetric with respect to 
%any 
permutations of $i_1, \cdots, i_{n-1}$. 
%Using this abbreviation, 
The last subscript $i_n$ denotes 
%designates 
%a number assigned to 
%the index of 
%each 
%soliton itself, 
the 
soliton's number, which is, of course, time independent. 
The significance of the other subscripts, $i_1, \cdots, i_{n-1}$, 
in $\{ \; \}$ 
%will be seen 
is clarified 
in the next section. 
%later. 
%of the $N$-soliton solutions at $t \to \pm \infty$:
%in a compact form:
%

%\noindent
%
\clearpage
\section{Factorization of an $N$-soliton collision 
into a superposition of 
%two-soliton 
pair collisions}
\setcounter{equation}{0}
\label{Facto}

In this section, 
%we reach the main goal of this paper. 
on the basis of 
%based on 
%Applying 
%Using 
the collision laws of two solitons 
%obtained 
presented in \S \ref{Two-soliton}, 
%section~\ref{Two-soliton}, 
we examine 
%discuss 
the asymptotic behavior 
%consider the aymptotic behavior 
of the $N$-soliton solution obtained 
in \S \ref{Asymp-N}. 
%section~\ref{Asymp-N}. 
%We 
%It is proved 
%Then 
We 
%prove 
conclude that an $N$-soliton collision in the 
Manakov model (\ref{cNLS}) 
%can be 
%can be explained by pair collisions 
%is 
can be factorized into a nonlinear superposition of pair collisions 
in arbitrary order. 
%prove that 
%established that 
%We prove that 

%For this purpose, we need 
%later, 
%vectors appearing in (\ref{u-def3}), 
We first prove 
%begin with 
%the following 
a lemma 
%In order 
needed later to compute the Hermitian product of 
%between 
$\vt{u}_{\{i_1, \cdots, i_{n-1} \}, j}$ 
and $\vt{u}_{\{i_1, \cdots, i_{n-1} \}, k}$. 
%later, 
%how to calculate 
%in a special case. 
%when all subscripts but one coincide. 
%
\vspace{5mm}
\begin{lemma}
\label{Her-pro}
{\it 
For any set of unit vectors $\vt{u}_{i_1}, \cdots, \vt{u}_{i_{n-1}}, 
\vt{u}_j, \vt{u}_k$ and $d_{il}$ defined by $(\ref{cd-def})$, the 
following equality holds:
\bea
\espace
\left|
\begin{array}{cccc}
d_{i_1 i_1} & \cdots & d_{i_1 i_{n-1}} & \vecu_{i_1} \\
\vdots & \ddots & \vdots & \vdots \\
d_{i_{n-1} i_1} & \cdots & d_{i_{n-1} i_{n-1}} & \vecu_{i_{n-1}} \\
d_{j i_1} & \cdots & d_{j i_{n-1}} & \vecu_{j} \\
\end{array}
\right| 
%\, 
\hspace{1pt}
{\Large{\mbox{$\cdot$}}} 
\hspace{1pt}
%\, 
\left|
\begin{array}{cccc}
d_{i_1 i_1} & \cdots & d_{i_1 i_{n-1}} & \vecu_{i_1} \\
\vdots & \ddots & \vdots & \vdots \\
d_{i_{n-1} i_1} & \cdots & d_{i_{n-1} i_{n-1}} & \vecu_{i_{n-1}} \\
d_{k i_1} & \cdots & d_{k i_{n-1}} & \vecu_{k} \\
\end{array}
\right|^{\large{\mbox{$\dagger$}}}
\nn \\
\eq 
\frac{\z_k -\z_j^\ast}{\i} 
\times
\left|
\begin{array}{ccc}
d_{i_1 i_1} & \cdots & d_{i_1 i_{n-1}} \\
\vdots & \ddots & \vdots \\
d_{i_{n-1} i_1} & \cdots & d_{i_{n-1} i_{n-1}} \\
\end{array}
\right| 
\times
\left|
\begin{array}{cccc}
d_{i_1 i_1} & \cdots & d_{i_1 i_{n-1}} & d_{i_1 k} \\
\vdots & \ddots & \vdots & \vdots \\
d_{i_{n-1} i_1} & \cdots & d_{i_{n-1} i_{n-1}} & d_{i_{n-1} k} \\
d_{j i_1} & \cdots & d_{j i_{n-1}} & d_{jk} \\
\end{array}
\right|. 
\label{dd-dd}
\eea
}
\end{lemma}
\vspace{5mm}
\noindent
{\bf Proof.\ }
%For brevity, 
%Throughout 
In the proof of this lemma, we 
%fix $i_1, \cdots, i_{n-1}, j, k$ and 
use $D$ to denote the last determinant in (\ref{dd-dd}): 
%define an $n \times n$ determinant $D$ by
\[
D \equiv 
\left|
\begin{array}{cccc}
d_{i_1 i_1} & \cdots & d_{i_1 i_{n-1}} & d_{i_1 k} \\
\vdots & \ddots & \vdots & \vdots \\
d_{i_{n-1} i_1} & \cdots & d_{i_{n-1} i_{n-1}} & d_{i_{n-1} k} \\
d_{j i_1} & \cdots & d_{j i_{n-1}} & d_{jk} \\
\end{array}
\right|.
\]
%Moreover, we write 
We express 
%denote 
%express 
%$(n-1) \times (n-1)$ 
minor determinants obtained 
by 
%removing 
deleting one row and 
one 
column 
%from 
of $D$ 
as
%, {\it e.g.}\ 
\[
D \hspace{-1mm}
\left[
\begin{array}{c}
i_l \\
 k  \\
\end{array}
\right]
=
\left|
\begin{array}{ccc}
d_{i_1 i_1} & \cdots & d_{i_1 i_{n-1}} \\
\vdots  &   & \vdots \\
d_{i_{l-1} i_1} & \cdots & d_{i_{l-1} i_{n-1}} \\
d_{i_{l+1} i_1} & \cdots & d_{i_{l+1} i_{n-1}} \\
\vdots  &   & \vdots \\
d_{i_{n-1} i_1} & \cdots &  d_{i_{n-1} i_{n-1}} \\
d_{j i_1} & \cdots & d_{j i_{n-1}} \\
\end{array}
\right|,
\]
\[
D \hspace{-1mm}
\left[
\begin{array}{c}
 j \\
 i_l  \\
\end{array}
\right]
=
\left|
\begin{array}{ccccccc}
d_{i_1 i_1} & \cdots & d_{i_1 i_{l-1}} & d_{i_1 i_{l+1}} 
	& \cdots & d_{i_1 i_{n-1}} & d_{i_1 k} \\
\vdots &  & \vdots & \vdots & & \vdots & \vdots \\
d_{i_{n-1} i_1} & \cdots & d_{i_{n-1} i_{l-1}} & d_{i_{n-1} i_{l+1}} 
	& \cdots & d_{i_{n-1} i_{n-1}} & d_{i_{n-1} k} \\
\end{array}
\right|, 
\]
\[
D \hspace{-1mm}
\left[
\begin{array}{c}
 j \\
 k  \\
\end{array}
\right]
=
\left|
\begin{array}{ccc}
d_{i_1 i_1} & \cdots & d_{i_1 i_{n-1}} \\
\vdots  & \ddots  & \vdots \\
d_{i_{n-1} i_1} & \cdots & d_{i_{n-1} i_{n-1}} \\
\end{array}
\right|.
%, \; \; \; {\rm etc.}
\]
Using these abbreviations and the Laplace expansion 
of determinants, 
%for 
we can 
%calculate 
rewrite the left-hand side of (\ref{dd-dd}) as
\bea
{\rm l.h.s.}
\eq 
\left\{ 
\sum_{p=1}^{n-1} (-1)^{n+p} 
D \hspace{-1mm}
\left[
\begin{array}{c}
i_p \\
 k  \\
\end{array}
\right] \vt{u}_{i_p} +
D \hspace{-1mm} \left[
\begin{array}{c}
 j \\
 k \\
\end{array}
\right] \vt{u}_{j} \right\} \cdot
\left\{ 
\sum_{q=1}^{n-1} (-1)^{n+q} 
D \hspace{-1mm}\left[
\begin{array}{c}
 j \\
 i_q \\
\end{array}
\right] \vt{u}_{i_q}^\dagger +
D \hspace{-1mm}\left[
\begin{array}{c}
 j \\
 k \\
\end{array}
\right] \vt{u}_{k}^\dagger \right\} 
\nn \\
\eq 
\sum_{p=1}^{n-1} \sum_{q=1}^{n-1} (-1)^{p+q} 
D \hspace{-1mm}
\left[
\begin{array}{c}
i_p \\
 k  \\
\end{array}
\right] 
D \hspace{-1mm}
\left[
\begin{array}{c}
 j \\
 i_q  \\ 
\end{array}
\right] \times \left\{ \frac{(\z_{i_q}-\z_{j}^\ast) 
	+ ( \z_{j}^\ast -\z_{i_p}^\ast)}{\i}  \right\} d_{i_p i_q} 
\nn \\
\espace 
\mbox{}+ 
\sum_{q=1}^{n-1} (-1)^{n+q} 
D \hspace{-1mm}
\left[
\begin{array}{c}
 j \\
 k  \\
\end{array}
\right] 
D \hspace{-1mm}
\left[
\begin{array}{c}
 j \\
 i_q  \\ 
\end{array}
\right] \times \frac{(\z_{i_q}-\z_{j}^\ast)}{\i} d_{j i_q} 
\nn \\
\espace 
\mbox{}+ 
\sum_{p=1}^{n-1} (-1)^{n+p} 
D \hspace{-1mm}
\left[
\begin{array}{c}
i_p \\
 k  \\
\end{array}
\right] 
D \hspace{-1mm}
\left[
\begin{array}{c}
 j \\
 k \\ 
\end{array}
\right] \times \left\{ \frac{(\z_{k}-\z_{j}^\ast) 
	+ ( \z_{j}^\ast -\z_{i_p}^\ast)}{\i}  \right\} d_{i_p k} 
\nn \\
\espace 
\mbox{}+ 
D \hspace{-1mm}
\left[
\begin{array}{c}
 j \\
 k  \\
\end{array}
\right] 
D \hspace{-1mm}
\left[
\begin{array}{c}
 j \\
 k  \\ 
\end{array}
\right] \times \frac{(\z_{k}-\z_{j}^\ast)}{\i} d_{j k} 
\nn \\
\eq 
\sum_{q=1}^{n-1} 
(-1)^{n+q}
D \hspace{-1mm}
\left[
\begin{array}{c}
 j \\
 i_q \\
\end{array}
\right] 
\frac{(\z_{i_q}-\z_{j}^\ast)}{\i} 
\left\{ 
\sum_{p=1}^{n-1} (-1)^{n+p} 
D \hspace{-1mm}
\left[
\begin{array}{c}
 i_p \\
 k  \\ 
\end{array}
\right] d_{i_p i_q} 
+
D \hspace{-1mm}
\left[
\begin{array}{c}
 j  \\
 k  \\ 
\end{array}
\right] d_{j i_q} 
\right\} 
\nn \\
\espace \mbox{} +
\sum_{p=1}^{n-1} 
(-1)^{n+p}
D \hspace{-1mm}
\left[
\begin{array}{c}
 i_p \\
 k \\
\end{array}
\right] 
\frac{(\z_{j}^\ast - \z_{i_p}^\ast)}{\i} 
\left\{ 
\sum_{q=1}^{n-1} (-1)^{n+q} 
D \hspace{-1mm}
\left[
\begin{array}{c}
 j \\
 i_q  \\ 
\end{array}
\right] d_{i_p i_q} 
+
D \hspace{-1mm}
\left[
\begin{array}{c}
 j  \\
 k  \\ 
\end{array}
\right] d_{i_p k} 
\right\} 
\nn \\
\espace \mbox{} +
D \hspace{-1mm}
\left[
\begin{array}{c}
 j \\
 k \\
\end{array}
\right] 
\frac{(\z_{k} - \z_{j}^\ast)}{\i} 
\left\{ 
\sum_{p=1}^{n-1} (-1)^{n+p} 
D \hspace{-1mm}
\left[
\begin{array}{c}
 i_p \\
 k  \\ 
\end{array}
\right] d_{i_p k} 
+
D \hspace{-1mm}
\left[
\begin{array}{c}
 j  \\
 k  \\ 
\end{array}
\right] d_{j k} 
\right\} 
\nn \\
\eq 
\sum_{q=1}^{n-1} 
(-1)^{n+q}
D \hspace{-1mm}
\left[
\begin{array}{c}
 j \\
 i_q \\
\end{array}
\right] 
\frac{(\z_{i_q}-\z_{j}^\ast)}{\i} 
\left|
\begin{array}{cccc}
d_{i_1 i_1} & \cdots & d_{i_1 i_{n-1}} & d_{i_1 i_q} \\
\vdots & \ddots & \vdots & \vdots \\
d_{i_{n-1} i_1} & \cdots & d_{i_{n-1} i_{n-1}} & d_{i_{n-1} i_q} \\
d_{j i_1} & \cdots & d_{j i_{n-1}} & d_{j i_q} \\
\end{array}
\right|
\nn \\
\espace \mbox{} +
\sum_{p=1}^{n-1} 
(-1)^{n+p}
D \hspace{-1mm}
\left[
\begin{array}{c}
 i_p \\
 k \\
\end{array}
\right] 
\frac{(\z_{j}^\ast - \z_{i_p}^\ast)}{\i} 
\left|
\begin{array}{cccc}
d_{i_1 i_1} & \cdots & d_{i_1 i_{n-1}} & d_{i_1 k} \\
\vdots & \ddots & \vdots & \vdots \\
d_{i_{n-1} i_1} & \cdots & d_{i_{n-1} i_{n-1}} & d_{i_{n-1} k} \\
d_{i_p i_1} & \cdots & d_{i_p i_{n-1}} & d_{i_p k} \\
\end{array}
\right| 
\nn \\
\espace \mbox{} 
+ D \hspace{-1mm}
\left[
\begin{array}{c}
 j \\
 k \\
\end{array}
\right] 
\frac{(\z_{k} - \z_{j}^\ast)}{\i} D. 
\nn
\eea
It is easily seen that 
in the last expression, only the last term remains. 
%, which 
%coincides with 
%surely 
This 
is the right-hand side of (\ref{dd-dd}). \hfill $\Box$
%\\
%Thanks to the orthogonality relation between rows of a matrix and 
%those of the cofactor matrix, only the last term remains and we get
%
%\[
%D \hspace{-1mm}
%\left[
%\begin{array}{c}
% j \\
% k  \\ 
%\end{array}
%\right] 
%\frac{(\z_{k} - \z_{j}^\ast)}{\i}  D,
%\]
%which is precisely the right-hand side o f
\vspace{5mm}
\begin{corollary}
\label{unit}
{\it 
The vector $\vecu_{\{ i_1, \cdots, i_{n-1} \}, i_n}$ 
defined for distinct positive integers $i_1, \cdots, i_{n-1}, i_n$ 
by $(\ref{u-def3})$ with $(\ref{phi-def3})$ is a unit vector, i.e.
\[
\left| \! \left| \vecu_{\{ i_1, \cdots, i_{n-1} \}, i_n} 
	\right| \! \right| =1.
\]
}
\end{corollary}
\vspace{5mm}
\noindent
{\bf Proof.\ } 
%With the help of 
%We 
%
%
%
Using Lemma~\ref{Her-pro} in the special case $j=k\hspace{2pt}(\equiv i_n)$, 
we 
%obtain that 
have 
%that 
%
%
%that 
\bea
\left| \! \left| \vecu_{\{ i_1, \cdots, i_{n-1} \}, i_n} 
	\right| \! \right|^2 
\eq \vecu_{\{ i_1, \cdots, i_{n-1} \}, i_n} \cdot 
	\vecu_{\{ i_1, \cdots, i_{n-1} \}, i_n}^\dagger
\nn \\
\eq 
\prod_{l=1}^{n-1} 
\left| \frac{\z_{i_l} - \z_{i_n}^\ast}{\z_{i_l} - \z_{i_n}} \right|^2
\times
\frac{ d_{i_n i_n} \left|
\begin{array}{ccc}
d_{i_1 i_1} & \cdots & d_{i_1 i_{n-1}} \\
\vdots & \ddots & \vdots \\
d_{i_{n-1} i_1} & \cdots & d_{i_{n-1} i_{n-1}} \\
\end{array}
\right|}{ \left|
\begin{array}{ccc}
d_{i_1 i_1} & \cdots & d_{i_1 i_n} \\
\vdots & \ddots & \vdots \\
d_{i_n i_1} & \cdots & d_{i_n i_n} \\
\end{array}
\right|}
\nn \\
\espace \mbox{} \times
\prod_{l=1}^{n-1} 
\left|
\frac{\z_{i_l}^\ast - \z_{i_n}^\ast}{\z_{i_l} - \z_{i_n}^\ast} \right|^2
\times 
\frac{ \Ds \frac{\z_{i_n} - \z_{i_n}^\ast }{ \i}
\left|
\begin{array}{ccc}
d_{i_1 i_1} & \cdots & d_{i_1 i_{n-1}} \\
\vdots & \ddots & \vdots \\
d_{i_{n-1} i_1} & \cdots & d_{i_{n-1} i_{n-1}} \\
\end{array}
\right|
\left|
\begin{array}{ccc}
d_{i_1 i_1} & \cdots & d_{i_1 i_{n}} \\
\vdots & \ddots & \vdots \\
d_{i_{n} i_1} & \cdots & d_{i_{n} i_{n}} \\
\end{array}
\right| }
{ \left|
\begin{array}{ccc}
d_{i_1 i_1} & \cdots & d_{i_1 i_{n-1}} \\
\vdots & \ddots & \vdots \\
d_{i_{n-1} i_1} & \cdots & d_{i_{n-1} i_{n-1}} \\
\end{array}
\right|^2 }
\nn \\
\eq 1.
\hspace{123mm} 
%\mbox{\hfill}
\Box 
\nn 
\\
\nn
\eea
%\hfill $\Box$
%\vspace{5mm}
%\hfill $\Box$
%\noindent
%Thanks to Corollary~\ref{unit}, 
%

\begin{figure}[t]
%\vspace{20mm}
  \begin{center}
    \includegraphics[width=14cm]{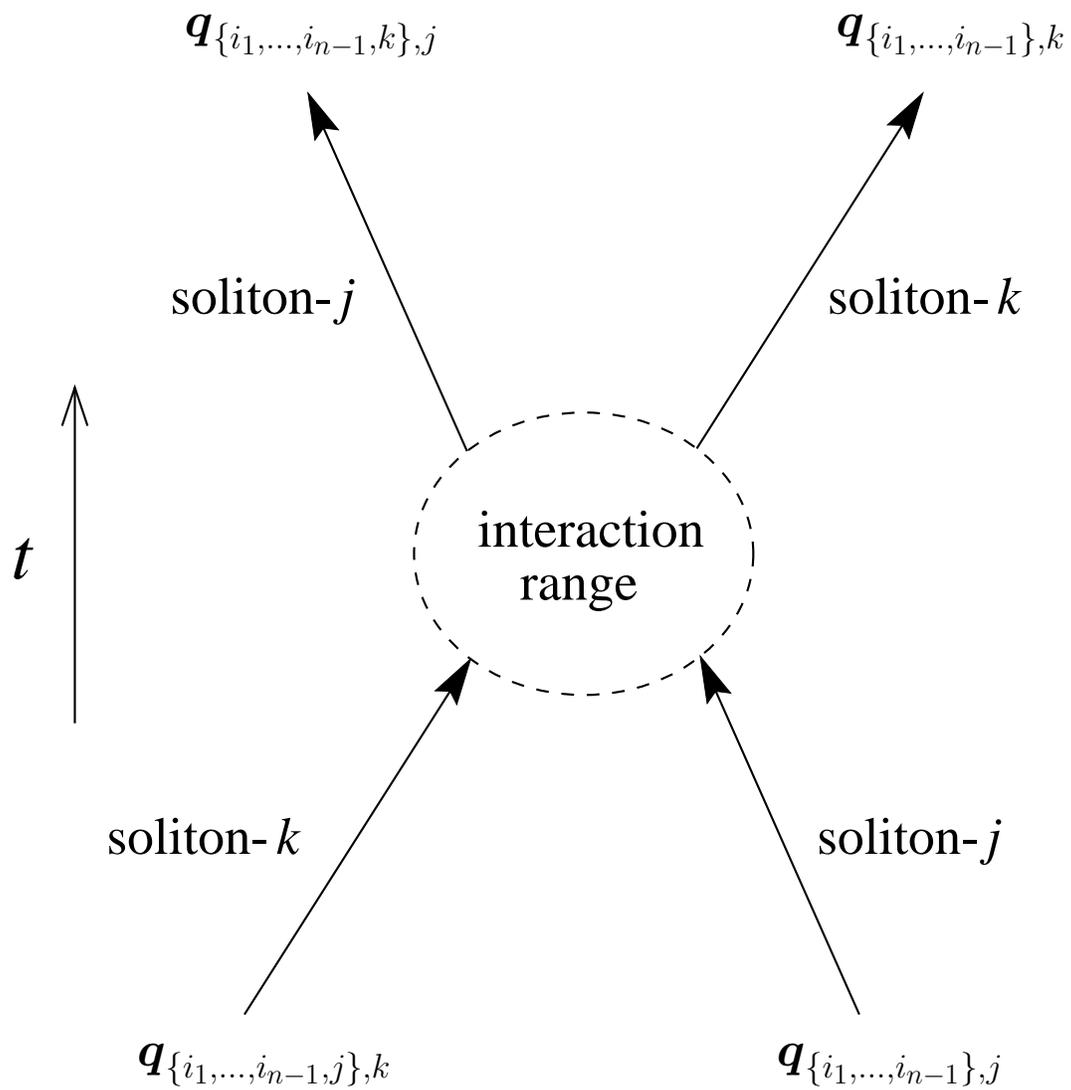}
  \\
    \vspace*{10mm}
    \caption{
%Scattering 
	Two-soliton collision in the presence of other solitons.}
%    \caption{Two-soliton scattering}
    \label{scatter-rule}
%  \nonumber
%\noindent
  \end{center}
\end{figure}

We are now able to apply the 
%two-soliton 
collision laws 
%of two solitons 
%collisions 
%describes by 
%(see Theorem~\ref{two-rule}) to the case 
defined by Theorem~\ref{two-rule} to the 
%case 
%collision 
%where 
two-soliton collision in which 
%that 
soliton $\vt{q}_{ \{i_1, \cdots, i_{n-1}, j \},k}$ 
overtakes soliton $\vt{q}_{ \{i_1, \cdots, i_{n-1} \},j}$ 
[\hspace{0pt}cf.\ the definition (\ref{q-def})]. 
Here, $i_1, \cdots, i_{n-1},j,k$ are distinct positive integers. 
%
%we can show the following 
%proposition. 
\vspace{5mm}
\begin{proposition}
\label{joverk}
{\it 
%we can prove (by very tiresome calculations) that 
%%scattering 
The two-soliton collision in which 
%that 
$\vt{q}_{ \{i_1, \cdots, i_{n-1},j \},k}$ 
overtakes $\vt{q}_{ \{i_1, \cdots, i_{n-1} \},j }$ 
changes these solitons to 
$\vt{q}_{ \{i_1, \cdots, i_{n-1} \},k}$ 
and $\vt{q}_{ \{i_1, \cdots, i_{n-1},k \},j}$, 
%respectively, 
%obeys the 
%scattering 
%laws 
as shown in Fig.~$\ref{scatter-rule}$. 
%If we express the laws in terms of equations, they are 
According to Theorem~$\ref{two-rule}$, 
this 
%statement 
is equivalent to the following set of 
equalities: 
%equivalent to the following: 
% set of equations: 
%equalities: 
\bseq
\bea
\espace
\e^{-2 \left( \phi_{ \{i_1, \cdots, i_{n-1}, j \},k} 
	- \phi_{ \{i_1, \cdots, i_{n-1} \},k} \right) } 
= \e^{-2 \left( \phi_{ \{i_1, \cdots, i_{n-1},k  \}, j} 
	- \phi_{ \{i_1, \cdots, i_{n-1} \}, j} \right) }  
\label{pro1-1}
\\
\eq \left|\frac{\z_j - \z_k}{\z_j - \z_k^\ast} \right|^2
\left\{ 
1 + \frac{(\z_j -\z_j^\ast)(\z_k -\z_k^\ast)}
{|\z_j - \z_k^\ast|^2} \left| 
\vecu_{ \{i_1, \cdots, i_{n-1} \}, j}  \cdot 
\vecu_{ \{i_1, \cdots, i_{n-1} \}, k}^\dagger \right|^2 
\right\},
\label{pro1-2}
\eea
\label{pro-1}
\eseq
\bea
\vecu_{ \{i_1, \cdots, i_{n-1},j \}, k} 
\eq
\e^{\phi_{ \{i_1, \cdots, i_{n-1}, j \},k} 
	- \phi_{ \{i_1, \cdots, i_{n-1} \},k} } 
\frac{\z_j^\ast -\z_k^\ast}{\z_j -\z_k^\ast}
\biggl\{
\vecu_{ \{i_1, \cdots, i_{n-1} \}, k} 
\nn \\ 
\espace 
\mbox{} - \frac{\z_j -\z_j^\ast}{\z_j -\z_k^\ast}
\left( \vecu_{ \{i_1, \cdots, i_{n-1} \}, k} 
	\cdot \vecu_{ \{i_1, \cdots, i_{n-1} \}, j}^\dagger \right) 
	\vecu_{ \{i_1, \cdots, i_{n-1} \}, j} \biggr\},
\label{pro-2}
\eea
\bea
\vecu_{ \{i_1, \cdots, i_{n-1}, k \}, j} 
\eq
\e^{\phi_{ \{i_1, \cdots, i_{n-1}, k \},j} 
	- \phi_{ \{i_1, \cdots, i_{n-1} \},j} } 
\frac{\z_k^\ast -\z_j^\ast}{\z_k -\z_j^\ast}
\biggl\{
\vecu_{ \{i_1, \cdots, i_{n-1} \}, j} 
\nn \\ 
\espace 
\mbox{} - \frac{\z_k -\z_k^\ast}{\z_k -\z_j^\ast}
\left( \vecu_{ \{i_1, \cdots, i_{n-1} \}, j} 
	\cdot \vecu_{ \{i_1, \cdots, i_{n-1} \}, k}^\dagger \right) 
	\vecu_{ \{i_1, \cdots, i_{n-1} \}, k} \biggr\}.
\label{pro-3}
\eea
}
\end{proposition}
\vspace{5mm}
\noindent
{\bf Proof.\ } 
%For brevity, 
Throughout 
%the 
this proof, 
we employ 
%use 
the following notation 
in order 
to express 
%write 
determinants compactly: 
%concisely: 
%throughout the proof: 
%abbreviation: 
%of $(d_{lk})$:
%we Throughout the proof, we use 
\[
d \left(
\hspace{-1mm}
\begin{array}{c}
j_1, j_2, \cdots, j_l \\
k_1, k_2, \cdots, k_l \\
\end{array}
\hspace{-1mm}
\right)
\equiv 
\left|
\begin{array}{cccc}
d_{j_1 k_1} & d_{j_1 k_2} & \cdots & d_{j_1 k_{l}} \\
d_{j_2 k_1} & d_{j_2 k_2} & \cdots & d_{j_2 k_{l}} \\
\vdots & \vdots & \ddots & \vdots \\
d_{j_{l} k_1} & d_{j_l k_2} & \cdots & d_{j_{l} k_{l}} \\
\end{array}
\right|.
\]
To prove (\ref{pro-1}), 
we first 
%prove equality (\ref{pro1-1}). 
%We 
rewrite the left-hand side of (\ref{pro1-1}) 
as 
%\beq
\begin{equation}
\e^{-2 \left( \phi_{ \{i_1, \cdots, i_{n-1}, j \},k} 
	- \phi_{ \{i_1, \cdots, i_{n-1} \},k} \right) } 
= 
\left| 
\frac{\z_j - \z_k}{\z_j -\z_k^\ast} 
\right|^2 \times 
\frac{
d \left(
\hspace{-1mm}
\begin{array}{c}
i_1, \cdots, i_{n-1}, j,k \\
i_1, \cdots, i_{n-1}, j,k \\
\end{array}
\hspace{-1mm}
\right)
d \left(
\hspace{-1mm}
\begin{array}{c}
i_1, \cdots, i_{n-1} \\
i_1, \cdots, i_{n-1} \\
\end{array}
\hspace{-1mm}
\right)
}{
d \left(
\hspace{-1mm}
\begin{array}{c}
i_1, \cdots, i_{n-1}, j \\
i_1, \cdots, i_{n-1}, j \\
\end{array}
\hspace{-1mm}
\right)
d \left(
\hspace{-1mm}
\begin{array}{c}
i_1, \cdots, i_{n-1}, k \\
i_1, \cdots, i_{n-1}, k \\
\end{array}
\hspace{-1mm}
\right) },
\label{e-jk}
%\eeq
\end{equation}
%by 
using the definition of 
$\phi_{\{ i_1, \cdots, i_{n-1} \}, i_n}$, (\ref{phi-def3}). 
%
%Since 
%We note that 
%Then, 
Obviously, the right-hand side of (\ref{e-jk}) is 
symmetric with respect to interchange of $j$ and $k$. 
%invariant under interchange of $j$ and $k$, 
%it is obvious that 
It follows that 
the 
equality (\ref{pro1-1}) holds. 

Next, we 
%shall 
prove the 
equality (\ref{pro1-2}).  
Using the definition of 
$\vecu_{ \{i_1, \cdots, i_{n-1} \}, i_n}$, (\ref{u-def3}), 
and Lemma~\ref{Her-pro}, we obtain 
%have 
\bea
%\hspace{-5mm}
\espace 
\vecu_{ \{i_1, \cdots, i_{n-1} \}, j}  \cdot 
\vecu_{ \{i_1, \cdots, i_{n-1} \}, k}^\dagger 
\nn \\
%\hspace{-5mm}
\eq 
\e^{\phi_{ \{i_1, \cdots, i_{n-1} \},j} 
	+ \phi_{ \{i_1, \cdots, i_{n-1} \},k} } 
\times
\frac{\z_k - \z_j^\ast}{\i}  
\times
\prod_{l=1}^{n-1} \frac{(\z_{i_l}^\ast -\z_j^\ast)(\z_{i_l} -\z_k)}
{(\z_{i_l} -\z_j^\ast)(\z_{i_l}^\ast -\z_k)}
%\nn \\
%\espace 
%\mbox{} 
%
\times 
\frac{
d \left(
\hspace{-1mm}
\begin{array}{c}
i_1, \cdots, i_{n-1}, j \\
i_1, \cdots, i_{n-1}, k \\
\end{array}
\hspace{-1mm}
\right)
}{
d \left(
\hspace{-1mm}
\begin{array}{c}
i_1, \cdots, i_{n-1} \\
i_1, \cdots, i_{n-1} \\
\end{array}
\hspace{-1mm}
\right) }.
\hspace{7mm}
%\nn \\
%\espace
\label{uj-uk}
\eea
%
%Then, recalling 
%Then we obtain 
%We can 
%Computing 
%Taking the norm of both sides of (\ref{uj-uk}), 
%Then, 
Multiplying (\ref{uj-uk}) by its complex conjugate on each side, 
we obtain, with the help of 
%the definition of $\phi_{ \{i_1, \cdots, i_{n-1} \}, i_n}$ 
(\ref{phi-def3}), 
%that 
%as 
%that 
%
\[
\left| 
\vecu_{ \{i_1, \cdots, i_{n-1} \}, j}  \cdot 
\vecu_{ \{i_1, \cdots, i_{n-1} \}, k}^\dagger \right|^2 
=
- \frac{\left| 
\z_k - \z_j^\ast
\right|^2 }{(\z_j -\z_j^\ast)(\z_k -\z_k^\ast)} 
\times 
\frac{
d \left(
\hspace{-1mm}
\begin{array}{c}
i_1, \cdots, i_{n-1}, j \\
i_1, \cdots, i_{n-1}, k \\
\end{array}
\hspace{-1mm}
\right)
d \left(
\hspace{-1mm}
\begin{array}{c}
i_1, \cdots, i_{n-1}, k \\
i_1, \cdots, i_{n-1}, j \\
\end{array}
\hspace{-1mm}
\right)
}{
d \left(
\hspace{-1mm}
\begin{array}{c}
i_1, \cdots, i_{n-1}, j \\
i_1, \cdots, i_{n-1}, j \\
\end{array}
\hspace{-1mm}
\right)
d \left(
\hspace{-1mm}
\begin{array}{c}
i_1, \cdots, i_{n-1}, k \\
i_1, \cdots, i_{n-1}, k \\
\end{array}
\hspace{-1mm}
\right) }.
\]
%
%Thus 
Then, 
we can rewrite the right-hand side of (\ref{pro1-2}) 
%is rewritten 
as 
%follows:  
\bea
\espace
\left|\frac{\z_j - \z_k}{\z_j - \z_k^\ast} \right|^2
\left\{ 
1 + \frac{(\z_j -\z_j^\ast)(\z_k -\z_k^\ast)}
{|\z_j - \z_k^\ast|^2} \left| 
\vecu_{ \{i_1, \cdots, i_{n-1} \}, j}  \cdot 
\vecu_{ \{i_1, \cdots, i_{n-1} \}, k}^\dagger \right|^2 
\right\}
\nn \\
\eq \left|\frac{\z_j - \z_k}{\z_j - \z_k^\ast} \right|^2
\left\{ 
1- 
\frac{
d \left(
\hspace{-1mm}
\begin{array}{c}
i_1, \cdots, i_{n-1}, j \\
i_1, \cdots, i_{n-1}, k \\
\end{array}
\hspace{-1mm}
\right)
d \left(
\hspace{-1mm}
\begin{array}{c}
i_1, \cdots, i_{n-1}, k \\
i_1, \cdots, i_{n-1}, j \\
\end{array}
\hspace{-1mm}
\right)
}{
d \left(
\hspace{-1mm}
\begin{array}{c}
i_1, \cdots, i_{n-1}, j \\
i_1, \cdots, i_{n-1}, j \\
\end{array}
\hspace{-1mm}
\right)
d \left(
\hspace{-1mm}
\begin{array}{c}
i_1, \cdots, i_{n-1}, k \\
i_1, \cdots, i_{n-1}, k \\
\end{array}
\hspace{-1mm}
\right) }
\right\}. 
\label{1-dd}
\eea
Here, thanks to the Jacobi formula for determinants, we have 
\bea
\espace
d \left(
\hspace{-1mm}
\begin{array}{c}
i_1, \cdots, i_{n-1}, j \\
i_1, \cdots, i_{n-1}, j \\
\end{array}
\hspace{-1mm}
\right)
d \left(
\hspace{-1mm}
\begin{array}{c}
i_1, \cdots, i_{n-1}, k \\
i_1, \cdots, i_{n-1}, k \\
\end{array}
\hspace{-1mm}
\right) 
-
d \left(
\hspace{-1mm}
\begin{array}{c}
i_1, \cdots, i_{n-1}, j \\
i_1, \cdots, i_{n-1}, k \\
\end{array}
\hspace{-1mm}
\right)
d \left(
\hspace{-1mm}
\begin{array}{c}
i_1, \cdots, i_{n-1}, k \\
i_1, \cdots, i_{n-1}, j \\
\end{array}
\hspace{-1mm}
\right)
\nn \\
\eq 
d \left(
\hspace{-1mm}
\begin{array}{c}
i_1, \cdots, i_{n-1} \\
i_1, \cdots, i_{n-1} \\
\end{array}
\hspace{-1mm}
\right)
d \left(
\hspace{-1mm}
\begin{array}{c}
i_1, \cdots, i_{n-1}, j, k\\
i_1, \cdots, i_{n-1}, j, k \\
\end{array}
\hspace{-1mm}
\right).
\label{Jacobi}
\eea
%Therefore, 
%and 
Thus, (\ref{1-dd}) 
%equals 
is equal to 
(\ref{e-jk}). 
%coincide, which 
This 
%which 
completes the proof of 
the 
equality (\ref{pro1-2}). 

To prove the 
equality (\ref{pro-2}), 
%for (\ref{pro-2}). 
%in parallel. 
we 
%prepare an extended version 
%consider 
need to extend 
%an extension of 
%modified version 
the Jacobi formula (\ref{Jacobi}). 
We remark that, 
although the matrix elements $d_{il}$ 
here 
%in our context take a special 
%peculiar 
%form 
are 
%defined 
given 
by (\ref{cd-def}), 
%both 
the two 
sides of 
%the formula 
(\ref{Jacobi}) are 
%in fact 
%identical 
equal as a polynomial 
%in fact hold 
for 
%any other definition 
%regardless of the definition 
%of 
general 
%matrix 
elements $d_{il}$. 
%the matrix elements 
%the equality 
%Hence, 
%Thus, 
Therefore, 
%keeping 
maintaining 
the validity of (\ref{Jacobi}), 
we can replace 
%in (\ref{Jacobi}) 
%a column in (\ref{Jacobi}) by a column consisting of vectors. 
the columns 
%of 
%of 
with index $k$ 
by 
%the 
columns 
consisting of the vectors 
%which is 
%composed of 
%which consist of 
%with elements 
%(some of) 
%a subset of 
%consist of 
%contain (
$\vt{u}_{i_1}, \cdots, \vt{u}_{i_{n-1}}, \vt{u}_j$ or $\vt{u}_{k}$:
%as
%${}^t (\vt{u}_{i_1}, \cdots, \vt{u}_{i_{n-1}}, \vt{u}_j, \vt{u}_{k})$: 
%to get 
%obtain the following formula: 
%
\bea
\espace
d \left(
\hspace{-1mm}
\begin{array}{c}
i_1, \cdots, i_{n-1}, j \\
i_1, \cdots, i_{n-1}, j \\
\end{array}
\hspace{-1mm}
\right)
\left|
\begin{array}{cccc}
d_{i_1 i_1} & \cdots & d_{i_1 i_{n-1}} & \vecu_{i_1} \\
\vdots & \ddots & \vdots & \vdots \\
d_{i_{n-1} i_1} & \cdots & d_{i_{n-1} i_{n-1}} & \vecu_{i_{n-1}} \\
d_{k i_1} & \cdots & d_{k i_{n-1}} & \vecu_{k} \\
\end{array}
\right| 
- d \left(
\hspace{-1mm}
\begin{array}{c}
i_1, \cdots, i_{n-1}, k \\
i_1, \cdots, i_{n-1}, j \\
\end{array}
\hspace{-1mm}
\right)
\nn \\
\espace 
\mbox{} \times
\left|
\begin{array}{cccc}
d_{i_1 i_1} & \cdots & d_{i_1 i_{n-1}} & \vecu_{i_1} \\
\vdots & \ddots & \vdots & \vdots \\
d_{i_{n-1} i_1} & \cdots & d_{i_{n-1} i_{n-1}} & \vecu_{i_{n-1}} \\
d_{j i_1} & \cdots & d_{j i_{n-1}} & \vecu_{j} \\
\end{array}
\right| 
%\nn \\
%\eq 
= d \left(
\hspace{-1mm}
\begin{array}{c}
i_1, \cdots, i_{n-1} \\
i_1, \cdots, i_{n-1} \\
\end{array}
\hspace{-1mm}
\right)
\left|
\begin{array}{ccccc}
d_{i_1 i_1} & \cdots & d_{i_1 i_{n-1}} & d_{i_1 j} & \vecu_{i_1} \\
\vdots & \ddots & \vdots & \vdots & \vdots \\
d_{i_{n-1} i_1} & \cdots & d_{i_{n-1} i_{n-1}} & d_{i_{n-1} j} 
	& \vecu_{i_{n-1}} \\
d_{j i_1} & \cdots & d_{j i_{n-1}} & d_{jj} & \vecu_{j} \\
d_{k i_1} & \cdots & d_{k i_{n-1}} & d_{kj} & \vecu_{k} \\
\end{array}
\right|.
\nn \\
\espace
\label{Jacobi2}
\eea
%
%Using the definitions (\ref{phi-def3}), (\ref{u-def3}) and 
%the relation (\ref{uj-uk}) (with $j \leftrightarrow k$), 
We rewrite the right-hand side of (\ref{pro-2}) 
using 
%the definitions 
(\ref{phi-def3}), (\ref{u-def3}) and 
%the relation 
(\ref{uj-uk}) (with $j \leftrightarrow k$) as 
\bea
\espace 
\e^{\phi_{ \{i_1, \cdots, i_{n-1}, j \},k} 
	- \phi_{ \{i_1, \cdots, i_{n-1} \},k} } 
\frac{\z_j^\ast -\z_k^\ast}{\z_j -\z_k^\ast}
\biggl\{
\vecu_{ \{i_1, \cdots, i_{n-1} \}, k} 
 - \frac{\z_j -\z_j^\ast}{\z_j -\z_k^\ast}
\left( \vecu_{ \{i_1, \cdots, i_{n-1} \}, k} 
	\cdot \vecu_{ \{i_1, \cdots, i_{n-1} \}, j}^\dagger \right) 
\nn \\ 
\espace 
\hspace{60mm}
\mbox{} \times
	\vecu_{ \{i_1, \cdots, i_{n-1} \}, j} \biggr\}
\nn \\
\eq 
\e^{\phi_{ \{i_1, \cdots, i_{n-1}, j \},k} } 
\times \frac{\z_j^\ast -\z_k^\ast}{\z_j - \z_k^\ast} 
\prod_{l=1}^{n-1} 
	\frac{\z_{i_l}^\ast -\z_k^\ast}{\z_{i_l} - \z_k^\ast} 
\times 
\frac{1}{
d \left(
\hspace{-1mm}
\begin{array}{c}
i_1, \cdots, i_{n-1}, j \\
i_1, \cdots, i_{n-1}, j \\
\end{array}
\hspace{-1mm}
\right)
d \left(
\hspace{-1mm}
\begin{array}{c}
i_1, \cdots, i_{n-1} \\
i_1, \cdots, i_{n-1} \\
\end{array}
\hspace{-1mm}
\right)}
\nn \\
\espace \mbox{}
\times \left\{
d \left(
\hspace{-1mm}
\begin{array}{c}
i_1, \cdots, i_{n-1}, j \\
i_1, \cdots, i_{n-1}, j \\
\end{array}
\hspace{-1mm}
\right)
\left|
\begin{array}{cccc}
d_{i_1 i_1} & \cdots & d_{i_1 i_{n-1}} & \vecu_{i_1} \\
\vdots & \ddots & \vdots & \vdots \\
d_{i_{n-1} i_1} & \cdots & d_{i_{n-1} i_{n-1}} & \vecu_{i_{n-1}} \\
d_{k i_1} & \cdots & d_{k i_{n-1}} & \vecu_{k} \\
\end{array}
\right| 
- d \left(
\hspace{-1mm}
\begin{array}{c}
i_1, \cdots, i_{n-1}, k \\
i_1, \cdots, i_{n-1}, j \\
\end{array}
\hspace{-1mm}
\right)
\right.
\nn \\
\espace 
\hspace{8mm}
\left.
\mbox{} 
\times
\left|
\begin{array}{cccc}
d_{i_1 i_1} & \cdots & d_{i_1 i_{n-1}} & \vecu_{i_1} \\
\vdots & \ddots & \vdots & \vdots \\
d_{i_{n-1} i_1} & \cdots & d_{i_{n-1} i_{n-1}} & \vecu_{i_{n-1}} \\
d_{j i_1} & \cdots & d_{j i_{n-1}} & \vecu_{j} \\
\end{array}
\right| 
\right\}.
\nn 
\eea
Owing to the extended Jacobi formula (\ref{Jacobi2}), 
this is equal to 
%equals 
$\vecu_{ \{i_1, \cdots, i_{n-1}, j \}, k}$ 
%Thanks to 
%and the definition (\ref{u-def3}). 
[\hspace{0pt}cf.\ the definition 
(\ref{u-def3})]. 
%the left-hand side of (\ref{pro-2}), 
%Thus we could prove the 
%This completes 
Now, the proof of the equality (\ref{pro-2}) is complete. 
%have proven 
%which completes the proof of 
%A 
The proof of the equality (\ref{pro-3}) is obtained 
%accomplished 
%achieved 
%simply 
by interchanging $j$ and $k$ 
%in the above proof. 
in the proof 
%that 
of (\ref{pro-2}). \hfill $\Box$
%\\

\vspace{5mm}
\noindent
%
%These laws are 
%We can use 
%We note that 
%We can apply 
Proposition~\ref{joverk} 
is applicable to 
%can be applied to 
%any 
%any 
two-soliton collisions 
%in the $N$-soliton solution 
%gives the solution of collision problems 
%to estimate the effects of two-soliton collisions 
%in the $N$-soliton solution. 
%applies to 
%is applicable to 
%(or Fig.\ \ref{scatter-rule}) 
%is applicable to the 
%case when 
%two-soliton collisions such that before the collision 
%the case when
%if 
%\begin{enumerate}
%\item[]
%in which 
%which satisfies 
%under 
satisfying the following condition: 
%that 
%
\begin{itemize}
\item
Before the collision, 
the 
%subscript 
set of subscripts in the 
%left-side 
left-hand soliton's $\{ \; \}$ 
is equal to 
%equals 
%all 
the 
%whole 
entire set of subscripts 
%set of 
of the 
%right-side 
right-hand soliton: 
%including the last subscript outside $\{ \; \}$
%(both inside and outside $\{ \; \}$)
%$ \oplus $ 
%behind $\{ \; \}$
%as a set: 
$\{i_1, \cdots, i_{n-1},j  \} = \{i_1, \cdots, i_{n-1} \} \cup \{ j \}$. 
\end{itemize}
We also 
%should 
note the following properties 
%of the collision 
%see 
(cf.\ Fig.~\ref{scatter-rule}):
%can read from Fig.~\ref{scatter-rule} the following properties: 
%\begin{enumerate}
\begin{itemize}
%
%\item[(i)]
\item
%Even 
After the collision, 
the 
%subscript 
set of subscripts 
in the left-hand soliton's $\{ \; \}$ 
is still equal to 
%equals 
the entire 
%whole 
%subscript 
set of subscripts of 
the right-hand soliton: 
%relation 
%is preserved under the collision: 
$\{i_1, \cdots, i_{n-1}, k \} 
= \{i_1, \cdots, i_{n-1} \} \cup \{ k \}$. 
%\item[(ii)]
\item
The 
%whole 
%subscript 
entire set of subscripts 
of the left-hand soliton 
%all the 
is unchanged: 
%conserved: 
%invariant 
%for the left-hand solitons: 
$\{i_1, \cdots, i_{n-1},j \} \cup \{ k \} = 
\{i_1, \cdots, i_{n-1}, k \} \cup \{ j \}$. 
%on the left side. 
%\item[(iii)]
\item
The 
%subscript 
set of subscripts 
in the right-hand soliton's $\{ \; \}$ 
is unchanged: 
%conserved: 
%invariant 
%for the right-hand solitons: 
$\{i_1, \cdots, i_{n-1} \} = \{i_1, \cdots, i_{n-1} \}$. 
%on the right side. 
%\item[(iv)]
\item
%The subscript falls off 
%When the faster soliton overtakes the slower soliton, 
%In the collision, 
%index 
%the number designating 
The overtaken soliton's number 
%label 
%assigned to 
%, $k$, 
%comes out of 
%falls off 
%drops 
is removed from the overtaking soliton's $\{ \; \}$, 
%subscripts 
while the overtaking soliton's number 
%number designating the 
is 
%label 
%assigned to 
%comes out of 
added to the overtaken soliton's $\{ \; \}$. 
%subscript 
%\end{enumerate}
%\\
\end{itemize}
%

%from the result in \S 2:

\clearpage
\noindent
%
%Noting these properties, 

%\vspace{5mm}
We are now able to 
%now can 
%reach the mail goal of this paper. 
%prove the main theorem 
%obtain 
state 
the main result of this paper. 
%can prove 
%
\vspace{5mm}
\begin{theorem}
\label{main_theorem}
{\it 
%The 
An $N$-soliton collision in the Manakov model $(\ref{cNLS})$ 
%is 
can be 
factorized into 
%the
%(nonlinear) 
a nonlinear superposition of 
%${}_N \rm{C}_2$ 
$N \choose 2$ pair 
%wise 
collisions 
in arbitrary order. 
%, regardless of 
%the 
%collision order.
}
\end{theorem}

\vspace{5mm}
\noindent
{\bf Proof.\ } 
%\noindent
%As was shown, 
According to 
the asymptotic behavior 
%form 
of the $N$-soliton solution 
as $t \to - \infty$ (see Fig.~\ref{Asym-N}), 
%shows that 
%solitons-$N, \cdots, 1$ 
solitons-$N$, $\cdots$, $1$ 
%line up 
are initially 
%at first 
distributed along the $x$-axis as 
%(see Fig.~\ref{Asym-N})
\[
\vecq_{\{1, \cdots, N-1 \}, N}, \cdots, 
\vecq_{\{1, \cdots, n-1 \}, n}, \cdots, 
\vecq_{\{ \;\}, 1}.
\]
%on $-\infty < x < \infty$ 
%along the $x$-axis 
%(see Fig.~\ref{Asym-N}). 
We take this initial state 
%asymptotic form 
as 
%a 
the point of departure and 
%Let us 
assume that the solitons collide pairwise 
in a given order. 
Then, 
%the 
a pair collision 
%must 
%have to 
takes place 
%in a given order. 
%two-soliton collisions take place 
%occur 
%\[
${N \choose 2 }
%{}_N \mbox{C}_2 
= \frac{N (N-1)}{2} 
%= N (N-1)/2
\, \, \mbox{times.}$ 
%\]
%two-soliton 
%pair collisions take place in a given order. 
%in arbitrary order. 
%Then, 
What will 
%is 
%%asymptotic form 
%To determine the 
the final state be 
%like
under this assumption? 
%$\,$? 
%at $t \to +\infty$ ? 
To answer this, we 
%We 
%should 
note the following two points: 
%facts: 
%should remark the following facts: 
%two observations:
%\begin{enumerate}
\begin{itemize}
%
%\item[(i)]
\item
%At all times, 
The 
%subscript 
set of subscripts 
in each soliton's $\{ \; \}$ 
is 
always equal to 
%equals 
the entire 
%whole 
%subscript 
set of subscripts of 
the next soliton to the right.
%on the next right. 
%Thus, we can apply 
This ensures that 
Proposition~\ref{joverk} is 
%always 
applicable 
%can be applied to 
%for 
to every pair collision. 
%to be considered. 
%is applicable to 
%\item[(ii)]
\item
Soliton-$n$ will 
overtake solitons-$1$, $\cdots$, $n-1$ and 
will be 
%is 
overtaken by solitons-$n+1$, $\cdots$, $N$. 
%\end{enumerate}
\end{itemize}
%Then 
%Then, 
We see that, regardless of the 
%given 
order of the pair collisions, 
%we see that in the final state 
%at $t \to + \infty$, 
solitons-$1$, $\cdots$, $N$ are finally 
distributed along the $x$-axis as 
%line up as 
\[
\vecq_{\{2, \cdots, N \}, 1},  \cdots,
\vecq_{\{n+1, \cdots, N \}, n}, \cdots,
\vecq_{\{ \;\}, N}.
\]
%regardless of the order of pair collisions. 
%on $-\infty < x < \infty $, 
%along the $x$-axis, 
This final state 
%perfectly 
%entirely 
%surely 
%coincides with 
is exactly the same as 
the 
%actual 
asymptotic 
%form 
behavior of the 
$N$-soliton solution 
%as 
in the $t \to + \infty$ limit (see Fig.~\ref{Asym-N}). 

\hfill
{\bf Q.E.D.}\ 
%$\Box$
\vspace{6mm}
\\
\noindent
{\bf Remark~1.} 
%We remark that 
%note 
%It is now clear from 
Theorem~\ref{main_theorem}, 
%and 
%in combination 
together with 
%property of a two-soliton collision 
%the collision laws of two solitons 
(\ref{YBsol}), 
%(see \ref{YBsol}), 
%was not clear 
%evident 
%at the stage of 
%from Proposition~\ref{asym_N_sol} 
%shows 
demonstrates that 
%any 
the 
%an 
%given 
initial state 
of $N$ solitons 
uniquely determines 
%its 
%the 
%corresponding 
%the 
its final state. 
%after an $N$-soliton collision. 
%evident 
%at the stage of 
That is, an $N$-soliton collision 
described 
%determined 
%defined 
by Proposition~\ref{asym_N_sol} 
%completely 
%described 
%is expressible 
%in terms of 
%in the form of 
%by 
defines 
a mapping 
from the initial state 
%of $N$ solitons 
%into 
to the final state. 
%acting on 
%%the space of $N$-soliton states. 
%the state of $N$ solitons. 
%We note that 
This fact 
%does not seem to be evident 
is not evident 
from 
%in 
Proposition~\ref{asym_N_sol}. 
\vspace{5mm}
\\
%\newpage \noindent
{\bf Remark~2.} 
%It is now easy to see that 
We can now 
prove 
%verify 
that 
%Now we can show that 
%easily seen that 
the 
%pair-collision 
mapping 
%defined 
%given by 
(\ref{YBsol}) 
%describing the collision laws of two solitons 
%given by (\ref{})
%from collision laws of two solitons 
%is shown to 
satisfies the property (b) 
or, equivalently 
%and 
the Yang-Baxter property (b'), 
%mentioned 
as stated in the introduction. 
%Introduction. 
%In fact, 
%we only need 
%For this purpose, 
It 
%only remains 
%suffices 
is sufficient 
to verify 
%show 
%see only the following two points:
%claim 
%see 
%observe 
%say 
%demonstrate 
%note 
%only 
the following: 
%point: 
%which also is not evident from Proposition~\ref{asym_N_sol}: 
%two points: 
%facts: 
%
\begin{center}
\vspace{1.5mm}
%\hspace{5mm}
\begin{minipage}{14cm}
%
%\begin{itemize}
%the final state is uniquely determined by the initial state. 
%of them. 
%\item[]
The initial polarization vectors 
%state 
of $N$ solitons 
%set of 
%polarization 
%state 
%vectors 
%
%i.e.\ 
%of $N$ solitons 
%\phi_{\{1, \cdots, n-1 \}, n} \bigr) 
%\e^{-\i \Thet_n} \vecu_{\{1, \cdots, n-1\}, n}.
$(\vecu_{\{1, \cdots, N-1\}, N}, \cdots, 
%\vecu_{\{1, \cdots, n-1\}, n}, \cdots, 
\vecu_{\{ \;\}, 1})$ 
%expressed 
%given by 
given in 
%defined by 
%in 
Proposition~\ref{asym_N_sol} can 
%become 
%take 
%give 
%any polarization state, 
%vectors , up to a choice of 
%combination of 
%$N$ solitons, 
%i.e.\ 
%an arbitary combination of 
%y 
%be 
%set 
%take 
%admit 
be made to 
coincide with 
%reproduce 
any 
combination 
%choice 
%value in the space of the direct product 
of $N$ unit vectors in ${\mathbb C}^m$ 
%up to a choice of 
by appropriately 
%through 
%an appropriate choice of 
%changing 
varying the 
%%{\it bare} \/
bare polarization vectors 
$(\vecu_N, \cdots, \vecu_1)$. 
%$\vecu_1, \vecu_2, \cdots, \vecu_N$. 
%
%span the space obtained as the direct product of 
%$N$ one-soliton states of individual solitons. 
%\end{itemize}
%
\end{minipage}
\end{center}
\vspace{1.5mm}
%
%This is proved by induction if we consider 
%To show this, we have only to 
This is easily 
%shown 
verified 
%done 
%seen 
%clear 
if we 
%concentrate on 
consider 
%only need to 
%archieved 
%by considering 
the 
%particular 
%such 
order 
%sequence 
of the 
pair collisions 
%such 
such that 
%in which 
every soliton 
experiences 
%becomes 
%bears 
%at some stage 
%enjoys 
the bare polarization 
%state 
%, i.e.\ 
$\vt{u}_{\{ \; \}, j } (= \vt{u}_{j})$ 
once 
%once, 
%for soliton $j$ 
and 
%recall 
inductively 
use 
%invertibility 
%reversibility 
%property 
%of the pair-collision mapping (\ref{YBsol}) inductively. 
the fact 
that 
%for any 
%%fixed 
%unit vector $\vecu_{1}$, 
the mapping $\vecu_{2} \mapsto \vecu_{\{1\},2}$ 
in Theorem~\ref{two-rule} is a bijection 
on the unit sphere. 
%on the unit sphere. 
\vspace{5mm}
\\
{\bf Remark~3.} 
%Consequently, 
%Specifically, 
%Remark~2 
The validity 
%effectiveness 
of the Yang-Baxter property 
allows us to extract 
%(b') 
%means 
%%says 
%that the pair-collision mapping 
%%(\ref{}) 
%%in 
%of the Manakov model 
%(\ref{cNLS}) 
%gives 
%is 
a new ``set-theoretical'' solution to the 
quantum Yang-Baxter equation (cf.\ Refs.~\citen{Dri} and~\citen{Ves}). 
%In particular, 
To be more specific, 
%a ``set-theoretical'' solution 
%from (\ref{}) 
%the mapping 
%which maps 
%definite, 
%i.e.\ 
%just by 
%viewing 
regarding 
(\ref{YBsol}) 
as the 
%a 
mapping 
%from 
$(\vecu_{\{1\},2}, \vecu_1) \mapsto 
%into 
%\mapsto 
%\longmapsto 
%\phi_{12}, 
(\vecu_{2}, \vecu_{\{2\},1})$, 
we 
%can extract 
obtain 
%it 
%as 
%from the initial pair of 
%defined by 
%it 
%gives a parameter-dependent 
a nontrivial 
%``set-theoretical'' 
solution to the 
%qunatum 
parameter-dependent Yang-Baxter equation 
%defined 
for 
%the 
mappings that 
%acting 
%which 
act on the direct product of 
two (complex) unit 
%polarization 
vectors. 
%of unit length, namely, 
%Needless to say, 
Naturally, 
%It goes without saying that 
we can 
%slightly 
extend 
%generalize 
this solution further 
%so that 
%incorporating 
%by extending the domain of the mapping 
%in such a way that 
by 
%including 
adding 
%the 
information concerning 
%on 
%space 
%adding 
%so that 
%the position of the soliton centers 
the center positions of solitons. 
%a displacement of the soliton centers 
%of 
%is also included. 
%in the mapping. 
However, this 
extension 
%does not seem to be 
is not very 
%interesting
intriguing, because 
%since
%the position of solitons 
%centers 
the net change 
%variation 
%caused 
wrought by 
the mapping 
%itself 
does not depend on 
%is independent of 
the 
%newly 
added information. 
%the added information doed not play any essential role in the mapping. 
%the mapping itself is not 
%added information do not play any role in the mapping itself. 
%affect the mapping itself. 
%such exntension 

%
\section{Concluding remarks}
\setcounter{equation}{0}
\label{}

In this paper, we have investigated 
%elucidated 
%properties 
%the mechanism of 
soliton collisions 
%properties 
in the Manakov model 
%(\ref{cNLS}) 
%with 
for the 
general 
case of 
$m$ components 
(\ref{cNLS}) 
%We took the most 
%by 
using a straightforward approach. 
We first derived 
%an exaxt formula for 
the general 
$N$-soliton solution of the Manakov model 
from that of the matrix NLS equation (\ref{mNLS}) 
through a 
simple 
%natural 
reduction.\cite{PhD} \ We considered the limits $t \to \mp \infty$ 
for the $N=2$ case
%with the 
and obtained the collision laws of two solitons in the Manakov model. 
Next, we considered the same limits for the 
case of general $N$ 
%case 
and obtained the asymptotic 
%forms 
behavior of the $N$-soliton solution. 
%as $t \to \mp \infty$. 
%for the first time. 
%The asymptotic forms were 
We were able to 
%could 
diagram 
%expressed 
the asymptotic 
%forms 
behavior in 
%a 
the simplest way 
%concisely 
%with the help of 
%by introducing 
%using 
in terms of the quantity 
%abbreviation 
$\vecq_{\{i_1, \cdots, i_{n-1} \}, i_n}$ 
defined by (\ref{q-def}) (see Fig.~\ref{Asym-N}). 
Taking 
%the maximum 
advantage of this, 
%abbreviation, 
we proved 
with a simple combinatorial treatment 
%discussion 
that an $N$-soliton collision in the Manakov model can be 
%is 
%in the Manakov model 
%nothing but the superposition of 
factorized into a nonlinear superposition of 
$N \choose 2$ 
%$\frac{N (N-1)}{2}$ 
%$N (N-1)/{2}$ 
pair collisions in arbitrary order. 
This 
%finally 
%can 
%now 
%removes 
clears up 
the longtime 
%longstanding 
misunderstanding that 
%an $N$-soliton collision does not reduce to a pair collision \cite{}. 
%suspicion that 
%or 
%This establishes that 
multi-particle effects 
%may 
exist in the Manakov model. 

%For the bright $N$-soliton solutions of the Manakov model (\ref{cNLS}) 
%derived through the ISM, 
%%via the inverse scattering method, 
%we investigated 
%%considered 
%the asymptotic forms 
%at $t \to \mp \infty$. 
%%\item[(ii)]
%We mention that 
%
This result 
%conclusion 
%The main result of this paper 
is far from 
%being 
trivial 
%at all 
%not 
%obvious 
in the 
%general 
$m \ge 2$ 
%-component
case. 
%the following reason. 
%
%factorizabiitiy 
%main result of this paper
%\item
% 
In the 
%simplest 
%single-component 
$m=1$ 
%(usual 
case (scalar NLS), 
%Thus 
all the soliton parameters 
%which 
that play an essential role 
%determine 
in the collision laws 
%the effect of a pair collision 
(in the notation of this paper, 
$\z_1, \z_2, \cdots, \z_N$) 
are invariant 
%do not change 
in time. 
%of the collisions. 
%are invariant by the collision. 
%The 
%Indeed, 
A pair collision results only in 
a displacement of the soliton centers 
and a shift of the phases, 
%a change of phases, 
%the initial position and initial phase
which will not 
%influence 
%affect 
%play a role in 
%the laws 
change the effects 
of 
%other 
%later 
future pair collisions. 
%by a 
%%any 
%pair collision. 
%Therefore, 
Thus, 
%Then, 
%in the single-component case, 
%it is obvious that 
%the total effects 
%the sum of 
%a 
a superposition of 
$N \choose 2$ 
%$\frac{N (N-1)}{2}$ 
pair collisions 
%in any 
%does not depend on the order of pair collisions. 
%is independent of the collision order. 
gives the same results for every order of the pair collisions. 
%and 
%Then 
%which seems to makethe proof of 
%Then, 
%Consequently, 
%In such a case, 
It is not 
%a hard task 
difficult to prove 
%know whether 
in this case 
that an $N$-soliton collision 
%in fact 
reduces to 
%is factorized into 
a pair collision.\cite{ZS,Novikov,FT} \ In contrast, in the 
%the factorization theorem. 
%rather easier. 
%quite easy. 
%much easier. 
%although it only 
%does not directly 
%proves a part of the factorization theorem. 
%in the single-component case. 
%gives the same result. 
%
%in an $N$-soliton collision is taken as 
%additive. 
%multi-component (
%general 
$m \ge 2$ case, 
a pair 
%(two-soliton?) 
collision results in a change of 
%soliton parameters 
the polarization vectors. 
%which will 
%will 
This 
changes the effects of 
%other 
%later 
future pair collisions completely. 
%essentially. 
Therefore, 
%Thus, 
it was not 
%evident 
obvious 
%until 
before the present 
%this 
work that 
%whether 
%at all obvious that 
%a sum of 
a nonlinear superposition of 
$N \choose 2$ 
%$\frac{N (N-1)}{2}$ 
pair collisions 
%in any 
%does not depend on the order of pair collisions. 
%leads to 
%yields 
gives the same results for every order of 
the pair collisions 
or that it exactly coincides with 
%the effect of 
an $N$-soliton collision. 
%effect, 
%regardless of the collision order. 
%\item
The key 
to proving 
%which we 
%used to prove 
these facts 
is 
%facts was 
%complete the proof of 
%arrive at 
%For proving 
%To complete 
%the factorization theorem, 
%main 
%of this paper, 
%we need to show 
%encountered 
%the 
a highly nontrivial relation 
%(\ref{dd-dd}) 
%for (generalized) 
among determinants and extended determinants, 
given in 
Lemma~\ref{Her-pro}. 
%where $d_{il}$ are defined by (\ref{cd-def}). 
This 
%also 
implies 
%a 
the possibility that 
%we obtain 
%find 
%that we may obtain 
%such 
some new relations similar to Lemma~\ref{Her-pro} 
%(\ref{dd-dd}) 
%for (generalized) determinants 
%are 
can be 
%will be 
obtained 
%determinants 
%or generalized determinants 
%of a particular form 
%by 
%studying 
through the investigation of 
%investigating 
%considering 
%the mechanism of 
soliton collisions 
%factorization of 
%problem for 
%an $N$-soliton collision 
%multi-soliton 
in multi-component integrable systems. 
%into pair collisions. 
%
%
%Finally, we 
%%would like to 
%briefly mention 
%%that 
%the potential applicability of 
%the results obtained in this paper. 
%%in nonlinear optics. 

%It has been pointed out by 
Very recently, 
Steiglitz and coworkers\cite{Steig1,Steig2} 
proposed 
%the idea 
%pointed out 
%theoretically 
%showed 
that 
%, in theory, 
soliton collisions in the Manakov model 
%(\ref{cNLS}) 
can be 
%used to 
%%a method for implementing 
%implement 
utilized to carry out 
%for performing 
any 
%arbitrary 
computation 
%efficiently 
with beams 
in a nonlinear optical medium 
%media 
(see 
%also 
Ref.~\citen{Steig3} for the experimental foundations). 
%medium 
We believe that 
the results 
%some 
obtained in 
%of 
this paper 
%can 
%are 
will be 
useful 
to refine and reinforce 
%in refining and reinforcing 
%strengthening 
%expected to 
%refine and strengthen 
%sophisticating 
their 
interesting 
idea. 
%constructing 
%realizing 
%about 
%for an efficient optical computer. 
%based on soliton collisions. 
%all-soliton 
%give 
In particular, 
Theorem~\ref{main_theorem} 
%proved 
%ensured the 
%verifies 
supports\footnote{To be 
%more 
precise, we 
%have 
need to extend our 
%analysis 
result 
to the case in which 
%where 
%oen or some 
%there are 
some 
%groups of 
solitons 
%among the $N$ solitons 
%propagate together, i.e.\ with 
%have the same velocity and 
%%thus 
%%copropagate. 
%propagate parallel with one another. 
propagate 
%with 
at the same velocity and 
%thus 
%copropagate. 
do not collide with one another. 
%in parallel. 
%ing with same velocity. 
This 
%case 
%problem 
is beyond the scope of this paper and 
%we leave it 
%thus left 
is left 
%for 
to the reader as a future problem.
%for a future study. 
} 
%
%certifies 
%the appropriateness of 
%correctness 
%of 
their 
%implicit 
%assumption 
%ansatz 
hypothesis on 
the pairwise nature of soliton collisions 
and 
%which was assumed from the beginning in \cite{Steig1,Steig2}. 
%implicitly 
%in the Manakov model
%proved in this paper 
Proposition~\ref{asym_N_sol}, together with 
Proposition~\ref{joverk}, 
%is 
%gives 
provides a 
%good 
hint 
%starting point 
%about 
on
%for 
%designing 
how to design optical 
%logic gates 
logic operations 
%simple but 
%and 
%in a more rigorous way 
more simply and reliably 
%rigorously 
%simpler and more rigorous 
than 
%that 
%in \cite{Steig1,Steig2}. 
%in 
the method proposed in Ref.~\citen{Steig2}. 
%finding new designs of optical 

\section*{Acknowledgements}
%\setcounter{equation}{0}
%One of the authors (TT) 
The author is grateful to Prof.\ Miki Wadati, 
Prof.\ Ryu Sasaki, 
Prof.\ Tetsuji Tokihiro, Prof.\ Jianke Yang, 
%and 
Dr.\ Ken-ichi Maruno and 
%the 
%two 
anonymous 
referees for 
their 
useful comments. 
%and helpful suggestions. 
%TT also apreciates 
This research was supported in part by 
a JSPS Research Fellowship for Young Scientists.


\begin{thebibliography}{99}

\bibitem{Meny}
C.~R.~Menyuk, 
%Pulse propagation in an elliptically birefringent Kerr medium, 
{IEEE J.\ Quantum Electron.} {\bf 25} (1989), 2674.
%--2682.

\bibitem{Agra}
G.~P.~Agrawal, {\it Nonlinear Fiber Optics}, 2nd ed.\ 
(Academic Press, San Diego, 1995). 

\bibitem{Akhm}
N.~N.~Akhmediev and A.~Ankiewicz, {\it Solitons: Nonlinear 
pulses and beams} (Chapman \& Hall, London, 1997).

\bibitem{Manakov}
S.~V.~Manakov, 
%On the theory of two-dimensional stationary self-focusing of 
%electromagnetic waves, 
{Sov.\ Phys.\hspace{2pt}-JETP} {\bf 38} (1974), 248.
%--253.

\bibitem{Makhankov}
V.~G.~Makhan'kov and O.~K.~Pashaev, 
%Nonlinear Schr\"{o}dinger equation with noncompact isogroup, 
{Theor.\ Math.\ Phys.} {\bf 53} (1982), 979.
%--987.

\bibitem{Rad}
R.~Radhakrishnan, M.~Lakshmanan and J.~Hietarinta, 
%Inelastic collision and switching of coupled bright solitons 
%in optical fibers, 
{Phys.\ Rev.} E {\bf 56} (1997), 2213.
%--2216.

\bibitem{Shche}
V.~Shchesnovich, 
%S.\ 
%Polarization scattering by soliton-soliton collisions, 
%{\em e-print arXiv} \/
solv-int/9712020.

\bibitem{Yang}
J.~Yang, 
%Multisoliton perturbation theory for the Manakov 
%equations and its applications to nonlinear optics, 
{Phys.\ Rev.} E {\bf 59} (1999), 2393.
%--2405.

\bibitem{Elgin}
J.~P.~Silmon-Clyde and J.~N.~Elgin, 
%Dynamics of collisions between soliton trains, 
{J.\ Opt.\ Soc.\ Am.} B 
{\bf 16} (1999), 1348.
%--1353. 

\bibitem{ZS}
V.~E.~Zakharov and A.~B.~Shabat, 
%Exact theory of two-dimensional self-focusing and 
%one-dimensional self-modulation of waves in nonlinear media, 
{Sov.\ Phys.\hspace{2pt}-JETP} {\bf 34} (1972), 62.
%--69.

\bibitem{Novikov}
S.~Novikov, S.~V.~Manakov, L.~P.~Pitaevskii and V.~E.~Zakharov, 
{\it Theory of Solitons: The Inverse Scattering Method}
(Consultants Bureau, New York, 1984).

\bibitem{FT}
L.~D.~Faddeev and L.~A.~Takhtajan, 
{\it Hamiltonian Methods in the Theory of Solitons} 
(Springer, Berlin, 1987).

\bibitem{Tsuchida1}
T.~Tsuchida and M.~Wadati, 
%The coupled modified Korteweg--de Vries equations, 
{J.\ Phys.\ Soc.\ Jpn.} {\bf 67} (1998), 1175.
%--1187.

\bibitem{PhD}
T.~Tsuchida, 
``Study of multi-component soliton equations based on the 
inverse scattering method'', Ph.D thesis, Department of Physics, 
University of Tokyo (2000), 
available at 

\qquad \quad
{\texttt http://poisson.ms.u-tokyo.ac.jp/\~{}tsuchida/thesis/}
%{\it http://poisson.ms.u-tokyo.ac.jp/\~{}tsuchida/thesis/}

\bibitem{Dri}
V.~G.~Drinfeld, 
%On some unsolved problems in quantum group theory, 
%in 
%{\it Quantum groups} 
%(Leningrad, 1990) 
{\it Lecture Notes in Math.}, Vol.\ 1510 (Springer, Berlin, 1992) p.\ 1.
%--8. 

\bibitem{Ves}
A.~P.~Veselov, 
%Yang--Baxter maps and integrable dynamics, 
{Phys.\ Lett.} A {\bf 314} (2003), 214.
%--221. 

\bibitem{Zak}
V.~E.~Zakharov, 
%Kinetic equation for solitons, 
{Sov.\ Phys.\hspace{2pt}-JETP} {\bf 33} (1971), 538.
%--541. 

\bibitem{Kul}
P.~P.~Kulish, 
%Factorization of the classical and the 
%quantum $S$ matrix and conservation laws, 
{Theor.\ Math.\ Phys.} {\bf 26} (1976), 132.
%--137. 

\bibitem{Gon}
V.~M.~Goncharenko, 
%Multisoliton solutions of the matrix KdV equation, 
{Theor.\ Math.\ Phys.} {\bf 126} (2001), 81.
%--91. 

\bibitem{Ohta}
Y.~Ohta, 
%Wronskian-type solutions of soliton equations, 
{RIMS k\^{o}ky\^{u}roku} {\bf 684} (1989), 1 
%--17 
[in Japanese].

\bibitem{Chow}
K.~W.~Chow and D.~W.~C.~Lai, 
%Coalescence of wavenumbers and exact solutions for a system of 
%coupled nonlinear Schr\"{o}dinger equations, 
{J.\ Phys.\ Soc.\ Jpn.} {\bf 67} (1998), 3721.
%--3728.

\bibitem{Trubatch1}
M.~J.~Ablowitz, Y.~Ohta and A.~D.~Trubatch, 
%On discretizations of the vector nonlinear Schr\"{o}dinger equation, 
{Phys.\ Lett.} A {\bf 253} (1999), 287.
%--304.

\bibitem{Nogami}
Y.~Nogami and C.~S.~Warke, 
%Soliton solutions of multicomponent nonlinear Schr\"{o}dinger equation, 
{Phys.\ Lett.} A {\bf 59} (1976), 251.
%--253.

\bibitem{Akhmed}
A.~A.~Sukhorukov and N.~N.~Akhmediev, 
%Soliton X-junctions with controllable transmission, 
%{\em e-print arXiv} \/
nlin.PS/0208043.

\bibitem{Park}
Q-H.~Park and H.~J.~Shin, 
%Systematic construction of vector solitons, 
{IEEE J.\ Sel.\ Top.\ Quantum Electron.} 
{\bf 8} (2002), 432.
%--439. 

\bibitem{Ghosh}
A.~Borah, S.~Ghosh and S.~Nandy, 
%Interaction of coupled higher order nonlinear Schr\"{o}dinger equation 
%solitons, 
{Eur.\ Phys.\ J.} B {\bf 29} (2002), 221.
%--225. 

%\bibitem{Kanna}
%T.\ Kanna and M. Lakshmanan: Effect of phase shift in 
%shape changing collision of solitons in coupled nonlinear 
%Schr\"{o}dinger equations, 
%{\em Eur.\ Phys.\ J.} \/B {\bf 29} (2002) 249--254. 

\bibitem{Kanna}
T.~Kanna and M.~Lakshmanan, 
%Exact soliton solutions of coupled nonlinear Schr\"{o}dinger 
%equations: Shape-changing collisions, logic gates, 
%and partially coherent solitons, 
{Phys.\ Rev.} E {\bf 67} (2003), 046617. 

\bibitem{Steudel}
H.~Steudel, 
%$N$-soliton solutions to degenerate self-induced transparency, 
{J.\ Mod.\ Opt.} {\bf 35} (1988), 693.
%--702.

\bibitem{Tsuchida2002}
T.~Tsuchida, 
%Integrable discretizations of 
%derivative nonlinear Schr\"{o}dinger equations, 
{J.\ of Phys.\ A} 
%: Math.\ Gen.} 
{\bf 35} (2002), 7827.
%--7847. 

\bibitem{GI2}
V.~S.~Gerdzhikov and  M.~I.~Ivanov, 
%Hamiltonian structure of multicomponent nonlinear 
%Schr\"{o}dinger equations in difference form, 
{Theor.\ Math.\ Phys.} {\bf 52} (1982), 676.
%--685. 

\bibitem{Tsuchida1999}
T.~Tsuchida, H.~Ujino and M.~Wadati, 
%Integrable semi-discretization 
%of the coupled nonlinear Schr\"{o}dinger equations, 
{J.\ of Phys.\ A} 
%: Math.\ Gen.} 
{\bf 32} (1999), 2239.
%--2262.

\bibitem{Vakh}
O.~O.~Vakhnenko, 
%Nonlinear beating excitations on ladder lattice, 
{J.\ of Phys.\ A} 
%: Math.\ Gen.} 
{\bf 32} (1999), 5735.
%--5748.

\bibitem{Kay}
I.~Kay and H.~E.~Moses, 
%Reflectionless transmission through dielectrics and scattering potentials, 
{J.\ Appl.\ Phys.} {\bf 27} (1956), 1503.
%--1508. 

\bibitem{WT}
M.~Wadati and M.~Toda, 
%The exact $N$-soliton solution of the Korteweg--de Vries equation, 
{J.\ Phys.\ Soc.\ Jpn.} 
{\bf 32} (1972), 1403.
%--1411.

\bibitem{GGKM2}
C.~S.~Gardner, J.~M.~Greene, M.~D.~Kruskal and R.~M.~Miura, 
%Korteweg--de Vries equation and generalizations.\ VI\@.\ Methods for
%exact solution, 
{Commun.\ Pure Appl.\ Math.} {\bf 27} (1974), 97.
%--133. 

\bibitem{Steig1}
M.~H.~Jakubowski, K.~Steiglitz and R.~Squier, 
%State transformations of colliding optical solitons and 
%possible application to computation in bulk media, 
{Phys.\ Rev.} E {\bf 58} (1998), 6752.
%--6758. 

\bibitem{Steig2}
K.~Steiglitz, 
%Time-gated Manakov spatial solitons are computationally universal, 
{Phys.\ Rev.} E {\bf 63} (2001), 016608. 

\bibitem{Steig3}
C.~Anastassiou, J.~W.~Fleischer, 
T.~Carmon, M.~Segev and K.~Steiglitz, 
%Information transfer via cascaded collisions of vector solitons, 
{Opt.\ Lett.} {\bf 26} (2001), 1498.
%--1500. 

\end{thebibliography}
\end{document}